\documentclass[3p]{elsarticle}
\usepackage{color}
\usepackage{xspace}
\usepackage{pdfwidgets}
\usepackage{amsmath}	
\usepackage{amssymb}	
\usepackage{multicol}        
\usepackage{bm}		
\usepackage{pdflscape}	
\usepackage{hyperref}
\usepackage{tikz} 
\usetikzlibrary{shapes,arrows,fit,backgrounds}
\usepackage{subfigure}
\usepackage{longtable}
\usepackage{array}
\usepackage{float}
\usepackage{enumerate}

\let\new=\newcommand
\new{\diff}{{\rm d}}

\begin{document}

\title{Dynamics of accretion and winds in tidal disruption events}

\author[1]{T. Mageshwaran
}
\ead{t.mageshwaran@tifr.res.in}
\author[2]{A. Mangalam\corref{cor1}
}
\ead{mangalam@iiap.res.in}
\cortext[cor1]{Corresponding author}

\address[1]{Tata Institute of Fundamental Research, Mumbai, India}
\address[2]{Indian Institute of Astrophysics, Bangalore, India}

\date{\today}

\begin{abstract}

We have constructed self-similar models of a time-dependent accretion disk in both sub and super-Eddington phases with wind outflows for tidal disruption events (TDEs). The physical input parameters are the black hole (BH) mass $M_{\bullet}$, specific orbital energy $E$ and angular momentum $J$, star mass $M_{\star}$ and radius $R_{\star}$. We consider the sub-Eddington phase to be total pressure (model A1) and gas pressure (model A2) dominated. In contrast, the super-Eddington phase is dominated by radiation pressure (model B) with Thomson opacity. We derive the viscosity prescribed by the stress tensor, $\Pi_{r\phi}\propto \Sigma_d^b r^d$ where $\Sigma_d$ is the surface density of the disk, $r$ is the radius and $b$ and $d$ are constants. The specific choice of radiative or $\alpha$ viscosity is motivated, and its parameters are decided by the expected disk luminosity and evolution time scale being in the observed range. The disk evolves due to mass loss by accretion onto the black hole and outflowing wind, and mass gain by fallback of the debris; this results in an increasing outer radius. We have simulated the luminosity profile for both sub and super-Eddington disks. As an illustrative example, we fit our models to the observations in X-ray, UV, and Optical of four TDE events and deduce the physical parameters above.

\end{abstract}


\maketitle

\section{Introduction}

A star is tidally disrupted if its pericenter $r_p\leq r_t$ where $r_t \sim R_{\star} (M_{\bullet}/ M_{\star})^{1/3}$ is the tidal radius for BH mass less than a critical mass $\sim~3 \times 10^8 M_{\odot} $ and these events are called TDEs \citep{1975Natur.254..295H,1988Natur.333..523R}. The disrupted debris is assumed to follow a Keplerian orbit around the BH and the mass infall rate depends on the internal structure and properties of the star and follows the $t^{-5/3}$ law only at the late stages of its evolution \citep{1989IAUS..136..543P,2009MNRAS.392..332L,2013ApJ...767...25G}. The debris experiences stream collision either due to incoming stream that intersects with the outflowing stream at the pericenter \citep{1994ApJ...422..508K} or due to relativistic precession at the pericenter \citep{2013MNRAS.434..909H}. These interactions result in the circularization of the debris to form an accretion disk \citep{2013MNRAS.434..909H,2015arXiv150104635B,2015ApJ...804...85S}.

\citet{2015ApJ...814..141M} (hereafter MM15) have constructed a detailed stellar dynamical model of TDEs using parameters that include BH mass $M_{\bullet}$, specific orbital energy $E$ and angular momentum $J$, star mass $M_{\star}$ and radius $R_{\star}$ to calculate the capture rate of stars to the black hole by solving the steady-state Fokker-Planck equation and integrating over the energy- angular momentum phase space. In addition, MM15 have calculated the mass fallback rate of the debris $\dot{M}_{fb}$ as a function of energy and angular momentum of the star which includes the case of the mass fallback rate obtained by \citet{2009MNRAS.392..332L} for a star on parabolic orbit with $E=0$ and $J=\sqrt{2GM_{\bullet}r_t}$. MM15 have used the steady accretion model of \citet{2009MNRAS.400.2070S} in the case of a super-Eddington disk and standard $\alpha$ disk model for a sub-Eddington disk with time-varying accretion rate and derived luminosity profiles which are then compared with the sensitivity of the detector to calculate the survey detection rate of TDEs. The inclusion of angular momentum and energy in the TDE dynamics to obtain pericenter radius has increased the number of free parameters to fit the observations but it provides a better understanding of disruption dynamics in terms of star's orbital parameters. 

\citet{2009MNRAS.400.2070S} proposed a steady accretion model with the disk edges fixed and the fraction of mass outflow caused by the strong radiative pressure in the super-Eddington phase constant. This model was used by MM15 to calculate the detectable rates by optical and X-ray surveys. \citet{2011MNRAS.413.1623D} have constructed a super-Eddington slim disk model by solving the steady disk equation with $\alpha$ viscosity where the vertical structure of the super-Eddington disk is divided into two regions, a hydrostatic region which includes the porous atmosphere, and the region of a continuum driven wind and obtained the fraction of mass outflowing wind $f_{out}$ in terms of accretion rate $\dot{M}_{a}$. We tried to fit the steady accretion model to the observations in various spectral bands and found that it does not give a good fit; this necessitates the treatment of time-evolving disks. We now discuss some time-dependent disks, used in previous models.

\citet{2011ApJ...736..126M} have numerically solved the axisymmetric disk equation with the time-dependent mass input at the outer radius $r_p$ due to fallback debris, without an outflowing wind and the viscosity prescription given by $\nu=2\pi r u_{\phi}/R_{y}$, where $R_y$ is the Reynolds number characterizing the flow and $u_{\phi}$ is the azimuthal velocity. They have considered the disk edges to be constant and showed that the mass accretion rate follows the fallback rate at the late stage. \citet{2013MNRAS.430L..45A} applied this model to the PS1-10jh observations and deduced a black hole mass of $M_{\bullet}=6.3 \times 10^6 ~ M_{\odot}$. \citet{2020MNRAS.496.1784M} constructed a time-dependent relativistic thin accretion model with an $\alpha-$viscosity and a mass input at the constant outer radius for both full and partial disruption TDEs. They obtained the late-time evolution of luminosity to be $L \propto t^{-1.8}$ for full disruption TDEs and $L \propto t^{-2.3}$ for partial disruption TDEs. The luminosity declines faster than that inferred from $L \propto \dot{M}_{fb}$. However, this model does not include the wind and we plan to explore relativistic time-dependent models with the wind in the future.

\citet{2001A&A...379.1138M} assumed a general viscosity prescription $\Pi_{r\phi} \propto \Sigma_d^b r^d$ where $\Sigma_d$ is the surface density of the disk, $r$ is the radius and $b$ and $d$ are constants and solved for a non-relativistic collapsing disk with total angular momentum constant and obtained a self-similar solution which subsumes and generalizes the earlier time-dependent self-similar disk solution \citep{1987MNRAS.225..607L,1981ARA&A..19..137P, 1990ApJ...351...38C}. \citet{2003BASI...31..207M} has obtained a self-similar solution for a non-relativistic super-Eddington collapsing disk with radiative viscosity. 

In this paper, we construct the self-similar models of time-dependent and non-relativistic accretion disk for both sub-Eddington (models A1 and A2) and super-Eddington disks (model B) with an outflowing wind with a general viscosity prescription. We consider a sub-Eddington disk with Thomson opacity and $\Pi_{r\phi}=\alpha_s P H$, where $H$ is disk scale height and $P$ is the total pressure in model A1 and gas pressure $P_g$ in model A2. We consider a radiation pressure dominated disk with Thomson opacity in the super-Eddington regime and use the radiative viscous stress to obtain the constants $b$ and $d$. In our case, the total angular momentum of the disk is a function of time due to the introduction of the fallback and thus our TDE solutions differ from the solution obtained by \citet{2001A&A...379.1138M}. We construct the super-Eddington model for a radiative slim disk whereas \citet{2014ApJ...784...87S} have constructed the model of TDE disk with and without fallback from disrupted debris by developing a self-similar structure of a non-radiative, advective disk with an outflowing wind and using the self-similar solution of \citet{1990ApJ...351...38C} for a radiative thin disk with total angular momentum constant. In this work, we calculate the structure of outflowing wind using the vertical momentum equation and obtained a relation between mass out flowing rate $\dot{M}_w$ and $\dot{M}_{a}$, which is different from \citet{2011MNRAS.413.1623D}.

We first discuss the mass fall-back rate of disrupted debris that follows a Keplerian orbit in terms of physical parameters black hole (BH) mass $M_{\bullet}$, specific orbital energy $E$ and angular momentum $J$, star mass $M_{\star}$ and radius $R_{\star}$. The vertical structure of the sub-Eddington disk is similar to standard thin disk whereas, the vertical structure of the super-Eddington disk is taken to be nearly in a quasi-hydrostatic equilibrium up to the photosphere above which there is an outflowing wind. The time scale of vertical equilibration is smaller than the radial inflow time scale. The vertical variation is therefore considered to be steady over the primary longer time scale of radial inflow which is described by a time-dependent disk. This is an improvement over previous studies (MM15, \citet{2009MNRAS.400.2070S}) where no time-varying structure was assumed. Here, the vertical structure is subject to variation as determined by the physical parameters of the time-dependent disk models taking into account the possible physical conditions in the sub and super-Eddington phases. The vertical structure is appropriately set up for the case of the super-Eddington with a built-in wind structure. The self-similarity assumption imposes some restrictions on the properties of the disk but in \S \ref{sssol}, we show that the choice of the viscous stress, energy flow, and vertical structure are self-consistent and post justified in terms of fits with observations. The insight offered by this advanced semi-analytic model outweighs any underlying approximations and can be verified by simulations that are planned in the future. By comparing the luminosity from super-Eddington disk with radiative viscosity (model B) and $\alpha$ viscosity, we show that the radiative viscosity dominates over the $\alpha$ viscosity when the disk is dominated by radiation pressure. The sub-Eddington disk is taken to have $\alpha$ viscosity and Thomson opacity with total pressure in model A1 and gas pressure in model A2. We then construct a self-similar solution for a disk where it is assumed that the time scale for redistribution of the infalling matter in the disk (assuming a pro-rated distribution) is smaller than the radial inflow such that the matter is instantaneously distributed in the disk. We also show that for a self-similar disk, the advective rate for a super-Eddington disk is smaller than the viscous heating rate. The light curve obtained for models A1, A2, and B is fit to the observations to derive the physical parameters. Given the complexity of the TDE system, the fact that the time-dependent models fit the observations well and hence are superior to the quasi-steady models [MM15, \citet{2009MNRAS.400.2070S}] is the key result of this paper.

\begin{figure*}
\begin{center}
\subfigure[]{\includegraphics[scale=0.40]{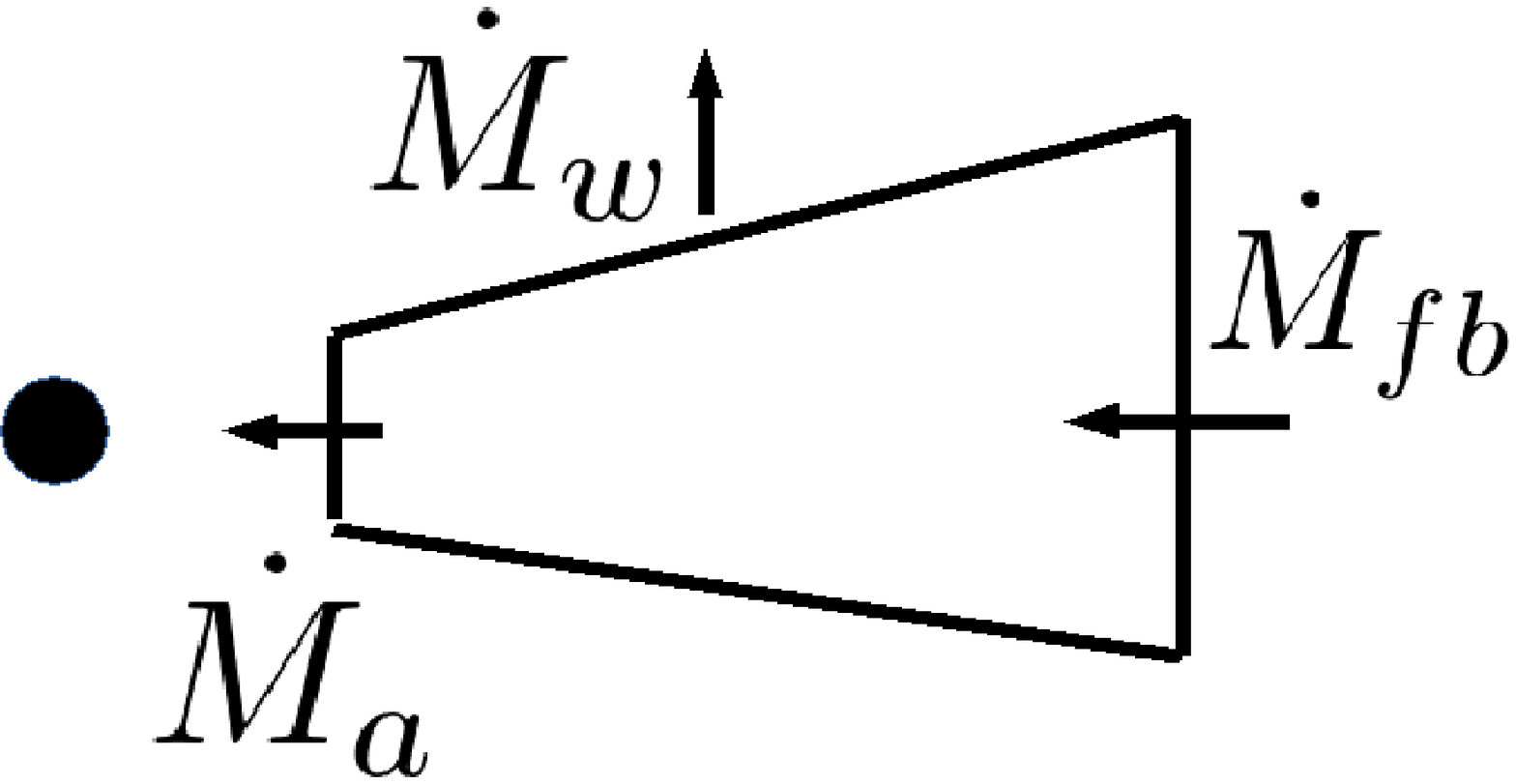}}
\subfigure[]{\includegraphics[scale=0.38]{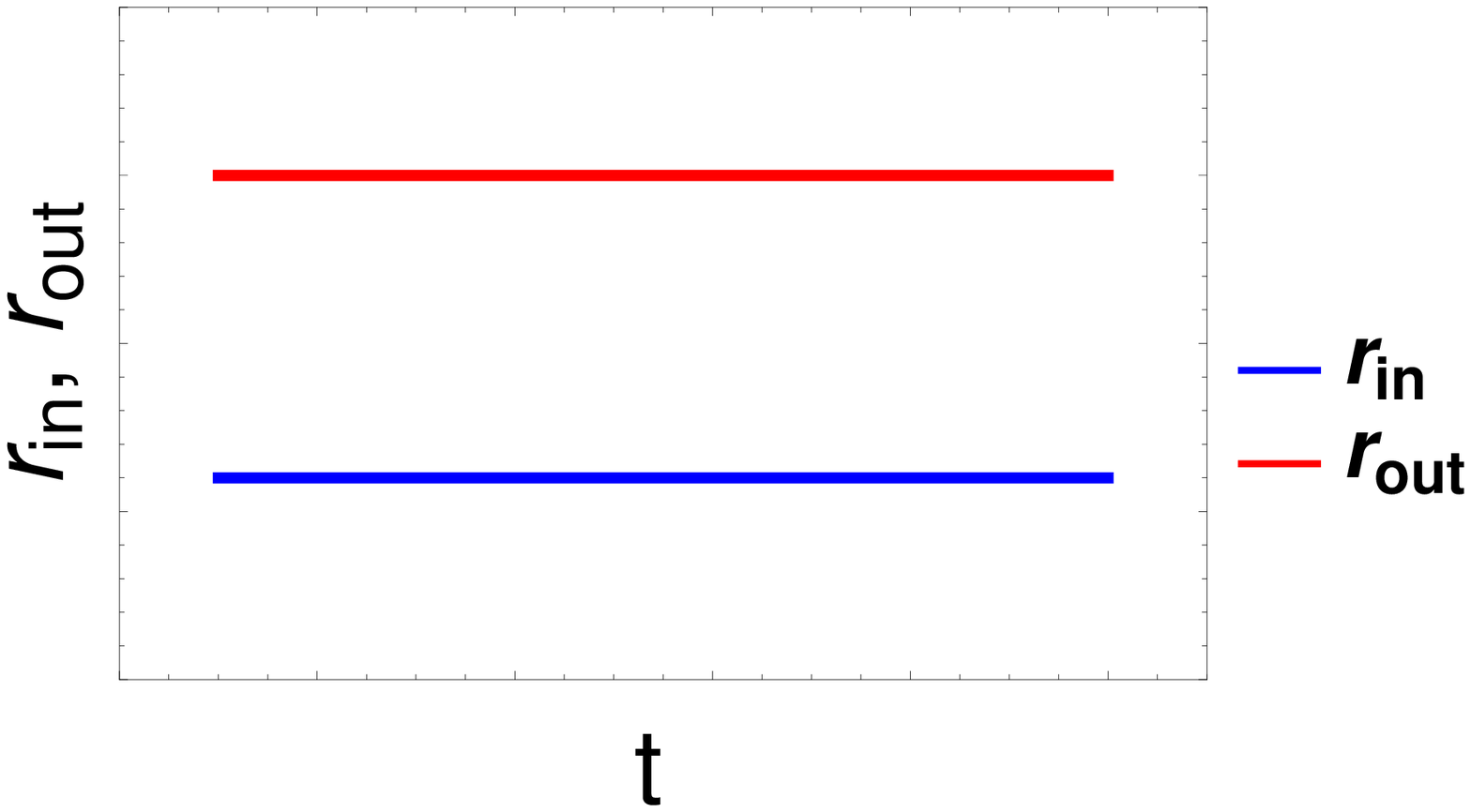}}
\subfigure[]{\includegraphics[scale=0.38]{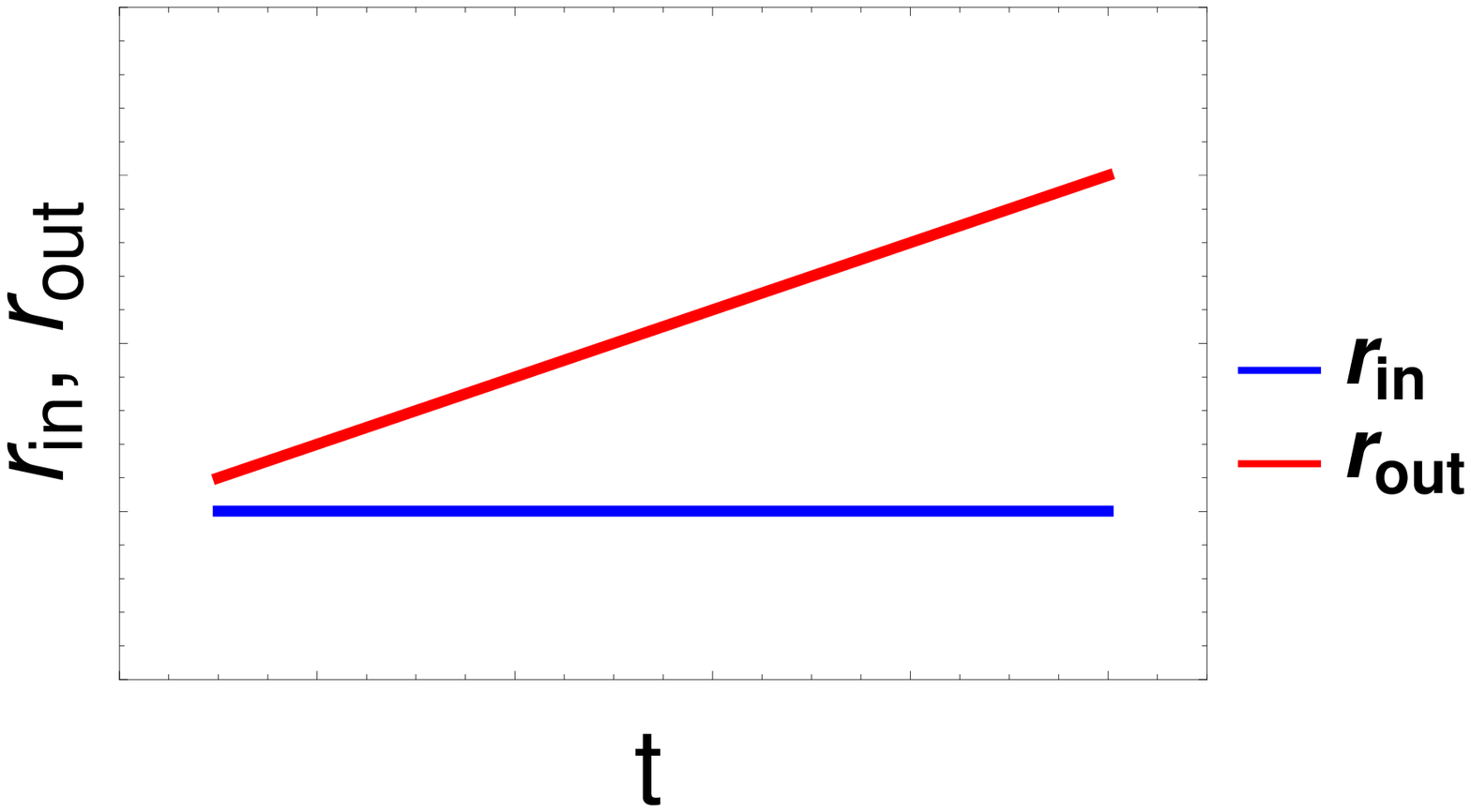}}
\end{center}
\caption{ (a) A schematic structure of accretion disk showing the accretion rate, $\dot{M}_a$, mass fallback rate, $\dot{M}_{fb}$, and mass outflow rate, $\dot{M}_w$. The evolution of the disk inner radius $r_{in}$ (blue) and outer radius $r_{out}$ (red) are shown in (b) and (c). These figures are used to compare the various accretion models as shown in Table \ref{compar}.}
\label{dcomp}
\end{figure*}

The plan of the paper is as follows: In \S \ref{mfr}, we discuss the physics of tidal disruption and the mass fallback rate of the disrupted debris. The time-dependent equations, the structure of accretion disks that include the accretion and wind, and the viscosity mechanism in the accretion disk will be discussed in \S \ref{ptdes}. The self-similar solutions of the sub and super-Eddington disks are discussed in \S \ref{sssol} and a more detailed calculation is given in the appendix \ref{sssolt}. A useful formulary is presented in Table \ref{modtab} of the models A1, A2 and B derived in (\S \ref{subres}, \ref{modelB}) that is easy to implement as compared to the less transparent and tedious numerical simulations. The key finding of the time-dependent behaviour is discussed in \S \ref{tbeh}. The fits to observations discussed in \S \ref{obsfits}, yield parameters of the star, and the black hole which can be a useful discriminant among the models and the derived parameters can be used for statistical studies of the demographics of black holes. The ongoing (ASAS-SN, iPTF, MAXI, ASTROSAT-SSM, OGLE, Pan-STARRS, eROSITA) and upcoming (LSST, ILMT) missions will be detecting a large sample of TDEs which will provide a rich source for the building of black hole distributions in mass and redshift space. In \S \ref{discussion}, we discuss our approach and its results in comparison with previous studies and point to the tangible advantages of our models A and B. We present a summary in \S \ref{summary} and the conclusions in \S \ref{conclusion}. A schematic representation of the disk geometry and the evolution of the outer radius is shown in Fig \ref{dcomp}. The various accretion models in the literature with and without steady structure are compared in Table \ref{compar}. The glossary of symbols used in the paper is given in Table \ref{gloss}.

\begin{landscape}
\begin{table}
\caption{Various accretion models in the literature with and without steady structure are compared. A schematic representation of the disk geometry and evolution of the outer radius is shown in Fig \ref{dcomp}. The accretion model of TDE by MM15 includes the dynamical parameters $E$ and $J$ whereas the other models of TDE have assumed the initial orbit of the disrupted star to be parabolic $E=0$ and have not included the angular momentum $J$.}
\label{compar}
\scalebox{0.8}{
\begin{tabular}{c c c c c c c c l}
\hline
&&&&&&&&\\
\# & Reference & Application  & Disk structure & $\dot{M}_{fb}$ & $\dot{M}_a$ & $\dot{M}_w$ & Edge radii &  Assumptions  \\
\hline
&&&&&&&&\\
1 &\vtop{\hbox{\strut Mangalam (2001)}\hbox{\strut \& (2003)}} &\vtop{\hbox{\strut BH formation}\hbox{\strut at $z=5$}} & \vtop{\hbox{\strut Time dependent disk}\hbox{\strut Sub-Eddington } \hbox{\strut Super-Eddington }} & \vtop{ \hbox{\strut}\hbox{\strut None}\hbox{\strut None}} & \vtop{ \hbox{\strut}\hbox{\strut None}\hbox{\strut None}} & \vtop{ \hbox{\strut}\hbox{\strut None}\hbox{\strut None}} & \vtop{\hbox{\strut Evolving} \hbox{\strut Fig \ref{dcomp}c}} & \vtop{\hbox{\strut Total disk angular momentum is constant}\hbox{\strut $\Pi_{r\phi} \propto \Sigma_d^b r^d$: $\alpha$ disk, gravitational Instability,}\hbox{\strut ~~~~~~~~~~~~~~~~ magnetic stress, radiative stress}\hbox{\strut Super-Eddington: self gravitating disk}} \\
\hline
&&&&&&&&\\
2 & \vtop{\hbox{\strut Strubbe \&}\hbox{\strut Quataert (2009)}\hbox{\strut }\hbox{\strut Lodato \& }\hbox{\strut Rossi (2011)}} &\vtop{\hbox{\strut TDE}} & \vtop{\hbox{\strut Steady structure disk}\hbox{\strut Sub-Eddington } \hbox{\strut Super-Eddington }} & \vtop{ \hbox{\strut}\hbox{\strut None}\hbox{\strut None}} & \vtop{ \hbox{\strut}\hbox{\strut $\propto t^{-5/3}$}\hbox{\strut $\propto t^{-5/3}$}} & \vtop{ \hbox{\strut}\hbox{\strut None}\hbox{\strut $\propto t^{-5/3}$}} & \vtop{\hbox{\strut Static} \hbox{\strut Fig \ref{dcomp}b}} & \vtop{\hbox{\strut $\alpha$ disk: Standard Shakura-Sunyaev viscosity}\hbox{\strut Sub-Eddington disk: radiative thin disk}\hbox{\strut Super-Eddington disk: slim disk with }\hbox{\strut ~~~adiabatic and spherical wind outflow}} \\
&&&&&&&&\\
\hline
&&&&&&&&\\
3 & \vtop{\hbox{\strut Armijo \&}\hbox{\strut Pacheco (2011)}} &\vtop{\hbox{\strut TDE}} & \vtop{\hbox{\strut Time dependent disk}\hbox{\strut Sub-Eddington }} & \vtop{ \hbox{\strut}\hbox{\strut $\propto t^{-5/3}$}} & \vtop{ \hbox{\strut}\hbox{\strut $\propto t^{-5/3}$}} & \vtop{ \hbox{\strut}\hbox{\strut None}} & \vtop{\hbox{\strut Static} \hbox{\strut Fig \ref{dcomp}b}} & \vtop{\hbox{\strut $\beta$ viscous model, $\nu=2\pi r u_{\phi}/R_{y}$ }\hbox{\strut Mass fallback at outer radius}\hbox{\strut Numerical simulation.}} \\
&&&&&&&&\\
\hline
&&&&&&&&\\
4 & \vtop{\hbox{Shen \& Matzner (2014)}} &\vtop{\hbox{\strut TDE}} & \vtop{\hbox{\strut Time dependent disk}\hbox{\strut Sub-Eddington } \hbox{\strut Super-Eddington }} & \vtop{ \hbox{\strut}\hbox{\strut $\propto t^{-5/3}$}\hbox{\strut $\propto t^{-5/3}$}} & \vtop{ \hbox{\strut}\hbox{\strut $\propto t^{-19/16}$}\hbox{\strut $\propto t^{-\eta}$}} & \vtop{ \hbox{\strut}\hbox{\strut None}\hbox{\strut $\propto t^{-\eta_1}$}} & \vtop{\hbox{\strut Evolving} \hbox{\strut Fig \ref{dcomp}c}} & \vtop{\hbox{\strut $\Pi_{r\phi} \propto \Sigma_d^b r^d$}\hbox{\strut Sub-Eddington with constant angular momentum}\hbox{\strut Super-Eddington: non radiative advective disk}\hbox{\strut $\eta$ and $\eta_1$ are function of ratio of wind to disk}\hbox{\strut angular momentum.}} \\
&&&&&&&&\\
\hline
&&&&&&&&\\
5 & \vtop{\hbox{MM15}} &\vtop{\hbox{\strut TDE}} & \vtop{\hbox{\strut Steady structure disk}\hbox{\strut Sub-Eddington } \hbox{\strut Super-Eddington }} & \vtop{ \hbox{\strut}\hbox{\strut None}\hbox{\strut None}} & \vtop{ \hbox{\strut}\hbox{\strut $\propto t^{-5/3}$}\hbox{\strut $\propto t^{-5/3}$}} & \vtop{ \hbox{\strut}\hbox{\strut None}\hbox{\strut $\propto f_{out}(t)t^{-5/3}$}} & \vtop{\hbox{\strut Static} \hbox{\strut Fig \ref{dcomp}b}} & \vtop{\hbox{\strut $\alpha$ disk: Standard Shakura-Sunyaev viscosity}\hbox{\strut Star's angular momentum is included}\hbox{\strut Fraction of mass outflow $f_{out}$ is obtained }\hbox{\strut from Dotan \& Shaviv (2011) }} \\
&&&&&&&&\\
\hline
&&&&&&&&\\
6 & \vtop{\hbox{Mageshwaran \&}\hbox{Bhattacharyya (2020)}} &\vtop{\hbox{\strut TDE}} & \vtop{\hbox{\strut Time dependent }\hbox{\strut relativistic thin disk. }\hbox{\strut Full disruption TDEs}\hbox{\strut Partial disruption TDEs}} & \vtop{ \hbox{\strut}\hbox{\strut}\hbox{\strut $\propto t^{-5/3}$}\hbox{\strut $\propto t^{-9/4}$}} & \vtop{ \hbox{\strut}\hbox{\strut}\hbox{\strut $\propto t^{-1.82}$}\hbox{\strut $\propto t^{-2.31}$}}  & \vtop{ \hbox{\strut}\hbox{\strut}\hbox{\strut None}\hbox{\strut None}} & \vtop{\hbox{\strut Static} \hbox{\strut Fig \ref{dcomp}b}} & \vtop{\hbox{\strut $\alpha$ disk: Standard Shakura-Sunyaev viscosity}\hbox{\strut Numerical simulation}} \\
&&&&&&&&\\
\end{tabular}
}
\end{table}
\end{landscape}

\begin{table*}
\caption{Glossary of symbols used in our calculations.}
\label{gloss}
\scalebox{0.7}{
\begin{tabular}{|l l | l l|}
\hline
{\bf Dynamical input parameters} & & & \\
\hline
&&&\\
$M_{\bullet}$ & black hole mass & $M_6$ & $M_{\bullet}/10^6 M_{\odot}$\\
$M_{\star}$ & star mass & $m$ & $M_{\star}/M_{\odot}$ \\
$E$ & orbital energy of star & $\bar{e}$ & $GM_{\bullet}/r_t$ \\
$J$ & orbital angular momentum & $\ell$ & $J/J_{lc}$ \\
$r_t$ & tidal radius & $J_{lc}$ & loss cone angular momentum \\
$R_{\star}$ & radius of star & $j$ & black hole spin \\
$r_{ISCO}$ & radius of innermost stable circular orbit & & \\
\hline
{\bf Disk structure parameters} & & & \\
\hline
$\rho$ & density of disk & $\kappa$ & opacity taken to be Thompson opacity \\
$\Sigma_d$ & surface density of disk & $H$ & disk height \\
$M_d$ & disk mass & $J_d$ & disk angular momentum \\
$r_{out}$ & outer radius of disk & $r_{in}$ & inner radius of disk  \\
$P_r$ & radiation pressure & $P_g$ & gas pressure \\
$\mathcal{K}$ & Constant of equation of state & $\dot{M}_a$ & accretion rate \\
\hline
{\bf Wind parameters} & & & \\
\hline
$c_1(r)$ & Bernoulli parameter & $c_2$ & $ c_1(r) \sqrt{r^2+z_{ph}^2} / (GM_{\bullet})$ \\
$\dot{\Sigma}_w$ & rate of surface density of wind & $\dot{M}_w$ & mass loss rate due to wind \\
$f_{out}$ & $\dot{M}_/\dot{M}_a$ & $z_{ph}$ & photosphere height \\
\hline
{\bf Accretion parameters} & & & \\
\hline
$R_{l}$ & maximum radius from star center to bound debris & $x_{l}$ & $R_l/R_{\star}$\\
$\varepsilon$ & $E_d/E_{dm}$ &$E_{\rm dm}$ &  energy of inner-most bound debris \\
$\mu_m$ & $M/M_{\star}$ & $M$ &  debris mass with energy $E_d$ \\
$\dot{M}_{fb}$ & fallback rate of debris & $\tau_m$ & $t/t_m$ \\
\hline
{\bf Viscosity parameters} & & & \\
\hline
$\Pi_{r\phi}$ & viscous stress & $\omega$ & angular frequency \\
$b$ & power index of $\Sigma_d$ & $d$ & power index of $r$ \\
\hline
{\bf Self similar quantities} & & & \\
\hline
$t_0$ & self similar time constant & $r_0$ & self similar radius taken to be outer radius at $t_0$ \\ 
$\Sigma_0$ & self similar surface density at $t_0$ & $\tau$ & $t/t_0$  \\ 
$v_r$ & radial velocity & & \\
\hline
{\bf Thermodynamical quantities} & & & \\
\hline
$T_e$ & effective temperature of disk & $L_b$ & bolometric luminosity \\
$T_h$ & hydrostatic temperature & $T_{ph}$ & photosphere temperature \\
$T_E$ & Eddington temperature & $L_E$ & Eddington luminosity  \\
$Q^{+}$ & viscous heating rate per unit area & $Q^{-}_{rad}$ & radiative loss rate per unit area \\
$Q^{-}_{adv}$ & energy loss rate per unit area due to advection & $L_{\nu}$ & luminosity in spectral band \\
\hline
\end{tabular}
}
\end{table*}


\section{Mass fallback rate}
\label{mfr}

A star with specific energy $E$ and angular momentum $J$ is tidally disrupted at the pericenter given by $\displaystyle{r_p=(GM_{\bullet}/(2E))\left[1-\sqrt{1-2E J^2/(G^2 M^2_{\bullet})}\right]}$. The stars are tidally captured if the angular momentum is $J\leq J_{lc}(E,r_{t})$ where $J_{lc}(E,r_{t})=\sqrt{2r^2_t(\Phi (r_t)-E)}$ is the loss cone angular momentum  \citep{1976MNRAS.176..633F}, the maximum value of $J$ is $J_{lc}(E,r_{t})$. As $J_{lc}(E,r_{t})\geq 0$, the maximum value of energy is $E_{m}= \Phi (r_t)$. We define the dimensionless energy $\bar{e}=E/E_m$ and angular momentum $\ell=J/J_{lc}$ and the constrained in energy and angular momentum phase space is given by $\bar{e}_h=r_t/r_h< \bar{e}<1$ and $0<\ell <1$ where $r_h=GM_{\bullet}/\sigma^2$ and $\sigma$ is the stellar velocity dispersion \citep{2005SSRv..116..523F}. 

The energy of the disrupted debris is given by $E_d(\bar{e},~\ell,~M_{\bullet},~m,~\Delta R) = \bar{e}E_{m} - 2 k G M_{\bullet}\Delta R/r^2_p$, where $\Delta R$ is the debris distance from the star center at the moment of breakup and $k$ is the spin up factor taken to be 3 \citep{2001ApJ...549..948A}. The time period of the most tightly bound debris is given by $t_{m}=2\pi GM_{\bullet}/[2E_d(\bar{e},~\ell,~M_{\bullet},~m,~-R_{\star})]^{3/2}$. The disrupted debris following a Keplerian orbit, returns to the pericenter with the mass fallback rate given by (MM15)

\begin{equation}
\dot{M}_{fb}=\frac{M_{\star}}{t_m} \frac{\diff \mu_m}{\diff \tau_m} 
\label{modlod}
\end{equation}

\noindent where $\tau_m=t/t_m$, $\mu_m=M/M_{\star}$, where $M$ is the debris mass and 

\begin{equation}
\varepsilon =\frac{x_{l}-x}{x_{l}+1},~~~~x =x_{l}-\tau_m^{-2/3}(1+x_{l}),~~~~\frac{\diff \mu_m}{\diff \varepsilon} = (1+x_{l}) \frac{\diff \mu_m}{\diff x},
\end{equation}

\begin{equation}
\frac{\diff \mu_m}{\diff \tau_m}=\frac{2}{3} \frac{\diff \mu_m}{\diff \varepsilon} \tau_m^\frac{-5}{3},~~~~\frac{\diff \mu_m}{\diff x} = \frac{3}{2} b_1 \int_{x}^1 \theta^{u_1} (x') x' \,\diff x'
\label{lod}
\end{equation} 

\noindent where $x_l \equiv x_l(E,~J,~M_{\bullet},~M_{\star})={\rm Min[}1,~r_p^2~E/(2k G M_{\bullet} R_{\star}) {\rm ]}$ is fractional radius of bound debris at the moment of breakup, $b_1$ is the ratio of central density $\rho_c$ to mean density $\overline{\rho_{\star}}=3M_{\star}/4\pi R_{\star}^3$ and $\theta$ is the solution of Lane--Emden equation for the given polytrope $u_1$ related to the density by $\rho=\rho_c \theta^{u_1}$. MM15 have shown that the peak of mass fallback rate increases with increase in $x_{\ell}$ implying that the more is mass bound to the black hole after disruption, higher is the peak mass fallback rate. 

The debris experiences stream collision either due to incoming stream that intersects with the outflowing stream at the pericenter \citep{1994ApJ...422..508K} or due to relativistic precession at the pericenter \citep{2013MNRAS.434..909H}. These interactions result in the circularization of the debris to form an accretion disk \citep{2013MNRAS.434..909H,2015ApJ...804...85S,2016MNRAS.455.2253B}. The hydrodynamical simulations by \citet{2009ApJ...697L..77R} have shown that the debris interactions result in the formation of an accretion disk with mass accretion rate showing deviation from \citet{2009MNRAS.392..332L} at early times and following $t^{-5/3}$ in the late stages. Very recently, \citet{2016MNRAS.455.2253B} have performed hydrodynamical simulations for a star on a highly elliptical orbit with the resulting debris undergoing apsidal precession; they found that the higher the eccentricity (and/or) the deeper the encounter, the faster is the circularization. For an efficient cooling, the debris forms a thin and narrow ring of gas. For an inefficient cooling, they settle in a thick and extended torus, mostly centrifugally supported against gravity. The general relativistic hydrodynamical simulation by \citet{2015ApJ...804...85S} have shown that the accretion rate still rises sharply and then decays as a power law. However, its maximum is 10 \% smaller than the previous expectation, and timescale of the peak accretion is longer than the previously predicted values. This is due to the mass accumulation at a higher radius because of angular momentum exchange at large radii. The overall conclusion is that the resulting debris will form an accretion disk. In MM15, we had developed a steady accretion model with edge radii constant and this model is based on a slim disk model given in \citet{2009MNRAS.400.2070S} with accretion rate following the eqn (\ref{modlod}). In the next section, we will construct the time-dependent accretion models.

\section{Physics of time-dependent accretion disks}
\label{ptdes}

The vertically integrated time-dependent equations of an axially symmetric disk are given by

\begin{subequations}
\begin{align}
\frac{\partial}{\partial t} \Sigma_d &=-\frac{1}{r}\frac{\partial}{\partial r}(r v_r \Sigma_d)-\dot{\Sigma}_w+\dot{\Sigma}_f \\
v_r \Sigma_d\frac{\partial}{\partial r}(r^2 \omega(r))&+\dot{\Sigma}_w r^2 \omega(r)=-\frac{1}{r}\frac{\partial}{\partial r}(r^2 \Pi_{r\phi})+\dot{j}_f,
\end{align}
\label{supdisk1}
\end{subequations} 

\noindent where $\omega(r)$ is the rotational velocity, $\Pi_{r\phi}$ is the viscous stress, $\dot{\Sigma}_w$ is the mass loss rate per unit area due to outflowing wind and $\dot{\Sigma}_f$ and $\dot{j}_f$ are the rates of mass and angular momentum per unit area added to disk by the fallback debris respectively.

Using eqn (\ref{supdisk1}), the reduced disk equation is given by

\begin{equation}
\dot{\Sigma}_d=\frac{1}{r}\frac{\partial}{\partial r}\left[\frac{\partial_r(r^2\Pi_{r\phi})}{\partial_r(r^2\omega(r))}\right]+\omega(r)r\frac{\partial}{\partial r}\left[\frac{\dot{\Sigma}_w r}{\partial_r(r^2\omega(r))}\right] +S(r,~t)
\label{diseqn}
\end{equation} 

\noindent where 

\begin{equation}
S(r,~t)=\dot{\Sigma}_f-\frac{1}{r} \frac{\partial}{\partial r}\left(\frac{\dot{j}_f r}{\partial_r(r^2\omega(r))}\right). 
\label{sfunc1}
\end{equation} 

\noindent Since the viscous stress in the accretion disk depends on the surface density and radius, we focus on calculating the viscous stress of the form $\Pi_{r\phi}=K \Sigma_d^b r^d$, where $b$ and $d$ are the constants. This structure is useful in obtaining the self-similar solutions of the TDE disks as constructed by \citet{2001A&A...379.1138M} for a collapsing disk and by \citet{1990ApJ...351...38C} for a TDE disk without fallback while taking the total angular momentum constant. Here we will construct a TDE disk model with the angular momentum of disk varying with time due to fallback of debris. 

The viscous stress results in the exchange of angular momentum and viscous heating in the disk. The heat generated due to viscous heating is emitted in the form of radiation in the sub-Eddington disk, whereas in a case of super-Eddington disk, some fraction of heat is advected to the black hole and the remaining is radiated. The heating flux in the disk is given by

\begin{equation}
Q^{+}=\frac{1}{4} r \frac{ \partial \omega}{\partial r} \Pi_{r\phi}.
\label{heat}
\end{equation}

\noindent The advection rate is given by $\displaystyle{Q_{adv}^{-}=[\dot{M}_a/(2 \pi r^2] T_c \diff S/(\diff \ln r)}$ \citep{2002apa..book.....F}, where $\dot{M}_a$ is the accretion rate, $T_c$ is the mid-plane temperature and $S$ is entropy per unit mass also called as specific entropy. The specific entropy is given by $T_c \diff S = \diff U + P \diff \left(1/\rho\right)$, where $\rho$ is the density, $P$ is the total pressure and $U$ is the internal energy per unit mass given by $U=U_g+U_R$, where $U_g=(3/2) k_B T_c/(\mu m_p)$ and $U_R=a T_c^4/\rho$ are the internal energy due to gas and radiation such that $\displaystyle{Q_{adv}^{-}=\dot{M}_a/(2 \pi r)\left[\diff U/\diff r+P \diff(1/\rho)/\diff r\right]}$. The flux radiated from the disk is given by

\begin{equation}
Q^{-}_{rad}=\sigma_{SB}T_e^4=Q^{+}-Q_{adv}^{-},
\label{radlum}
\end{equation}

\noindent where $T_e$ is the effective temperature of the disk and $\sigma_{SB}$ is the Stefan-Boltzmann constant. In the case of sub-Eddington disk, the advective rate is zero and the energy generated due to viscous heating is radiated in the form of radiation.

For a system with radiation $P_r=(1/3)aT^4$ and gas pressure $P_g=(R/V)T$, where $R$ is the gas constant, $V$ is the volume and $T$ is the temperature, the first law of thermodynamics gives \citep{1939isss.book.....C}

\begin{equation}
\diff Q=\frac{V}{T}\left(12 P_r+\frac{1}{\Gamma-1}\right) \diff T +\left(4 P_r+P_g\right)\diff V,
\label{cheq}
\end{equation} 

\noindent where $\Gamma=c_p/c_v$, $c_v=\diff Q/\diff T|_{{\rm constant~ V}}$ and $c_p=\diff Q/\diff T|_{{\rm constant~ P}}$ with $P=P_r+P_g$ as total pressure. 

We assume that $P_g = \beta_g P$ with constant $\beta_g$ and using eqn (\ref{cheq}), we obtained $P \propto \rho ^{\gamma}$ where $\gamma$ is given by \citep{1939isss.book.....C}

\begin{equation}
\gamma=\frac{\beta_g (\frac{C}{R}-\frac{\Gamma}{\Gamma-1})-16(1-\beta_g)}{\beta_g(\frac{C}{R}-\frac{1}{\Gamma-1})-12(1-\beta_g)} ,
\end{equation} 

\noindent where $C=\diff Q/\diff T$ is constant and $R$ is the gas constant. For $\beta_g \ll 1$, which is true for a radiation dominated model, $\gamma\approx 4/3$ ($\lim_{\beta_g\to 0} \gamma =4/3$).

The initial conditions and the formation of the disks are discussed in MM15. The sub and super-Eddington disks have different structures whose assumptions are given below:

\begin{itemize}

\item The sub-Eddington disk is considered to be the standard thin disk structure with $\alpha$ viscosity dominated by Thomson opacity. 

\item The super-Eddington disk has an outflowing wind. We considered the vertical structure of the disk to be nearly in hydrostatic equilibrium with a wind launched from the photosphere.

\item The super-Eddington disk is assumed to be dominated by radiation pressure and thus the prominent choice for the viscosity is the radiative viscosity with Thomson opacity which dominates when $\beta_g \ll 1$. 

\item The vertical structure of the atmosphere for a super-Eddington disk is assumed to be quasi-static and the photosphere height is calculated using the Eddington approximation.  

\end{itemize} 

\subsection{Sub-Eddington disk}
\label{subedd}

The vertical structure of sub-Eddington disk whose scale height $H \ll r$ (this is post-justified in \S \ref{subres} ) is

\begin{equation}
H^2 \frac{G M_{\bullet}}{r^3}=c_s^2=\left|\frac{\partial P}{\partial \rho}\right|_{z=0},
\label{hcwt}
\end{equation}
 
\noindent such that the radial momentum equation of the disk in the limit of $v_r \ll v_{\phi}$, where $v_r$ is the radial velocity and $v_{\phi}$ is the azimuthal velocity, results in $\omega(r) =\sqrt{GM_{\bullet}/r^3}$. The total pressure is $P=P_r+P_g$ with gas pressure $P_g=\rho k_ B T_c/\mu m_p$ and radiation pressure $P_r=aT_c^4/3$, where $\rho=\Sigma_d/(2H)$, $T_c$ is the mid-plane temperature of the disk, $m_p$ is the mass of the proton and $\mu$ is the mean molecular weight taken to be of solar metallicity equal to 0.65. 

\subsubsection{$\alpha$ viscous stress with total pressure}
\label{subtot}

The viscosity in a sub-Eddington disk is taken to be $\alpha$ viscosity so that the $\alpha$ viscous stress is given by \citep{1973A&A....24..337S}

\begin{equation}
\Pi_{\phi r}=\alpha_s P H,
\label{visstres}
\end{equation}

\noindent where $\alpha_s$ is the constant. Using eqn (\ref{radlum}) with zero advection, eqn (\ref{heat}) and the radiative loss given by $Q^{-}=\sigma T_e^4=(4 a c/3\kappa) (T_c^4/\Sigma_d)$, the viscous stress using $P=P_r/(1-\beta_g)$, is given by $\Pi_{\phi r}=K \Sigma_d^b r^d $ where 
	
\begin{equation}
b=-1,~~d=0,~~K=\frac{512}{9} \frac{(1-\beta_g)^2}{\alpha_s} \frac{c^2}{\kappa^2}.
\label{gascon}
\end{equation}

\subsubsection{$\alpha$ viscous stress with gas pressure } 
\label{subgas}

The viscous stress for a sub-Eddington disk with pressure dominated by gas pressure is given by \citep{1973A&A....24..337S}

\begin{equation}
\Pi_{\phi r}=-\alpha_s P_g H,
\label{visstres1}
\end{equation}

\noindent where $\alpha_s$ is the constant, $H$ is the scale height and $P_g$ is given below eqn (\ref{hcwt}). Using eqn (\ref{radlum}) with zero advection, eqn (\ref{heat}) and the radiative loss given by $Q^{-}=\sigma T_e^4=(4 a c/3\kappa) (T_c^4/\Sigma_d)$, the viscous stress is given by $\Pi_{\phi r}=K \Sigma_d^b r^d $ where 

\begin{equation}
b=\frac{5}{3},~~d=-\frac{1}{2},~~K=\left[\frac{9}{32} \frac{\kappa \sqrt{G M_{\bullet}}}{a c} \left(\frac{\alpha_s k_B}{2 \mu m_p}\right)^{4}\right]^{\frac{1}{3}}.
\label{gascon1}
\end{equation}

\subsection{Super-Eddington disk}

In the case of a super-Eddington disk, the pressure is dominated by radiation pressure which gives strong radiative outflows. The vertical momentum equation is given by

\begin{equation}
\frac{1}{2}\frac{\partial}{\partial z} v_z^2 =-\frac{1}{\rho} \frac{\partial}{\partial z} P -\frac{\partial}{\partial z} \Phi (r,z),
\label{vertt}
\end{equation}

\noindent where $P$ is the total pressure and $\Phi(r,~z)=-GM_{\bullet}/\sqrt{r^2+z^2}$. A radiation dominated disk is radiatively inefficient and the strong radiative pressure leads to an extended disk geometry, whose vertical structure is in hydrostatic equilibrium up to a height $z_{ph}$, from where the wind is launched. We consider a polytropic relation in which the  total pressure is given by $P=\mathcal{K}\rho^{\gamma}$ where $\mathcal{K}$ is a constant that is set by the entropy of the gas and by integrating eqn (\ref{vertt}), we obtained $\displaystyle{\frac{v_z^2}{2}+\frac{\gamma}{\gamma-1} \mathcal{K} \rho^{\gamma-1}+\Phi(r,~z)=c_1(r)}$, where $c_1(r)$ is constant of integration calculated at mid-plane. We assume that the vertical structure is in nearly hydrostatic equilibrium up to a photosphere height $z_{ph}$, which results in density structure given by

\begin{equation}
\frac{\gamma}{\gamma-1} \mathcal{K} \rho_0^{\gamma-1}=\frac{G M_{\bullet}}{r},~~~{\rm and}~~~ \rho=\rho_0 \left[\frac{r}{\sqrt{r^2+z^2}}\right]^{\frac{1}{\gamma-1}},
\label{densupt}
\end{equation}

\noindent where $\rho_0=B r^{-1/(\gamma-1)}$ is the density at mid-plane with $\displaystyle{B=\left[(\gamma-1)/\gamma\right]^{\frac{1}{\gamma-1}} \mathcal{K}^{-\frac{1}{\gamma-1}} (G M_{\bullet})^{\frac{1}{\gamma-1}}}$ and $c_1(r)=0$. The total pressure $P=P_r+P_g=(1/3)a T^4+ (k_B/\mu m_p)\rho T$ along with $P_g=\beta_g P$ gives $\gamma=4/3$ and $\mathcal{K}$ given by

\begin{equation}
\mathcal{K}=\left[\frac{3}{a}\right]^{\frac{1}{3}} \left[\frac{k_B}{\mu m_p}\right]^{\frac{4}{3}} (1-\beta_g)^{\frac{1}{3}} \beta_g^{-\frac{4}{3}}.
\label{kcc}
\end{equation}

The height of the disk $H$ using eqn (\ref{densupt}) and $\displaystyle{\Sigma_d= 2 \int_{0}^{H} \rho(r,z) \,\diff z =  2 (B/r^2) y_h/\sqrt{1+y_h^2}}$ where $y_h=H/r$; in the limit $y_h \ll 1$ which is justified a posteriori, we obtain $y_h=(1/2B) \Sigma_d r^2$.

The outflowing wind is launched from the photospheric height $z_{ph}$. Considering a plane parallel atmosphere, the temperature using Eddington approximation the photosphere temperature  is given by

\begin{equation}
T_{ph}^4=T_0^4~\frac{4}{3}~\frac{1}{\tau_{ph}+\frac{4}{3}},
\label{Eaprox}
\end{equation}

\noindent where $\tau_{ph} = \int_0^{z_{ph}} \rho \kappa \, \diff z= \rho_0~ \kappa~ r~ y_{ph}/\sqrt{1+y_{ph}^2}$, is the optical depth at the photosphere measured from the mid plane of the disk. We assume that the vertical structure of temperature given in eqn (\ref{Eaprox}) is similar to the temperature structure obtained using hydrostatic density structure which results in $y_{ph}$ given by 

\begin{equation}
y_{ph} (2+y_{ph}^2)\sqrt{1+y_{ph}^2}=\frac{3}{4} \rho_0 \kappa r,
\end{equation}

\noindent whose limiting solutions are

\begin{equation}
y_{ph} \approx \left\{
\begin{array}{ll}
\left(\frac{3 B \kappa }{4}\right)^{\frac{1}{4}} r^{-\frac{1}{2}}, & y_{ph} \gg 1 \\
& \\
\frac{3 B \kappa}{4} \frac{1}{r^2}, & y_{ph} \ll 1 
\end{array}
\right. 
\label{yphcal}
\end{equation}

The outflowing wind is launched at $z=z_{ph}$ and the velocity of the wind decreases for $z>z_{ph}$ due to gravity, thus $\diff v_z^2/\diff z|_{z_{ph}} <0$, so that the eqn (\ref{vertt}) reduces to  

\begin{equation}
v_z^2(z_{ph})=\frac{8}{\beta_g}\left[\frac{k_B}{\mu m_p}\right] (T_{ph}-T_E)+ 2 c_1(r),
\label{vzr}
\end{equation}

\noindent where $T_E$ is the Eddington temperature. Due to strong radiative pressure, the photospheric height $z_{ph}/r= y_{ph} \gg 1$ and thus using eqn (\ref{yphcal}), the Eddington temperature is given by 

\begin{equation}
T_E=\left[\frac{G M_{\bullet}}{a \kappa}\right]^{\frac{1}{4}} (1-\beta_g)^{\frac{1}{4}} r^{-\frac{1}{2}}.
\label{TEdd}
\end{equation}

The rate of surface density of the out flowing wind $\dot{\Sigma}_w^2=\rho^2(z_{ph}) v_z^2 (z_{ph})$ with $\rho(z_{ph})$ given in eqn (\ref{densupt}), is given by

\begin{equation}
\dot{\Sigma}_w= \frac{\sqrt{8}}{3} \left(\frac{\mu m_p}{k_B}\right)^{\frac{1}{2}} (G M_{\bullet})^{\frac{7}{8}} a^{\frac{1}{8}} \kappa^{-\frac{7}{8}} \beta_g^{\frac{1}{2}} (1-\beta_g)^{-\frac{1}{8}} r^{-\frac{7}{4}} \sqrt{\frac{T_{ph}}{T_E}-1+c_2},
\label{sigwcal}
\end{equation}

\noindent where $c_2=c_1(r)/(2 \Phi(r,~z_{ph}))$ is taken to be a constant. This simplifying assumption ensures the existence of the wind as decided by 

\begin{equation}
\frac{T_{ph}}{T_E}-1+c_2 \geq 0,
\label{tphcond}
\end{equation}

\noindent which is taken to be only a function of time and independent of radius. For a TDE disk, the viscous heating decreases with time after the wind is switched on which results in a decrease in the radiative pressure and thus a decrease in mass outflow rate. Since mass outflow rate decreases, $\dot{\Sigma}_w$ decreases, so we assume

\begin{equation}
\sqrt{\frac{T_{ph}}{T_E}-1+c_2}=\mathcal{W} \left(\frac{t}{t_0}\right)^{\delta},
\label{tphc}
\end{equation}

\noindent where $\mathcal{W}$ is a constant, such that $\dot{\Sigma}_w$ is given by

\begin{equation}
\dot{\Sigma}_w=W r^{-\frac{7}{4}} \left(\frac{t}{t_0}\right)^{\delta},
\label{sigw}
\end{equation}

\noindent where 

\begin{equation}
W=\frac{\sqrt{8}}{3} \left(\frac{\mu m_p}{k_B}\right)^{\frac{1}{2}} (G M_{\bullet})^{\frac{7}{8}} a^{\frac{1}{8}} \kappa^{-\frac{7}{8}} \beta_g^{\frac{1}{2}} (1-\beta_g)^{-\frac{1}{8}} \mathcal{W}.
\label{wcal}
\end{equation}

\noindent The eqn (\ref{tphc}) results in $T_{ph}/T_E$ being purely a function of time and independent of the radius and along with the assumption of constant $c_2$, this ensures that the entire disk is super-Eddington with wind outflow at all radii, provided eqn (\ref{tphcond}) is satisfied. 

The rotational velocity of the gas in a super-Eddington disk is non-Keplerian as the radiation pressure is significant. We adopt a solution of a thick disk along the lines of [see appendix B of \citet{1992ApJ...384..115L} and our appendix \ref{drv}] for a radiative viscosity $\eta_{\gamma}=8\epsilon_{\gamma}/(27 n_e \sigma_T c)$ \citep{1968ApJ...151..431M,1971ApJ...168..175W}, where $\epsilon_{\gamma}$ is the photon energy density, $n_e$ is electron density and $\sigma_T$ is Thompson scattering coefficient, the velocity $v_{\phi}(r)$ depends on the ratio $\delta_p=\epsilon_{\gamma}/(\rho c^2)$ which is taken to be $\delta_p=(27/8)\delta_0 \left(c^2 r/GM_{\bullet}\right)^{-s}$  as shown in appendix \ref{drv}. The azimuthal velocity, $v_{\phi}(r)$, is nearly constant if $\delta_p \ll 1$ which is valid in TDE disks for $r/(GM_{\bullet}) < (1/c^2) (v_0/c)^{-2/(1-\delta_0)}$. The TDE disks are evolving from the initial radii ratio $q=r_{out}/r_{in}$, which is slightly higher than the unity. We assume the angular frequency to be $\omega =\omega_s (r/r_s)^{-e}$, where 

\begin{equation}
\omega_s=\frac{v_0}{r_s}~~~{\rm and}~~~e=1+\frac{\delta_0}{2},
\end{equation}

\noindent where $r_s$ is taken to be ISCO radius, as obtained in the appendix \ref{drv} for the super-Eddington thick disk. Here, we consider $\delta_0$ as a free parameter in the range $0.018-0.11$. While this parameterization is justified within the context of a thick disk (with properties averaged for the mid-plane) as proposed by \citet{1992ApJ...384..115L}, a full quasi-spherical model will be presented in a subsequent paper.

The radiative viscous stress is given by $\displaystyle{\Pi_{r\phi}=\eta_{\gamma} r H~ \diff \omega/\diff r}$ \citep{1968ApJ...151..431M,1971ApJ...168..175W,2003BASI...31..207M}, where $\eta_{\gamma}=(8/27)(\epsilon_{\gamma}/\sigma_T n_e c)$. Following \citet{1992ApJ...384..115L}, the $\eta=(m_p c/\sigma_T)(c^2/(G M_{\bullet}))^s \delta_0 r^{-s}$, the viscous stress is given by

\begin{equation}
\Pi_{r\phi}=\frac{96e}{a}\frac{m_p}{\sigma_T c}\delta_0 \left(\frac{c^2}{G M_{\bullet}}\right)^{-s} \left[\frac{k_B}{\mu m_p}\right]^{4} (G M_{\bullet})^{-3} \omega_s r_s^e (1-\beta_g) \beta_g^{-4} \Sigma_d r^{3-e-s}.
\label{radviscosity}
\end{equation}

\begin{figure}
\begin{center}
\subfigure[Sub-Eddington]{\includegraphics[scale=0.5]{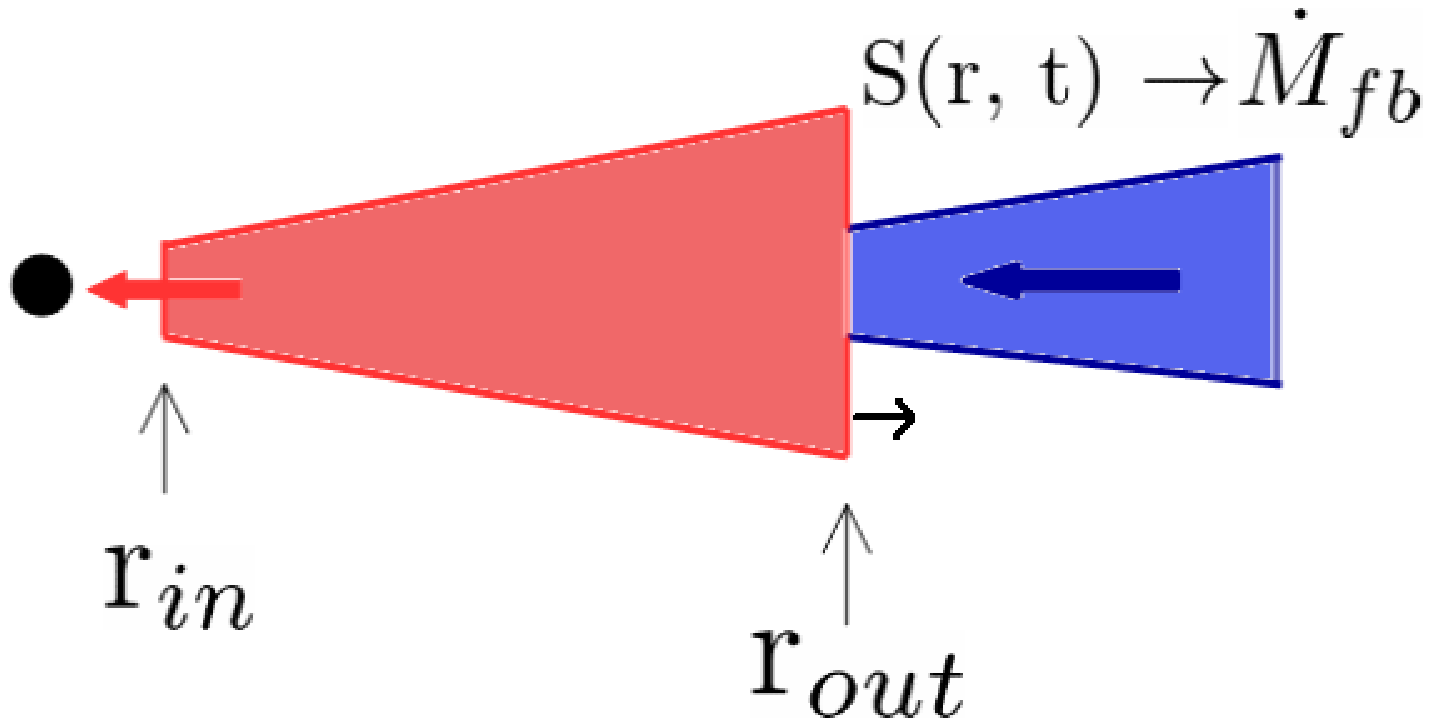}}
\subfigure[Super-Eddington]{\includegraphics[scale=0.5]{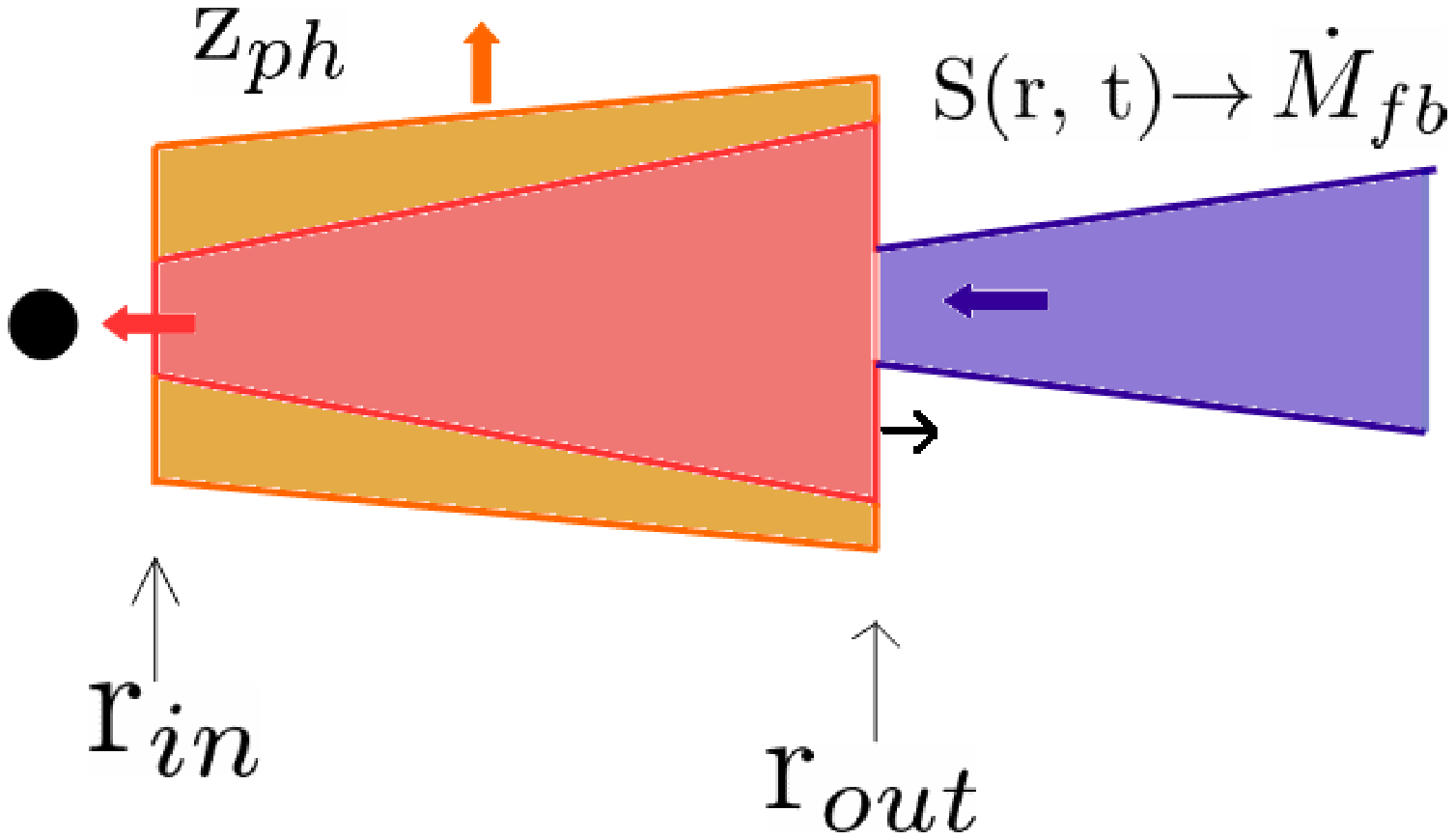}}
\end{center}
\caption{The schematic representation of the disk structure for the sub-Eddington and super-Eddington phase is shown. The blue, red, and orange shaded regions show the mass fallback of debris, the disk structure, and the wind structure respectively. The debris falls all over the disk so that the source function is continuous over the disk. The thick black arrows at the outer radius show the evolution of the outer radius.}
\label{visp}
\end{figure}

The smaller the value of $\beta_g$, the higher is the dominance of radiation pressure and the luminosity which implies that the radiation pressure dominated disk is more luminous compared to gas pressure dominated disk. The luminosity from the radiative disk is higher than $\alpha$ disk when the radiation pressure is dominated in the disk (see appendix \ref{vitress}). Thus, we have taken the super-Eddington disk to have radiative viscosity. The geometrical representation of sub and super-Eddington disks are shown in Fig \ref{visp}.

\section{Self-similar disk solution}
\label{sssol}

The self-similar form of surface density is given by

\begin{equation}
\Sigma_d=\Sigma_0\left(\frac{t}{t_0}\right)^{\beta}g(\xi),~~\xi=\frac{r}{r_0}\left(\frac{t}{t_0}\right)^{-\alpha}.
\label{ssol}
\end{equation}

Using eqns (\ref{diseqn}), (\ref{ssol}) and $\Pi_{r\phi}=K \Sigma_d^b r^d (t/t_0)^{\delta_1}$, we obtain 

\begin{equation}
\beta g(\xi)-\alpha \xi \diff_{\xi} g(\xi)-\frac{1}{2-e}\frac{1}{\xi}\diff_{\xi}(\xi^{e-1}\diff_{\xi}(\xi^{2+d} g^b(\xi)))-\frac{e-7/4}{2-e} \xi^{-7/4}= S(r,~t) \frac{t_0}{\Sigma_0} \left(\frac{t}{t_0}\right)^{1-\beta},
\label{diss}
\end{equation}

\begin{equation}
\frac{K}{\omega_s r_s^e}\Sigma_0^{b-1} r_0^{d+e-2}t_0=1,~~~~~~\delta_1+\beta (b-1)+\alpha (d+e-2)+1=0,
\label{diss1}
\end{equation}

\begin{equation}
\frac{W r_0^{-7/4} t_0}{\Sigma_0}=1,~~~~~{\rm and}~~~~~~~~\delta-\frac{7}{4}\alpha+1-\beta=0.
\label{diss2}
\end{equation}

We consider a power law solution $g (\xi)=A \xi^p$ so that the eqn (\ref{diss}) reduces to

\begin{equation}
(\beta - \alpha p)A-\frac{A^{b}}{2-e}(p~b+d+2)(p~b+d+e)\xi^{p~(b-1)+d+e-2}-\frac{e-7/4}{2-e}\xi^{-7/4-p} =S(r,~t) \frac{t_0}{\Sigma_0} \left(\frac{t}{t_0}\right)^{1-\beta} \xi^{-p},
\label{acalt}
\end{equation}

\noindent which has a solution for $p=(2-e-d)/(b-1)$, $-7/4$ and $S(r,~t)(t_0/\Sigma_0) \left(t/t_0\right)^{1-\beta} \xi^{-p}=s_c$, where $s_c$ is a constant and using eqn (\ref{ssol}), $S(r,~t)=(s_c/A) \Sigma_d/t$. A detailed calculation of self-similar solution is given in appendix \ref{sssolt}.

We assume that the matter added is instantaneously (quickly compared to the disk evolution time) distributed on the disk so that the self-similar solution holds at all radii. We consider a seed disk whose initial mass is $M_d(t_0)$, and the mass conservation equation is given by

\begin{equation}
\dot{M}_d=\dot{M}_{fb}-\dot{M}_a-\dot{M}_w,
\label{mcons}
\end{equation} 

\noindent where $\dot{M}_{fb}$ is the mass fallback rate given by eqn (\ref{modlod}), $M_d$ is the disk mass, $\dot{M}_a$ is the accretion rate onto the black hole and $\dot{M}_w$ is the mass outflow rate leaving the disk (see appendix \ref{sssolt}) and this results in

\begin{multline}
\frac{A(\beta+2\alpha)}{2+p}\left[\xi_{out}^{2+p}-\xi_{in}^{2+p}+\frac{2+p}{\beta+2\alpha}t \left( \xi_{out}^{1+p} \dot{\xi}_{out}-\xi_{in}^{1+p}\dot{\xi}_{in}\right) \right] =\chi_m \frac{\diff \mu_m}{\diff \varepsilon}\left(\frac{t}{t_0}\right)^{-\frac{5}{3}-\beta-2\alpha+1}- \\ \frac{t_0}{2\pi \Sigma_0 r_0^2}(\dot{M}_a+\dot{M}_w)\left(\frac{t}{t_0}\right)^{-\beta-2\alpha+1}.
\label{mceqnt}
\end{multline}

\noindent where $\chi_m=(1/2\pi)(M_{\star}/\Sigma_0r_0^2)(t_0/t_m)^{-2/3}$ and $\diff \mu_m/\diff \varepsilon$ is given by eqn (\ref{lod}). As $\diff \mu_m/\diff \varepsilon$ is initially an increasing function of time which attains a steady value at late times, we assume 

\begin{equation}
\beta+2\alpha=-\frac{2}{3},
\label{mcas}
\end{equation}

\noindent so that $\dot{M}_d \propto t^{-5/3}$ at late times. The inner radius of the disk is taken to be ISCO radius such that the eqn (\ref{mceqnt}) gives the evolution of the outer radius. The effective temperature and the luminosity of the disk using eqn (\ref{radlum}) are given by

\begin{equation}
\sigma_{SB}T_e^4 = \frac{e}{4}\frac{\omega_s^2 r_s^{2e} \Sigma_0 r_0^{2-2e} A^b}{t_0}\xi^{p~b+d-e}\left(\frac{t}{t_0}\right)^{\beta-2\alpha(e-1)-1} \left\{
\begin{array}{ll}
{\rm sub-Eddington}:
& \\
&\\
1,\\
& \\
{\rm super-Eddington}:
& \\
&\\
1-\frac{3}{2} \frac{\beta_g}{e(2-e)} \frac{G M_{\bullet}}{\omega_s^2 r_s^{2e}} \left[\frac{9}{4}-e+\frac{1}{A}\right] r_0^{2e-3} \left(\frac{t}{t_0}\right)^{\alpha(2e-3)} \times \\
\xi^{2e-3} , 
\end{array}
\right. 
\label{etemp}
\end{equation}

\begin{equation}
L_b^d =2 \pi r_0^2 \left(\frac{t}{t_0}\right)^{2\alpha}\int_{\xi_{in}}^{\xi_{out}} \sigma_{SB} T_e^4 \xi \, \diff \xi,
\label{dislum}
\end{equation}

\noindent where $\omega_s=\sqrt{G M_{\bullet}/r_s^3}$ with $e=3/2$ for sub-Eddington disk. The out flowing wind starts from the photosphere $z_{ph}$ and using eqns (\ref{TEdd}, \ref{tphc}), temperature of the photosphere, wind luminosity and Eddington luminosity are

\begin{equation}
T_{ph}=T_E \left(\mathcal{W}^2 \left(\frac{t}{t_0}\right)^{2\delta}+1-c_2\right)
\end{equation}

\begin{equation}
L_b^w=\left(\mathcal{W}^2 \left(\frac{t}{t_0}\right)^{2\delta}+1-c_2\right)^4 L_E
\label{lbw}
\end{equation}

\begin{equation}
L_E=\frac{\pi}{2} \frac{G M_{\bullet} c}{\kappa} (1-\beta_g)~ \ln \left(\frac{r_{out}(t)}{r_{in}}\right)
\label{edlum}
\end{equation}

The luminosity in the given spectral band \{$\nu_l$, $\nu_h$\}, is given by 

\begin{equation}
L(T)=\int_{\nu_l (1+z)}^{\nu_h (1+z)} \diff \nu \, \int^{r_{out}}_{r_{in}} \diff r \, 2 \pi r B(T)
\end{equation}

\noindent where $B(T)$ is the intensity of blackbody emission corresponding to temperature $T$ and $z$ is the redshift. The spectral luminosity is given by

\begin{equation}
L = \left\{
\begin{array}{ll}
L(T_e), & {\rm sub-Eddington} \\
& \\
L(T_e)+L(T_{ph}), & {\rm super-Eddington} 
\end{array}
\right. 
\label{splum}
\end{equation}

The unknown parameters in our models are black hole mass $M_{\bullet}$ and spin $j$, star mass $M_{\star}$, dimensionless orbital energy $\bar{e}$ and angular momentum $\ell$, $r_0$, $t_0$, $\Sigma_0$ and $\mathcal{W}$. We consider at initial time that $\{t=t_0$, $r_0=q~r_{in} \}$, where $q$ is the free parameter. The procedure to calculate the self-similar constants, $t_0$ and $\Sigma_0$ are discussed in \S \ref{subres}, for sub-Eddington disks and in \S \ref{modelB}, for super-Eddington disks. Using eqns (\ref{diss1}) and (\ref{diss2}), we see that $\mathcal{W}$ and self-similar constant $t_0$ are related for model B. So, we take $\mathcal{W}$ to be a free parameter to obtain $t_0$ in model B; the value of $\mathcal{W}$ is decided by obtaining the maximum $\mathcal{W}_{\rm max}$ numerically for the given  physical parameters $\{\bar{e},~\ell,~M_{\bullet},~M_{\star},~j,~q\}$ (see \S\ref{modelB} for more details). Then, we normalize the $\mathcal{W}$  with respect to its maximum, $\mathcal{W}_n=\mathcal{W}/\mathcal{W}_{max}$, and $\mathcal{W}_n$ is a free parameter which is used to obtain $t_0$. The parameter sets used in our calculations are given in Table \ref{supar}.

\begin{table}
\caption{The parameter sets used for simulations in our sub and super-Eddington models A and B. In case of model A, we take $\alpha_s=0.1$ and in case of model B, $\delta_0=0.05$ and $s=1$ (see appendix \ref{drv}).}
\label{supar}
\scriptsize
\center
\scalebox{1.2}{
\begin{tabular}{|c|c|c|c|c|c|c|c|}
\hline
&&&&&&&\\
Set & $\bar{e}$ & $\ell$ & $M_6$ & $m$ & $j$ & $q$ & $\mathcal{W}_n$ \\
\hline
&&&&&&&\\
I1 &  0.01 & 1 & 1 & 1 & 0 & 2 & 0.01 \\ 
&&&&&&&\\
I2 & 0.01 & 1 & 1 & 10 & 0 & 2 & 0.01 \\ 
&&&&&&&\\
I3 & 0.01 & 1 & 10 & 1 & 0 & 2 & 0.01 \\ 
&&&&&&&\\
I4 & 0.01 & 1 & 1 & 1 & 0 & 2 & 0.1 \\ 
&&&&&&&\\
I5 & 0.01 & 1 & 1 & 1 & 0.5 & 2 & 0.01 \\ 
&&&&&&&\\
I6 & 0.01 & 1 & 1 & 10 & 0.5 & 2 & 0.01 \\ 
&&&&&&&\\
I7 & 0.01 & 1 & 10 & 1 & 0.5 & 2 & 0.01 \\ 
&&&&&&&\\
I8 & 0.01 & 1 & 1 & 1 & 0.5 & 2 & 0.1 \\ 
&&&&&&&\\
\hline
\end{tabular}
}
\end{table}

\section{Sub-Eddington disk with $\alpha$ viscous stress}
\label{subres}

We present a model for a sub-Eddington disk with the viscous stress assumed in \S\ref{subedd} and self-similar formulation constructed in \S\ref{sssol}. The accretion models with viscous stress due to total pressure (\S\ref{subtot}) and gas pressure (\S\ref{subgas}) are named as models A1 and A2 respectively.  

\subsection{\bf Model A1: sub-Eddington disk with total pressure}
\label{modelA1}

We consider that at initial time $t=t_0$, $r_0=q~r_{in}$ where $q$ is a free parameter, $\xi_{in}(t_0)=1/q$ and taking $\xi_{out}(t_0)=1$, the eqns (\ref{diss1}, \ref{masdis}) give

\begin{equation}
\frac{M_d(t_0)}{\sqrt{t_0}}=\frac{64\pi}{(2+p)\sqrt{18}}A \frac{r_0^{7/4}(1-q^{-2-p})}{(G M_{\bullet})^{1/4}}\frac{1-\beta_g}{\sqrt{\alpha_s}} \frac{c}{\kappa},
\label{t0subc}
\end{equation} 

\noindent where

\begin{equation}
M_d(t_0)= \frac{M_{\star}}{t_m} \int_1^{t_0/t_m} \frac{\diff \mu_m}{\diff \tau_m}(\tau_m) \, \diff \tau_m.
\label{mdt0}
\end{equation}

\begin{table}
\caption{The values of $t_0$, $\Sigma_0$ and $r_0$ in units of $R_s=2.9 \times 10^{11} M_6$ in cm with $M_6=1$ obtained for the sub-Eddington $\alpha$ disk model A1 and $\beta_g=0.01$. See  \S\ref{modelA1}.}
\label{svalsub}
\scriptsize
\center
\scalebox{1.2}{
\begin{tabular}{|c|c|c|c|}
\hline
&&&\\
Set & $t_{0}$ (days) & $\Sigma_0 (10^3{\rm g~cm^{-2}})$ & $r_0 (R_s)$ \\
\hline
&&&\\
I1 & 2.47 & 7.76 & 6  \\
&&&\\
I2 & 4.22 & 10.1 & 6 \\
&&&\\
I3 & 7.70 & 4.3 & 6 \\
&&&\\
I4 & 2.47 & 7.76 & 6 \\
&&&\\
I5 & 2.42 & 8.4 & 4.23 \\
&&&\\
I6 & 4.17 & 10.9 & 4.23 \\
&&&\\
I7 & 7.26 & 4.6 & 4.23 \\
&&&\\
I8 & 2.42 & 8.4 & 4.23 \\
&&&\\
\hline
\end{tabular}
}
\end{table}

The eqn (\ref{t0subc}) is solved to obtain $t_0$ which is then used in the eqn (\ref{diss1}) to calculate the $\Sigma_0$. Using eqns (\ref{hcwt}, \ref{visstres}), $H$ is given by

\begin{equation}
\frac{H}{r}=\frac{32}{3}\frac{c}{\alpha_s\kappa}\Sigma_d^{-1} \omega^{-1} r^{-1}=1.1 \times 10^{-4} \left(\frac{\alpha_s}{0.1}\right)^{-1}\left(\frac{\Sigma_0}{10^{6}~{\rm g~cm^{-2}}}\right)^{-1} \left(\frac{r_{0}}{R_s}\right)^{\frac{1}{2}} \left(\frac{t}{t_0}\right)^{-1}\xi^{\frac{3}{4}}.
\end{equation}

\begin{figure}
	\begin{center}
		\subfigure[]{\includegraphics[scale=0.39]{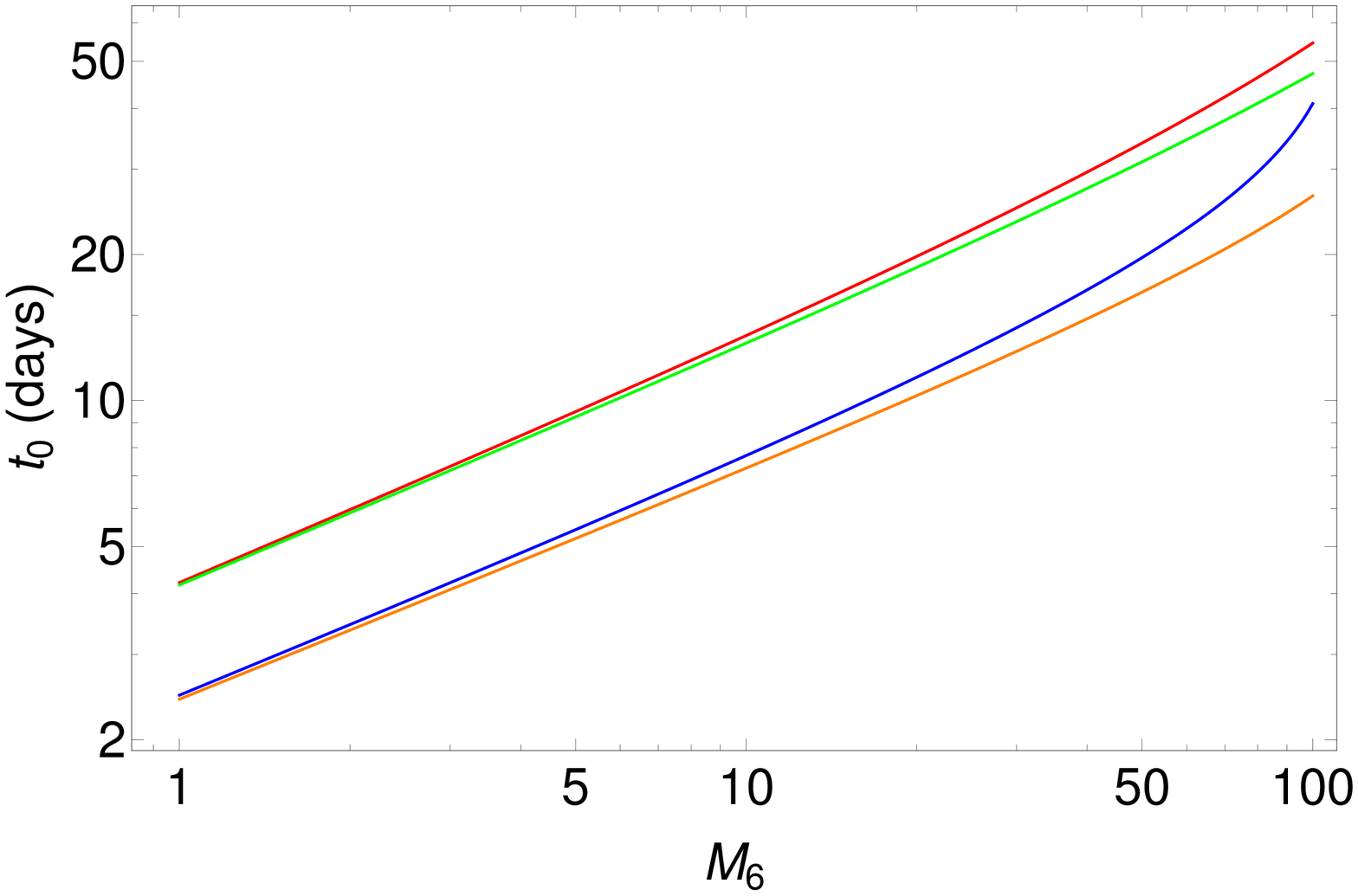}}
		\subfigure[]{\includegraphics[scale=0.41]{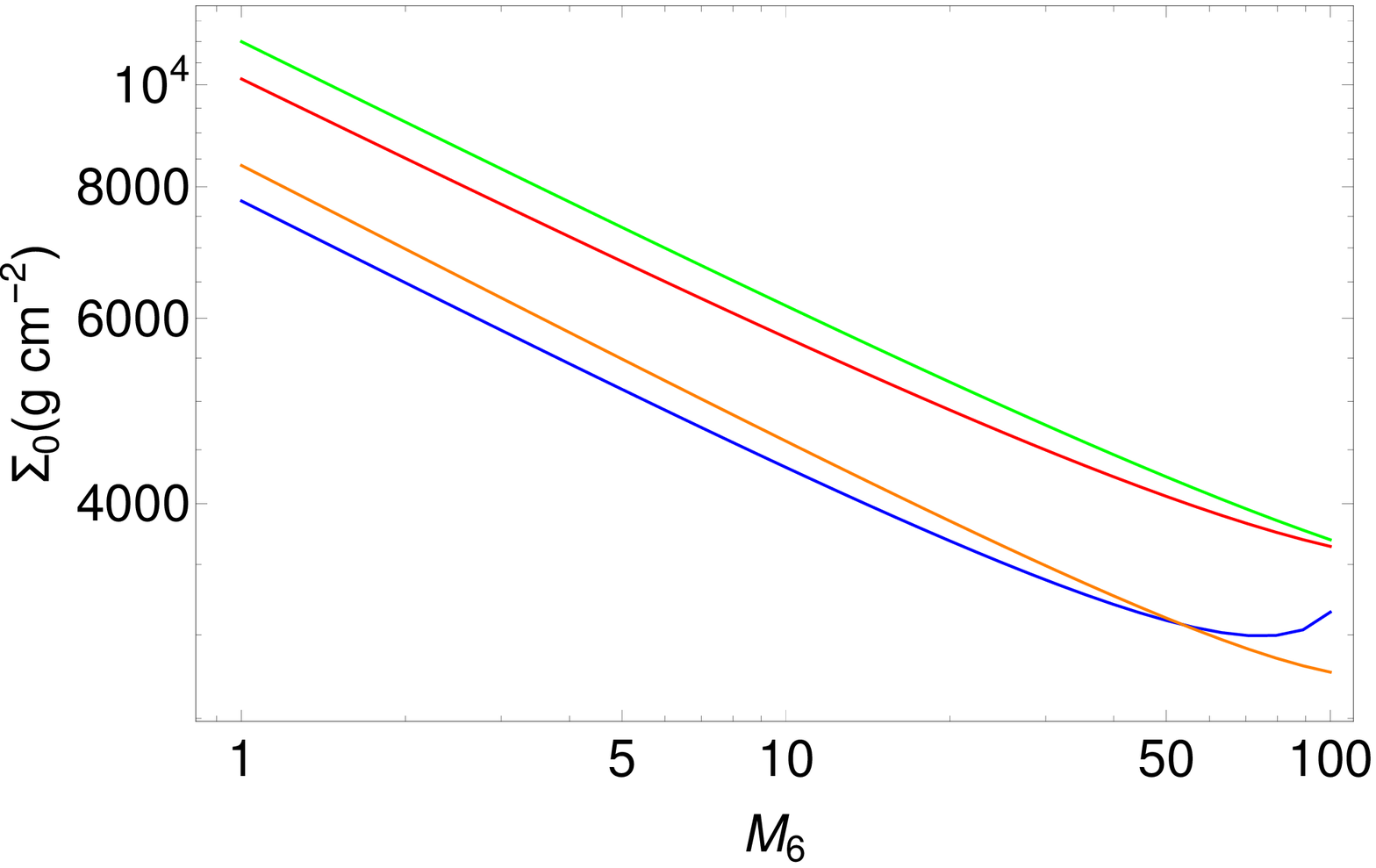}}
	\end{center}
	\caption{ (a) The self similar parameters $t_0$ in days and (b) $\Sigma_0$ as a function of black hole mass $M_6$ obtained for model A1 and parameter set I1 (blue), I2 (red), I5 (orange) and I6 (green). See \S\ref{modelA1}.}
	\label{t0plot}
\end{figure}

\begin{figure}
\begin{center}
\subfigure[]{\includegraphics[scale=0.39]{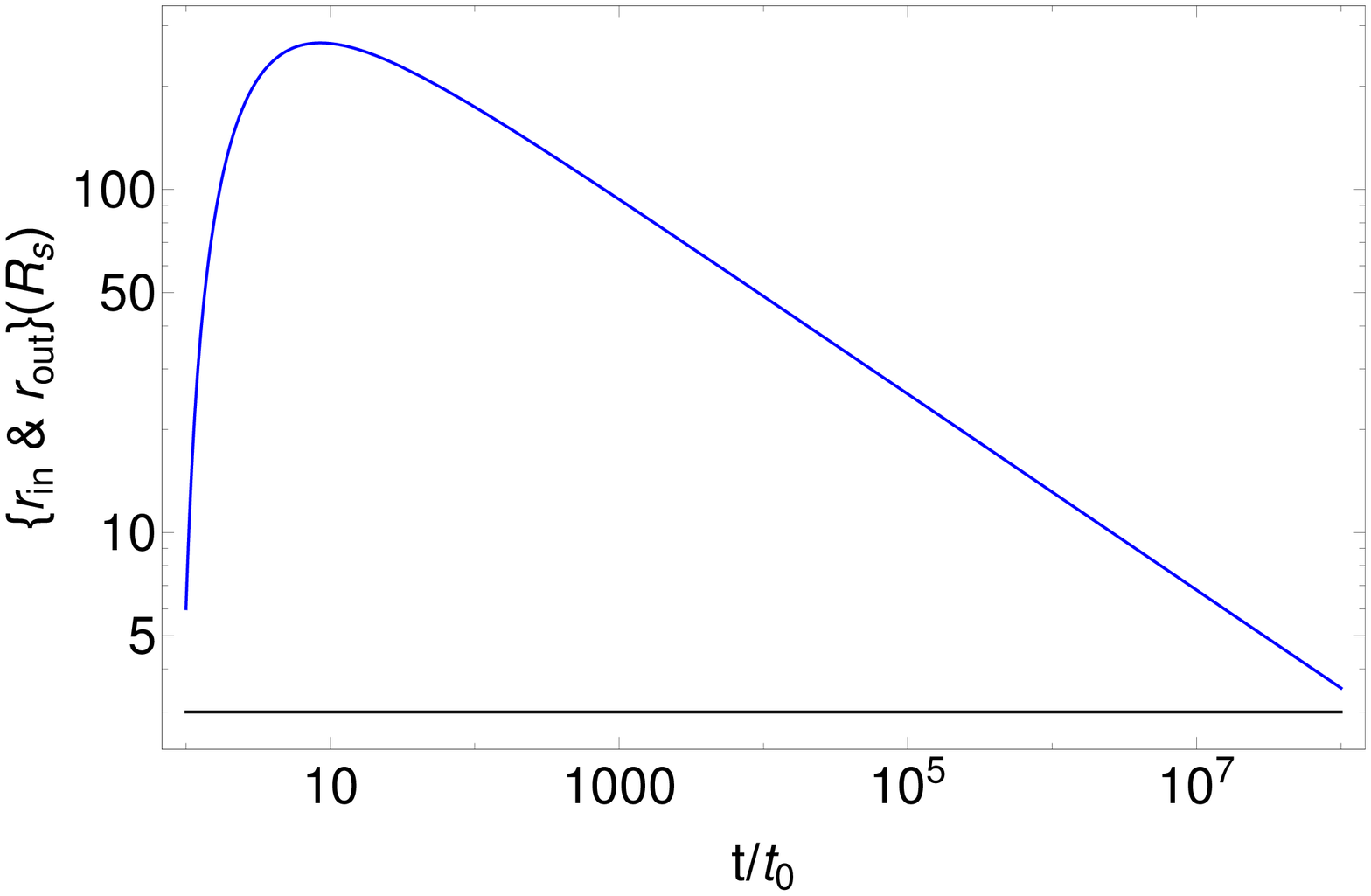}}~~~
\subfigure[]{\includegraphics[scale=0.48]{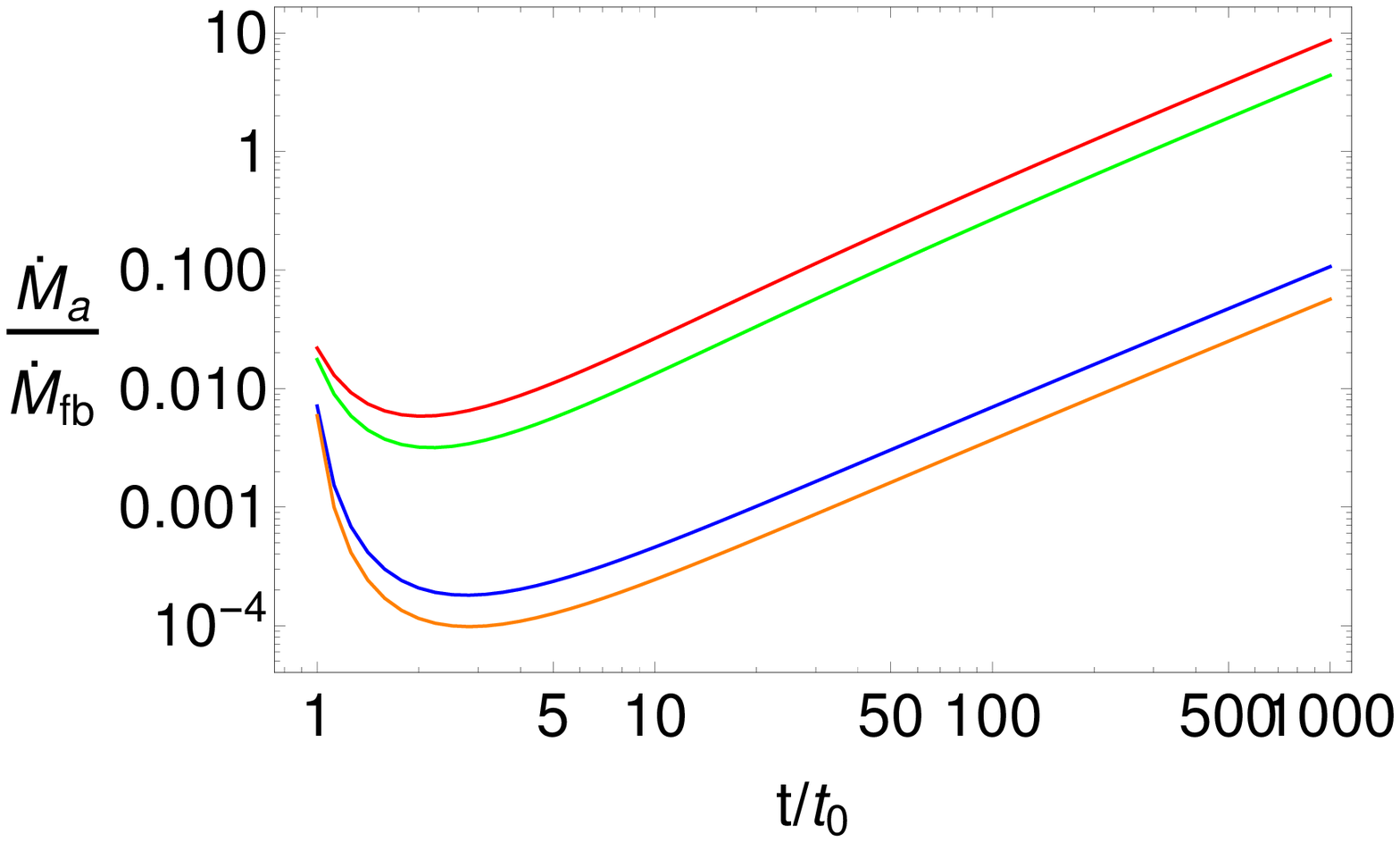}}
\end{center}
\caption{(a) The values of $r_{in}$ (black) and $r_{out}$ (red) in terms of $r_0$ for the parameter set I1 and model A1. (b) The ratio of accretion rate to mass fallback rate from the disrupted debris is shown for the parameter set I1 (blue), I3 (red), I5 (orange) and I7 (green). The increase in the ratio results in a decrease in the disk mass. See \S\ref{modelA1}.}
\label{xino}
\end{figure}

The free parameters are $\bar{e},~\ell,~M_6,~m,~\alpha,~e$ and $\beta_g$. We have solved eqn (\ref{t0subc}) numerically to obtain $t_0$ and $\Sigma_0$ shown in Fig \ref{t0plot}. The calculated values of $\Sigma_0$ and $r_0$ for each simulation set is given in Table \ref{svalsub}. The ratio of accretion rate of the black hole to the mass fallback rate is given by

\begin{equation}
\frac{\dot{M}_a}{\dot{M}_{fb}}=1.9 \frac{q^{-\frac{7}{4}}}{1-q^{-\frac{7}{4}}} \left(\frac{M_d(t_0)}{M_{\odot}}\right)\left(\frac{M_{\star}}{M_{\odot}}\right)^{-1}\left(\frac{t_0}{t_m}\right)^{\frac{2}{3}}\left[\frac{\diff \mu_m}{\diff \varepsilon}(t)\right]^{-1}\left(\frac{t}{t_0}\right)^{\frac{7}{6}}
\label{mamfeqn}
\end{equation}

\noindent where $\diff \mu_m/\diff \varepsilon$ is given by eqn (\ref{lod}). Since we assume the timescale for redistribution of the infalling matter in the disk (assuming a pro-rated distribution) is smaller than the radial inflow so that the self-similar structure remains same, the outer radius of the disk expands initially as the mass fallback is higher than the accretion and decreases at late times as accretion dominates over fallback as shown in Fig \ref{xino} obtained using eqn (\ref{mceqnt}). Since $t_0$ is smaller, the accretion dynamics is faster for high spin BHs which implies that the fallback rate is high, and hence more mass is added to the disk which causes the outer radius to move outward more. The accretion rate at late time decreases as $\dot{M}_a \propto t^{-1/2}$ as can seen from eqn (\ref{mamfeqn}) when $\dot{M}_f \propto t^{-5/3}$.

\begin{figure}
\begin{center}
\subfigure[]{\includegraphics[scale=0.4]{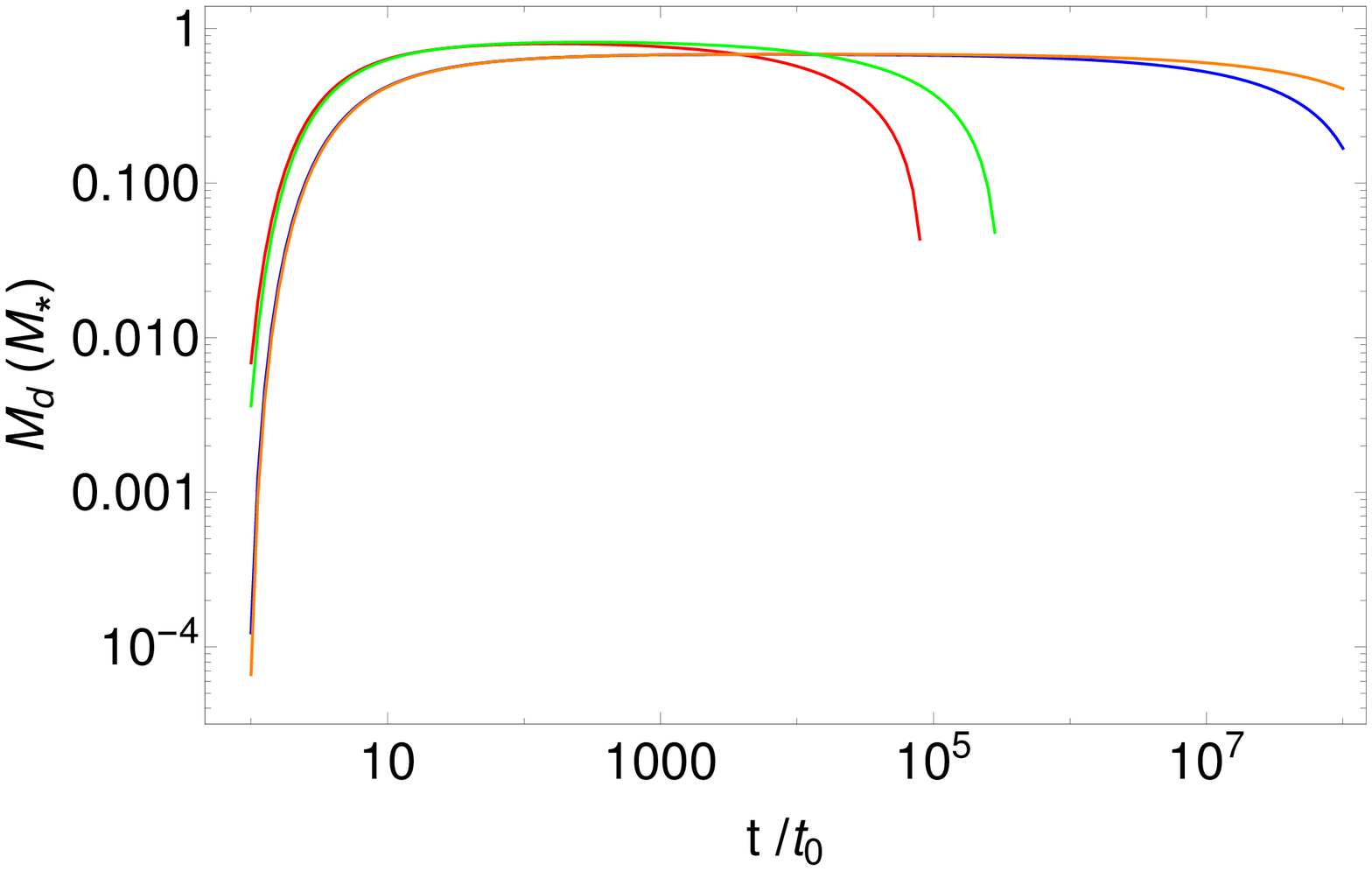}}~~~
\subfigure[]{\includegraphics[scale=0.4]{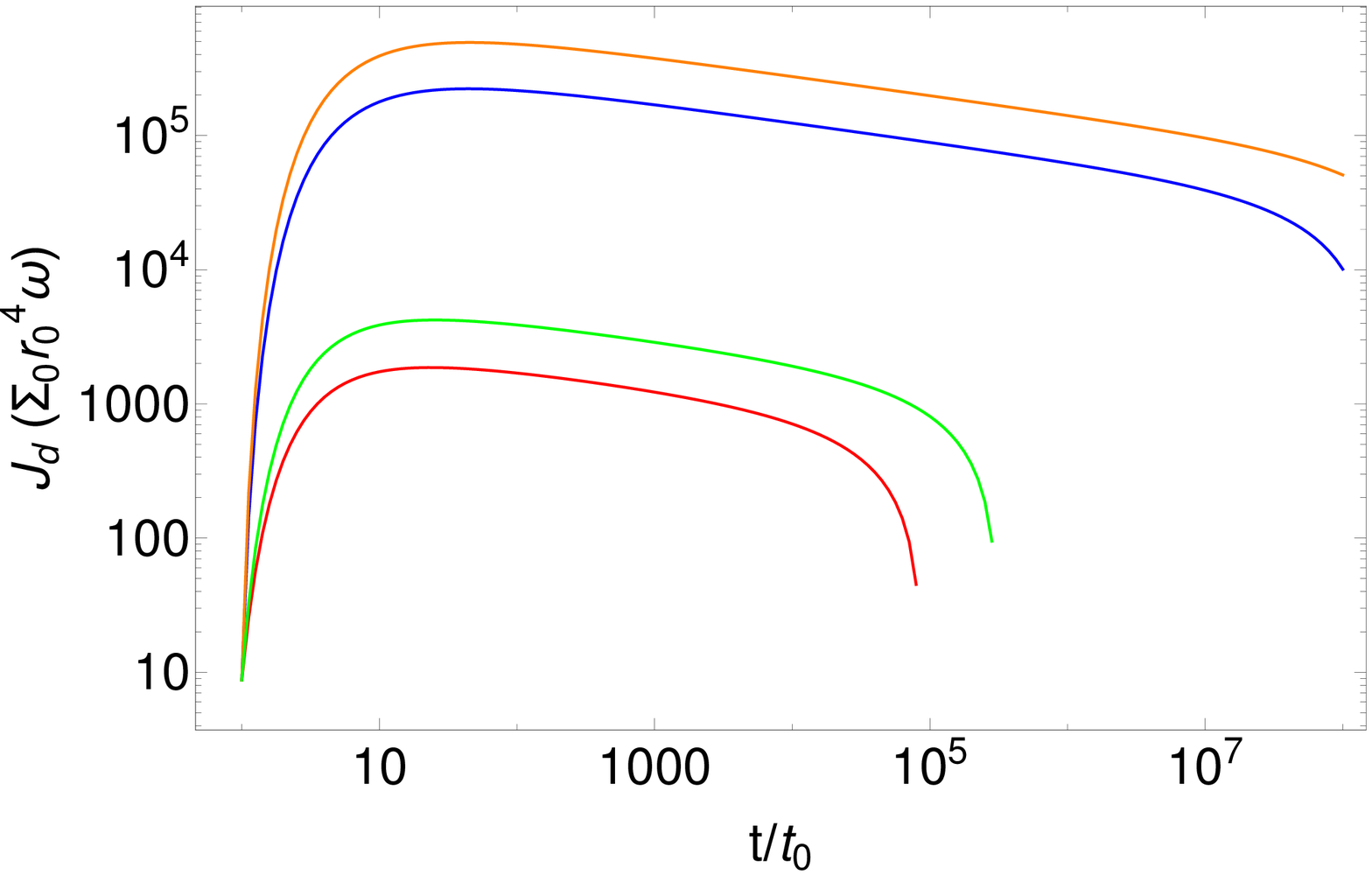}}
\end{center}
\caption{(a) The evolution of the mass of disk and (b) the angular momentum is shown for model A1 with the parameter set I1 (blue), I3 (red), I5 (orange) and I7 (green). See \S\ref{modelA1}. }
\label{mjds}
\end{figure}

\begin{figure}
\begin{center}
\subfigure[Bolometric]{\includegraphics[scale=0.41]{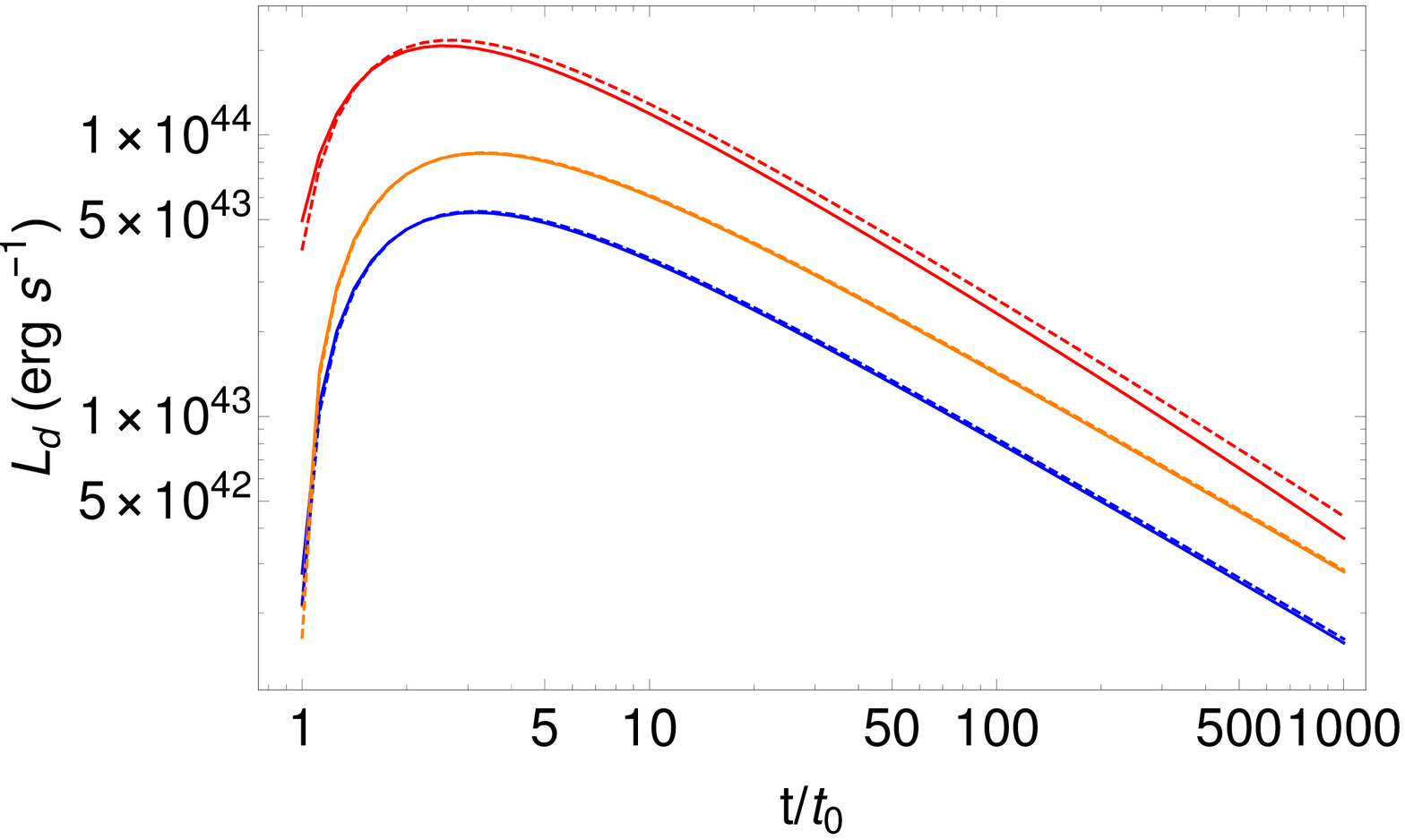}}
\subfigure[$n= \diff \ln L_d/\diff \ln t$ ]{\includegraphics[scale=0.44]{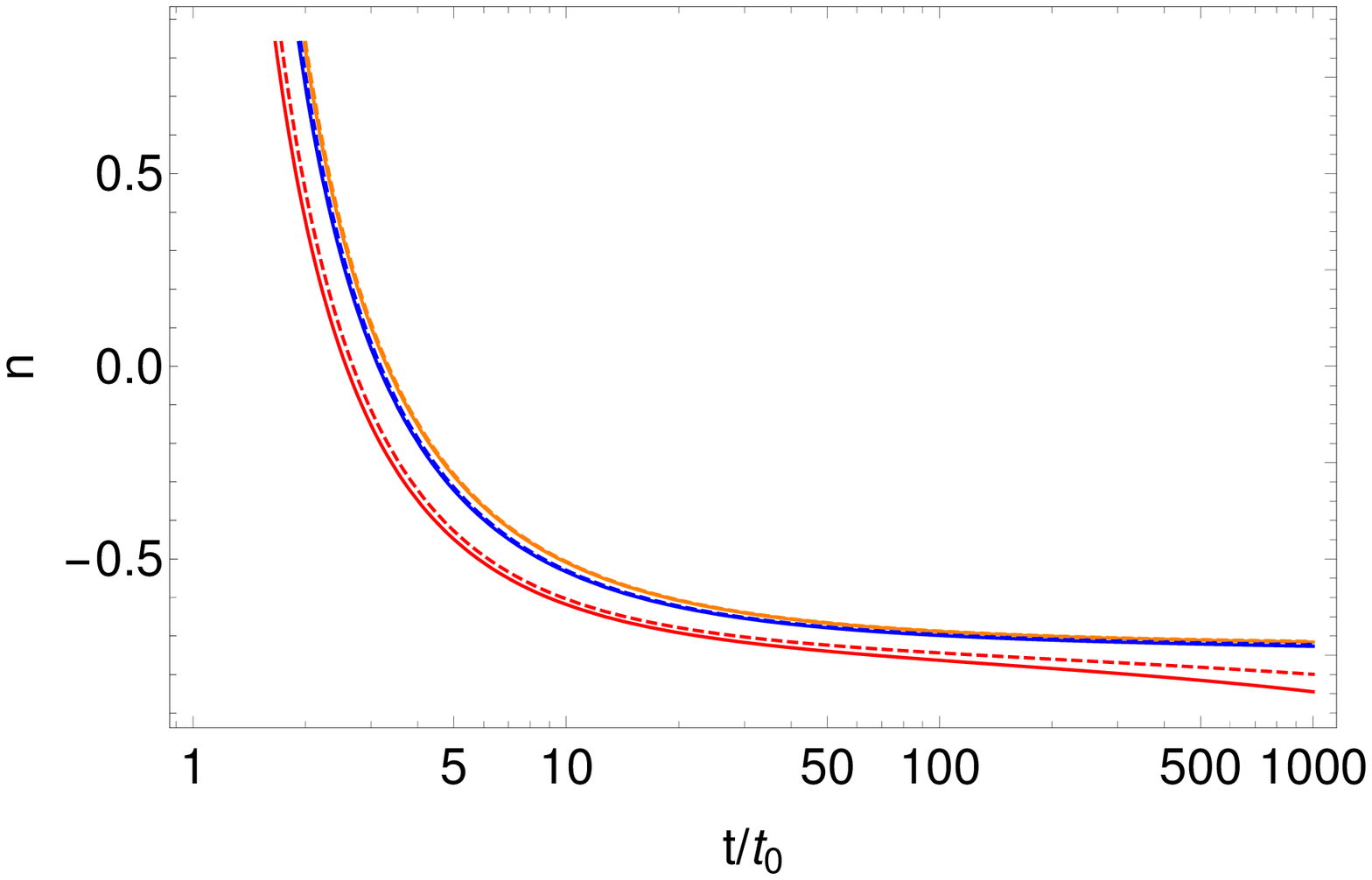}}
\subfigure[X-ray band]{\includegraphics[scale=0.4]{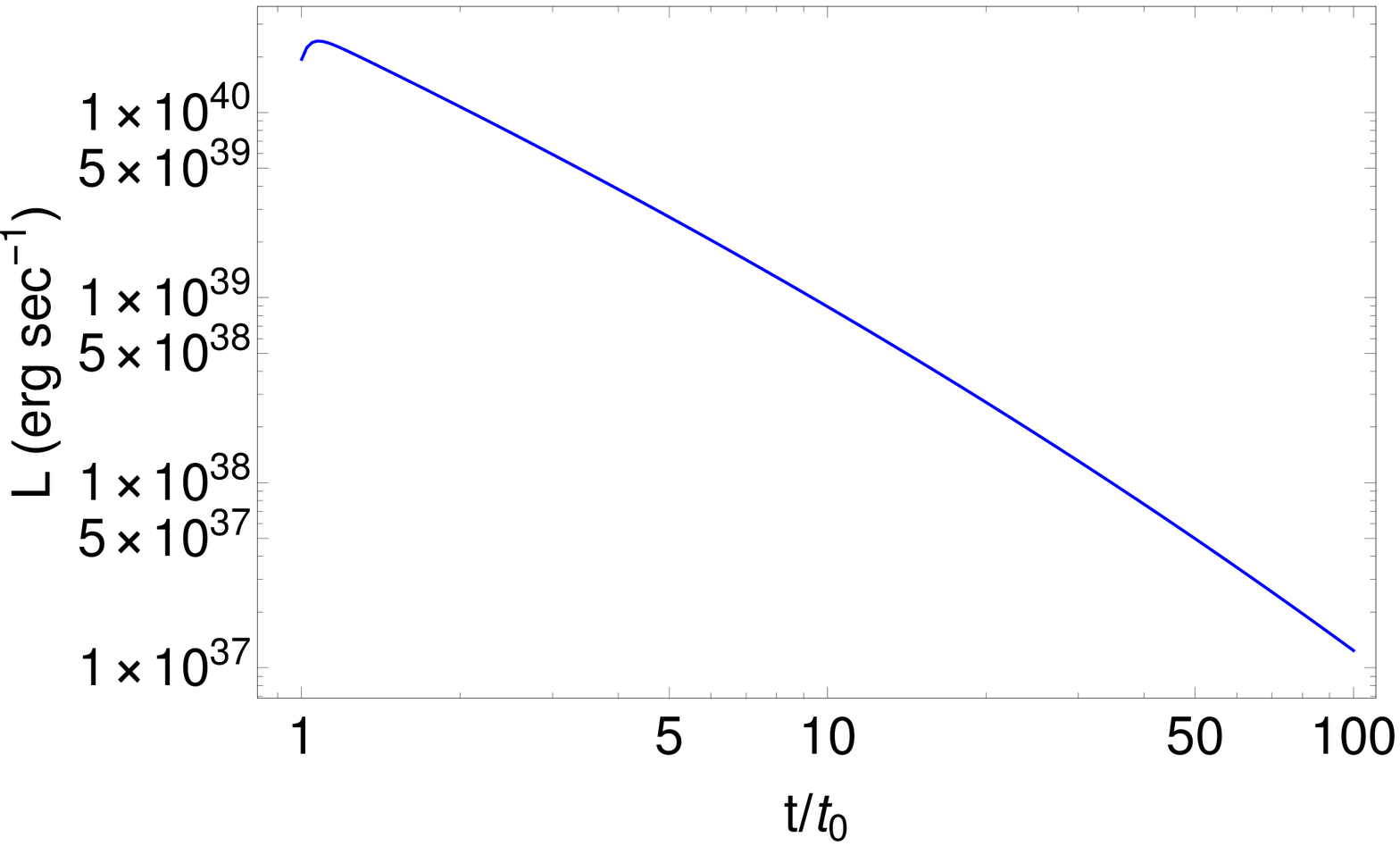}}
\subfigure[UV bands]{\includegraphics[scale=0.38]{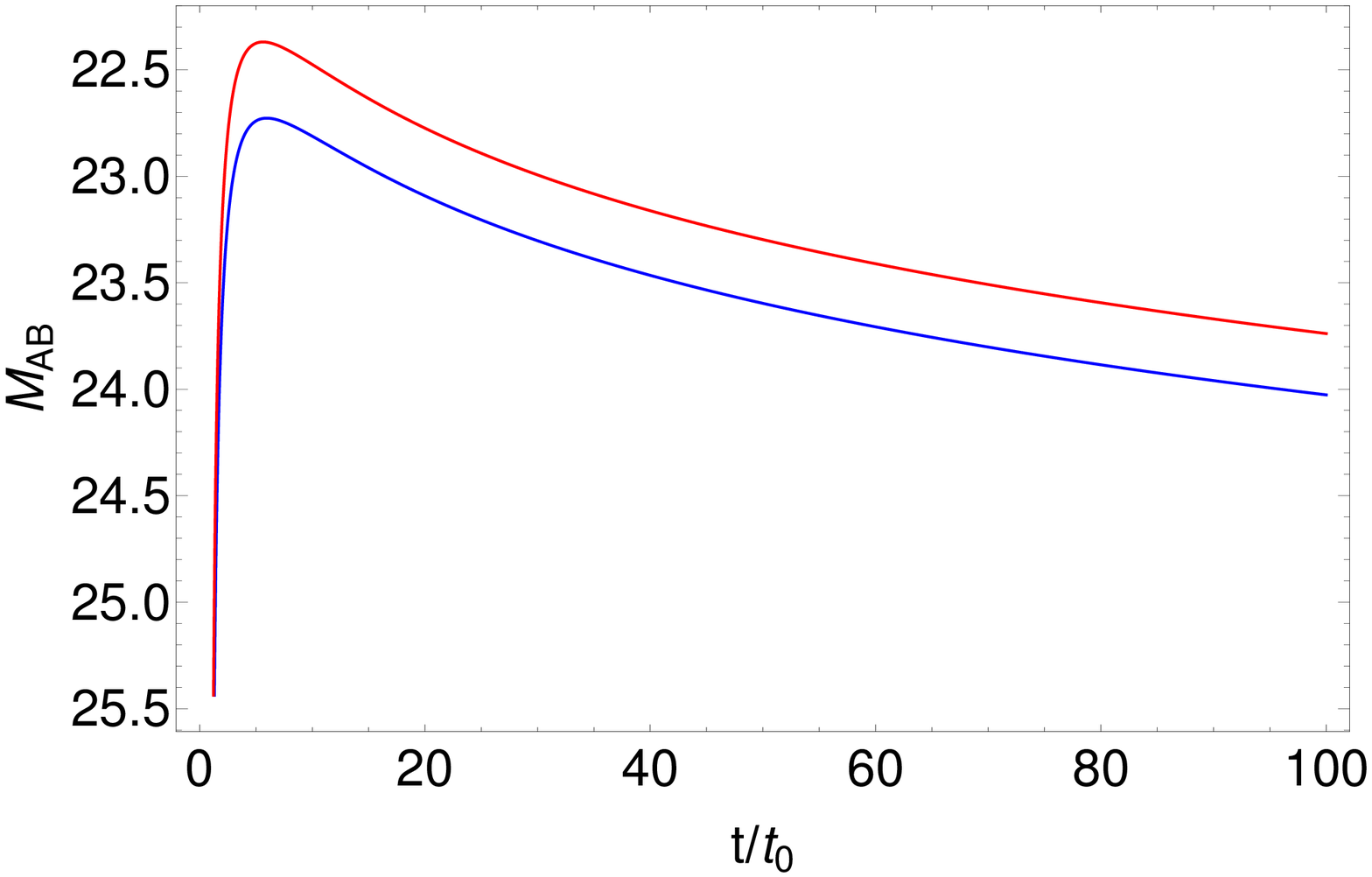}}
\subfigure[Optical bands]{\includegraphics[scale=0.38]{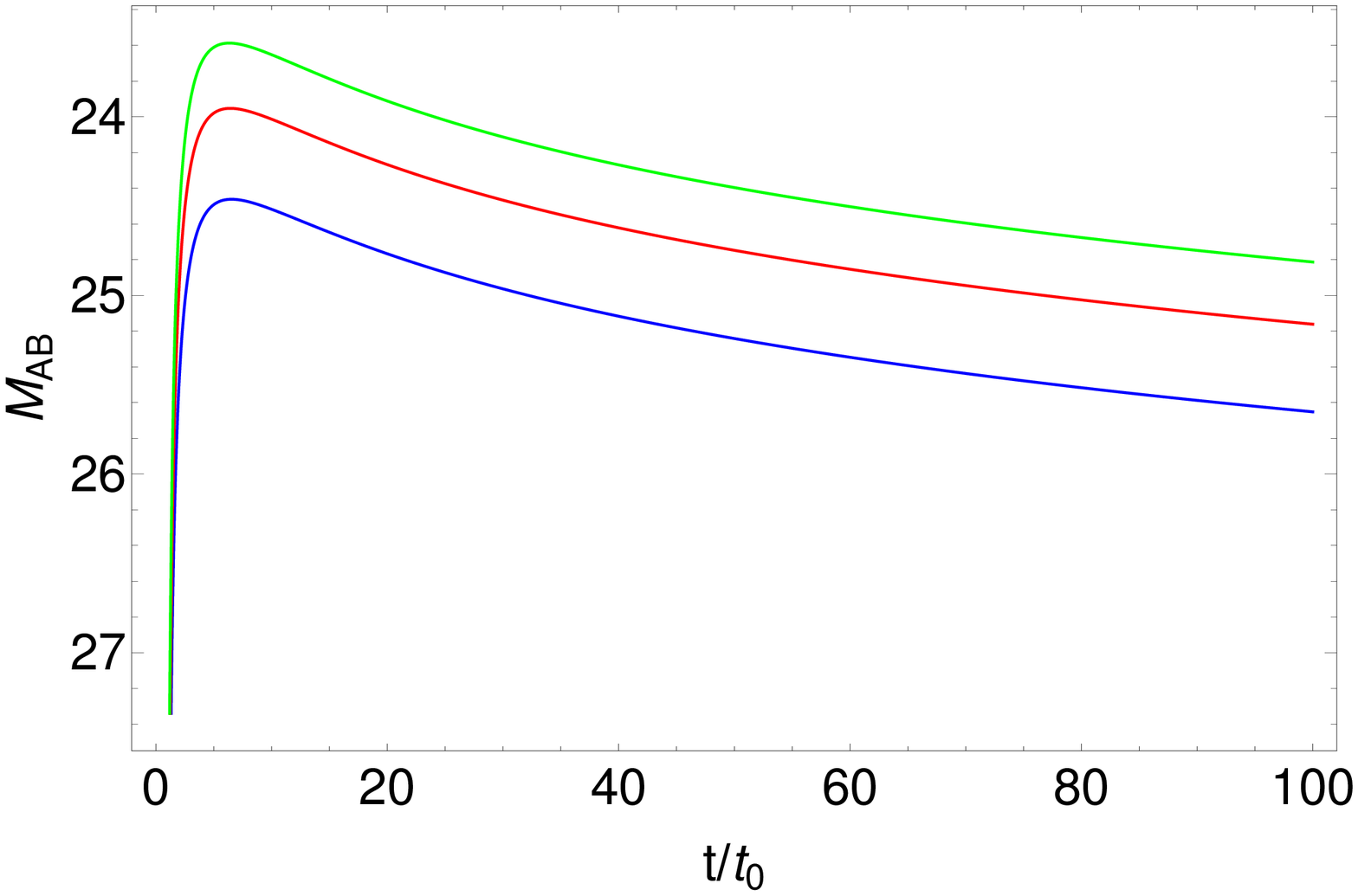}}
\end{center}
\caption{(a) The bolometric luminosity (eqn \ref{dislum}) as a function of $t$ is obtained for model A1 with the parameter set I1 (blue), I2 (orange), I3 (red), I5 (blue dashed), I6 (orange dashed) and I7 (red dashed). (b) The time evolution of $n= \diff \ln L_d/\diff \ln t$ is shown for the luminosity curves shown in (a). The spectral luminosity simulated in various bands for the run I1 and redshift $z=0.1$ for soft X-ray in (c), UV where Swift UVM2 (1800-3000 $A^{\circ}$) (blue) and UVW2 (1500-2500 $A^{\circ}$) (red) in (d) and optical in (e) where the curves for V band (blue), B Band (red) and U Band (green) are indicated. See \S\ref{modelA1}. }
\label{lsub}
\end{figure}

From eqn (\ref{mamfeqn}) and Fig \ref{xino}b, we see that the rate of mass loss by the disk due to accretion is higher than the mass gain by the addition of fallback debris at late times which results in the decline in the disk mass as shown in Fig \ref{mjds}. The overall angular momentum of the disk including the angular momentum loss due to accretion and the addition by fallback debris is shown in Fig \ref{mjds}. The Fig \ref{lsub}, shows the bolometric luminosity obtained using eqn (\ref{dislum}) as a function of $t$. The mass fallback rate causes an enhancement in surface density, viscous heating, effective disk temperature and luminosity. The spectral luminosity simulated in various spectral bands for the set I1 are also shown.

\subsection{\bf Model A2: sub-Eddington disk $\alpha$ viscous stress }
\label{modelA2}

\begin{figure}
\begin{center}
\subfigure[]{\includegraphics[scale=0.45]{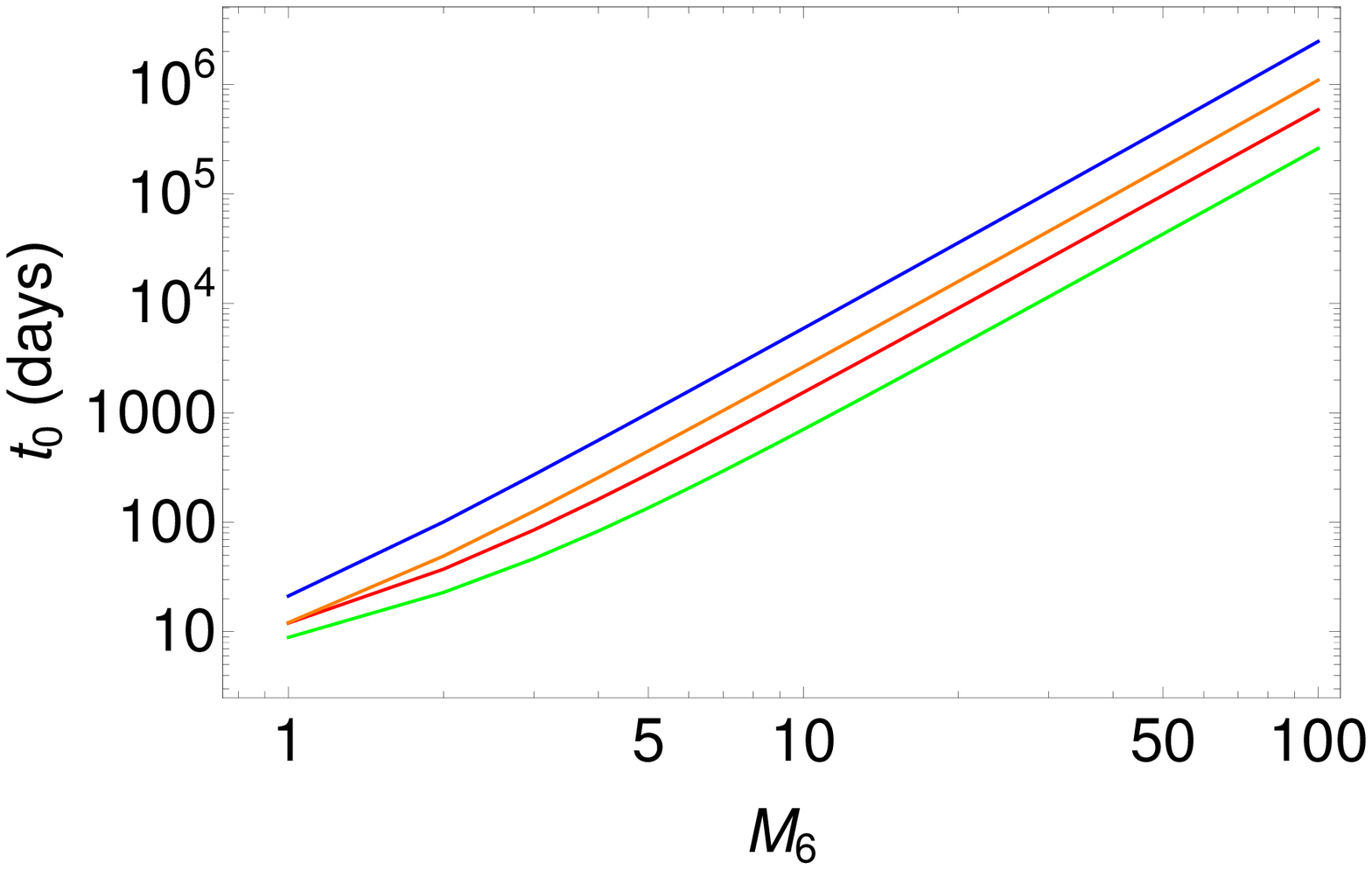}}
\subfigure[]{\includegraphics[scale=0.45]{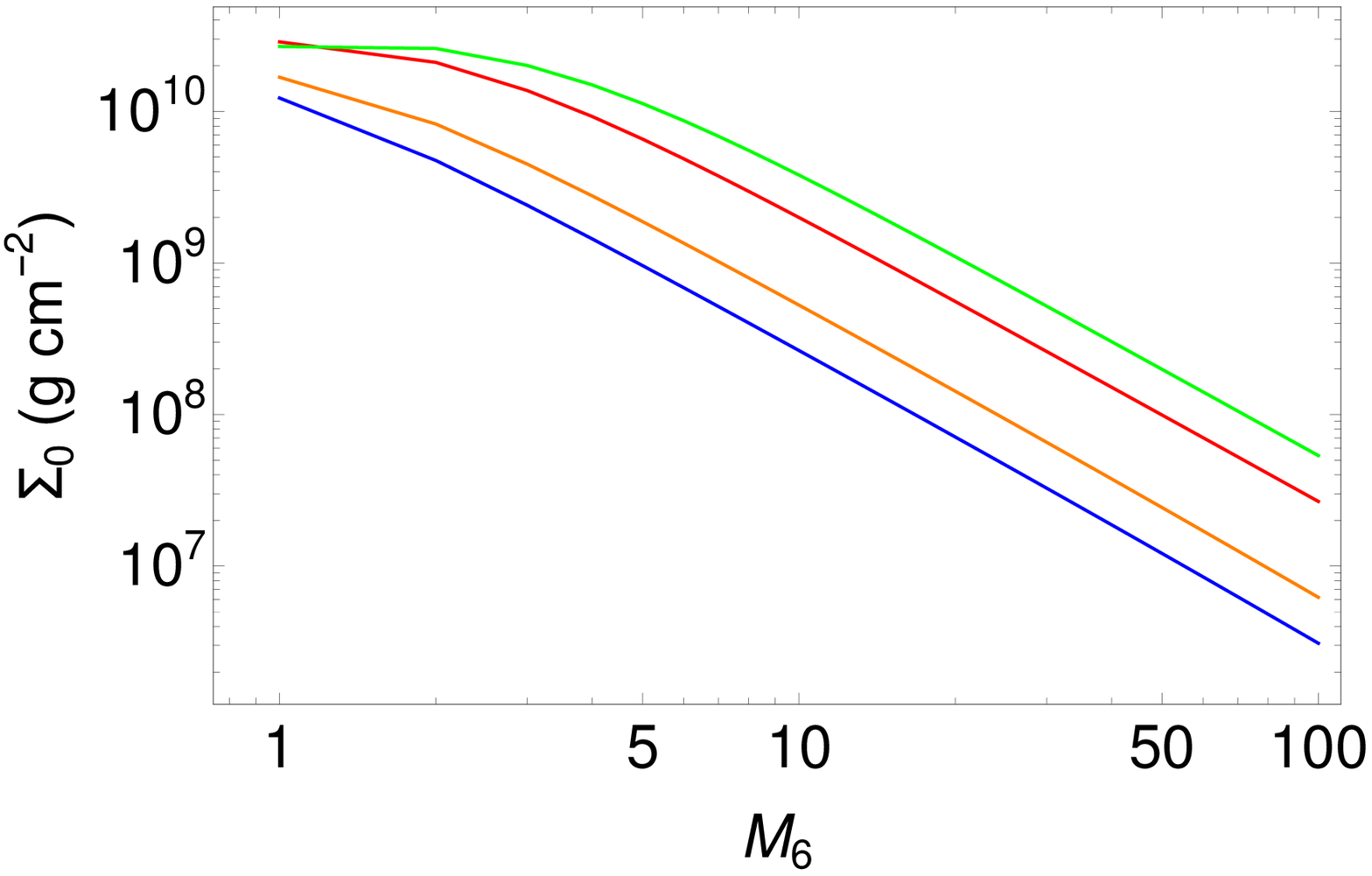}}
\end{center}
\caption{(a) The self similar parameters $t_0 (t_m)$ and (b) $\Sigma_0~{\rm (g~cm^{-2})}$ as a function of black hole mass $M_6$ for model A2 and the parameter sets I1 (blue), I2 (red), I5 (orange) and I6 (green). See \S\ref{modelA2}.}
\label{t0plotg}
\end{figure}

\begin{figure}
\begin{center}
\subfigure[]{\includegraphics[scale=0.33]{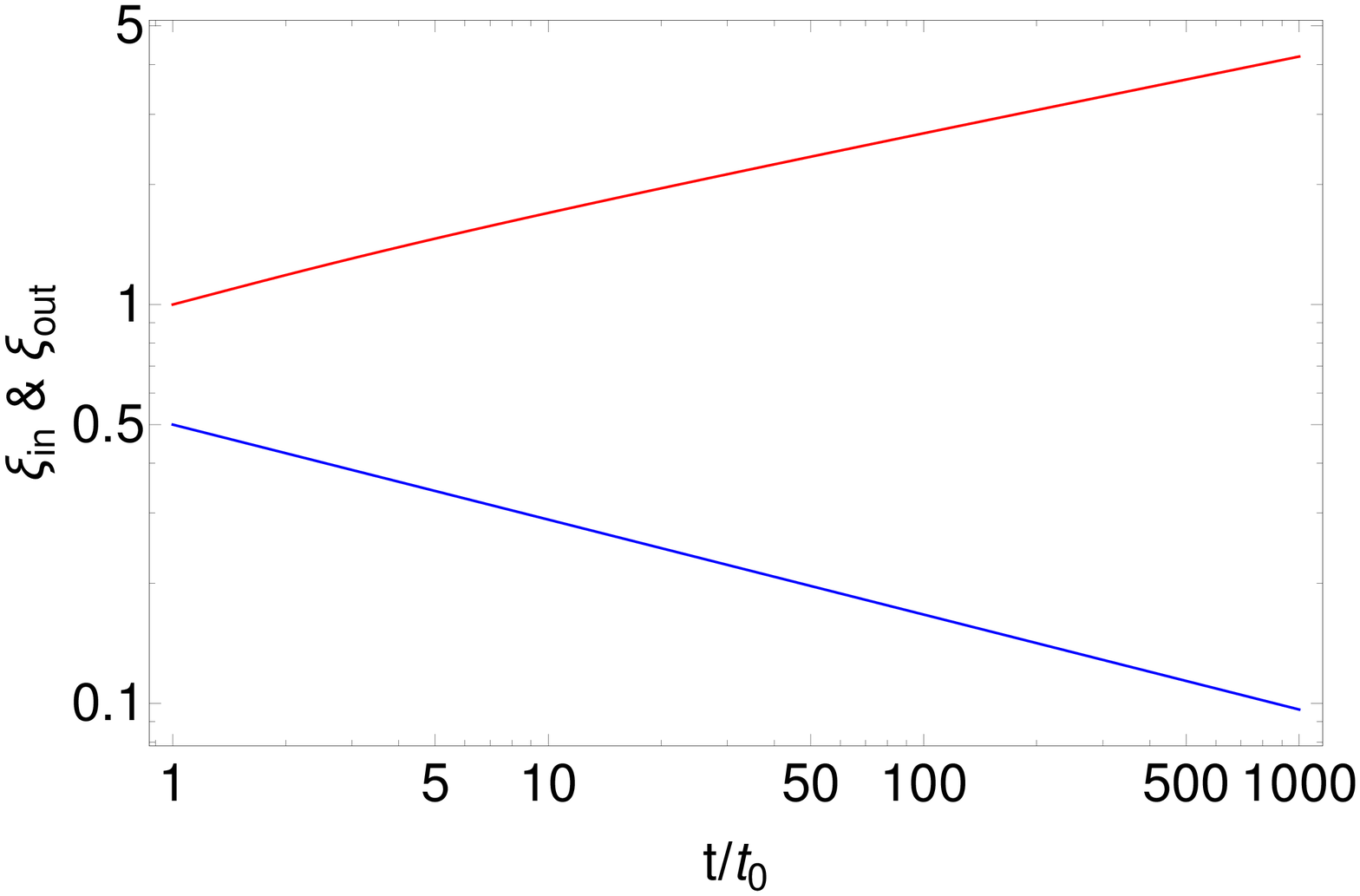}}
\subfigure[]{\includegraphics[scale=0.37]{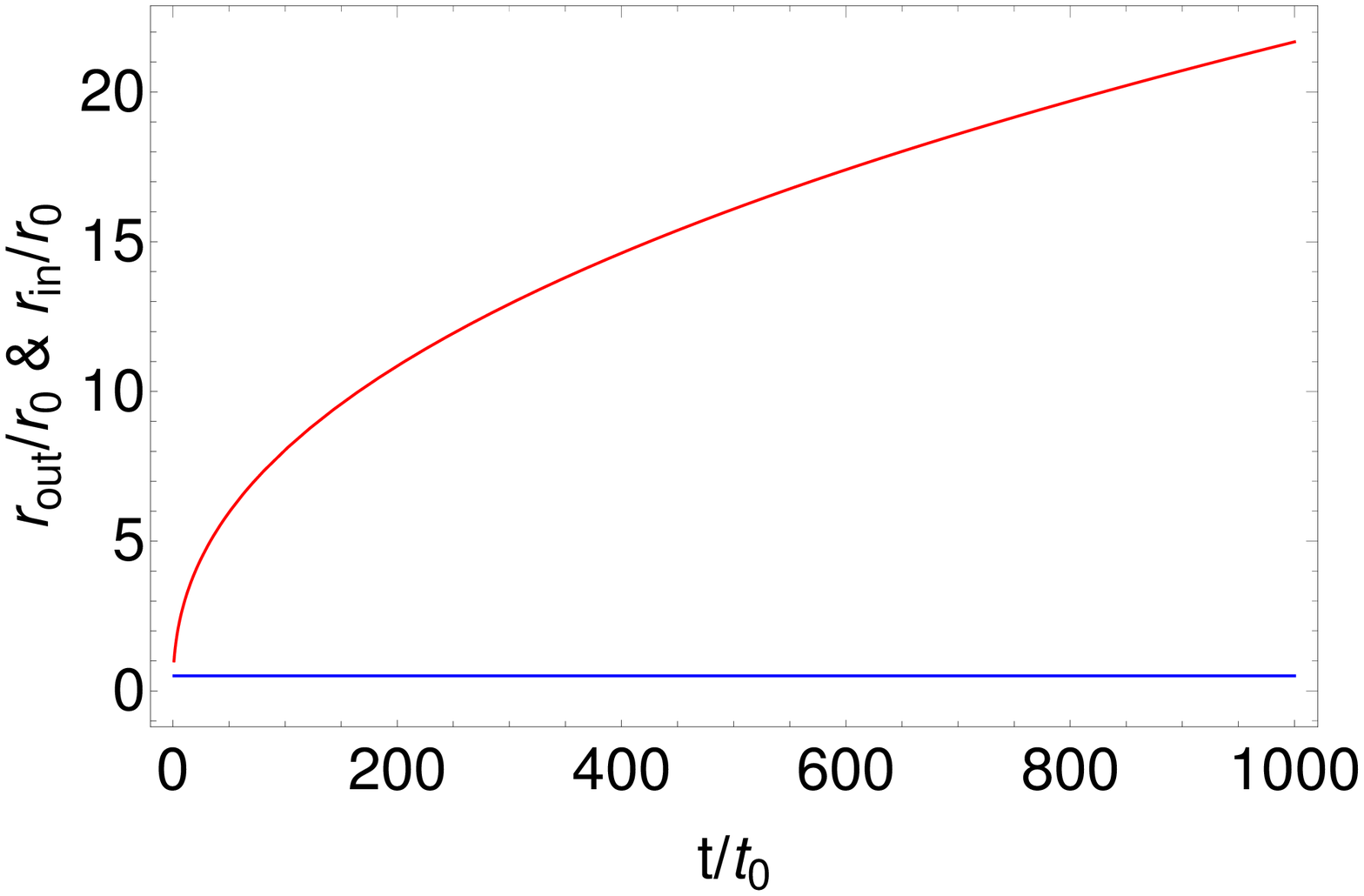}}
\end{center}
\caption{(a) The values of $\xi_{in}$ (blue) and $\xi_{out}$ (red) as a function of $t$ and (b) $r_{in}$ (blue) and $r_{out}$ (red) in terms of $r_0$ for model A2 and the parameter set I1. See \S\ref{modelA2}. }
\label{xinog}
\end{figure}

\begin{figure}
\begin{center}
\includegraphics[scale=0.45]{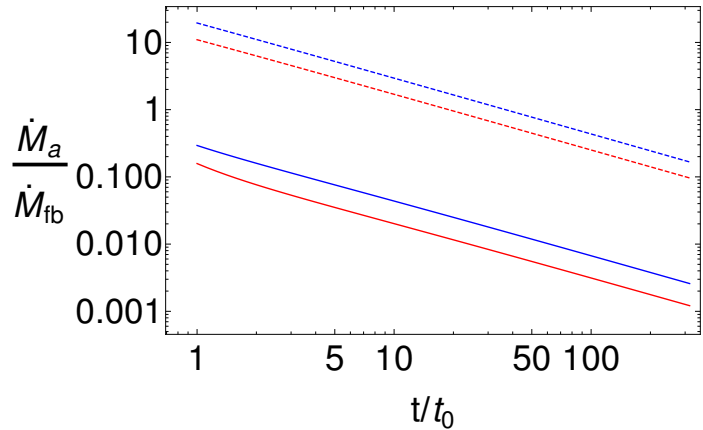}
\end{center}
\caption{The ratio of accretion rate to mass fallback rate from the disrupted debris is shown for model A2 with the parameter set I1 (blue), I3 (red), I5 (blue dashed) and I7 (red dashed). The late time decline is $\dot{M}_a/\dot{M}_{fb} \propto t^{-0.83}$ which results in mass accretion rate $\dot{M}_a \propto t^{-2.496}$ and is in agreement with $\dot{M}_a \propto t^{-5/2}$ obtained using eqn (\ref{mamfg}). The decline in the ratio results in an increase in the disk mass. See \S\ref{modelA2}. }
\label{mamfsubg}
\end{figure}

We consider that at initial time $t=t_0$, $r_0=q~r_{in}$ where $q$ is a free parameter, $\xi_{in}(t_0)=1/q$ and taking $\xi_{out}(t_0)=1$, the eqns (\ref{gascon1}, \ref{diss1}, \ref{masdis}) give

\begin{equation}
t_0 M_d(t_0)^{b-1}=\frac{\sqrt{GM_{\bullet}}}{K}\left[\frac{2+p}{2\pi A}\right]^{1-b} \frac{r_0^{1/2-d-2(1-b)}}{(\xi_{out}^{2+p}-\xi_{in}^{2+p})^{1-b}},
\label{t0subg}
\end{equation} 

\noindent where $K$ is given in eqn (\ref{gascon1}). The eqn (\ref{t0subg}) is solved to obtain $t_0$ which is then used in the eqn (\ref{diss1}) to calculate the $\Sigma_0$ and these are shown in Fig \ref{t0plotg}. The calculated values of $\Sigma_0$ and $r_0$ for each simulation set is given in Table \ref{svalsubg}. The ratio of accretion rate by the black hole and the mass fallback rate is given by

\begin{equation}
\frac{\dot{M}_a}{\dot{M}_{fb}}=1.5 \frac{q^{-\frac{7}{2}}}{1-q^{-\frac{7}{2}}} \left(\frac{M_d(t_0)}{M_{\odot}}\right)\left(\frac{M_{\star}}{M_{\odot}}\right)^{-1}\left(\frac{t_0}{t_m}\right)^{\frac{2}{3}} \left[\frac{\diff \mu_m}{\diff \varepsilon}(t)\right]^{-1}\left(\frac{t}{t_0}\right)^{-\frac{5}{6}}
\label{mamfg}
\end{equation}

\noindent where $\diff \mu_m/\diff \varepsilon$ is given in eqn (\ref{lod}). Since we assume that the matter added by the fallback debris is instantaneously distributed in the disk such that the self-similar structure remains the same, the outer radius of the disk expands as shown in Fig \ref{xinog}. The ratio of mass accretion rate to the mass fallback rate decreases with time as can be seen from Fig \ref{mamfsubg}. In eqn (\ref{mamfg}), the $\diff \mu_m(t)/\diff \varepsilon$ is nearly constant at late times and thus the mass accretion rate decreases as $\dot{M}_a \propto t^{-5/2}$. 

\begin{table}[h]
\caption{The values of $\Sigma_0$ and $r_0$ in units of $R_s=2.9 \times 10^{11} M_6$ in cm with $M_6=1$ obtained using equations given in Table \ref{modtab} for the sub-Eddington $\alpha$ disk model A2. See \S\ref{modelA2}.}
\label{svalsubg}
\scriptsize
\center
\scalebox{1.2}{
\begin{tabular}{|c|c|c|c|}
\hline
&&&\\
Set & $t_0 ({\rm days})$ & $\Sigma_0 (10^{10}{\rm g~cm^{-2}})$ & $r_0 (R_s)$ \\
\hline
&&&\\
I1 & 21.2 & $1.2$  & 6  \\
&&&\\
I2 & 12 & $2.8$ & 6 \\
&&&\\
I3 & 5891 & $0.026$ & 6 \\
&&&\\
I4 & 21.2 & $1.2$  & 6 \\
&&&\\
I5 & 12.12 & $1.7$  & 4.23 \\
&&&\\
I6 & 8.87 & $2.7$  & 4.23 \\
&&&\\
I7 & 2624.5 & $0.053$  & 4.23\\
&&&\\
I8 & 12.12 & $1.7$  & 4.23\\
&&&\\
\hline
\end{tabular}
}
\end{table}

\begin{figure}
\begin{center}
\subfigure[]{\includegraphics[scale=0.35]{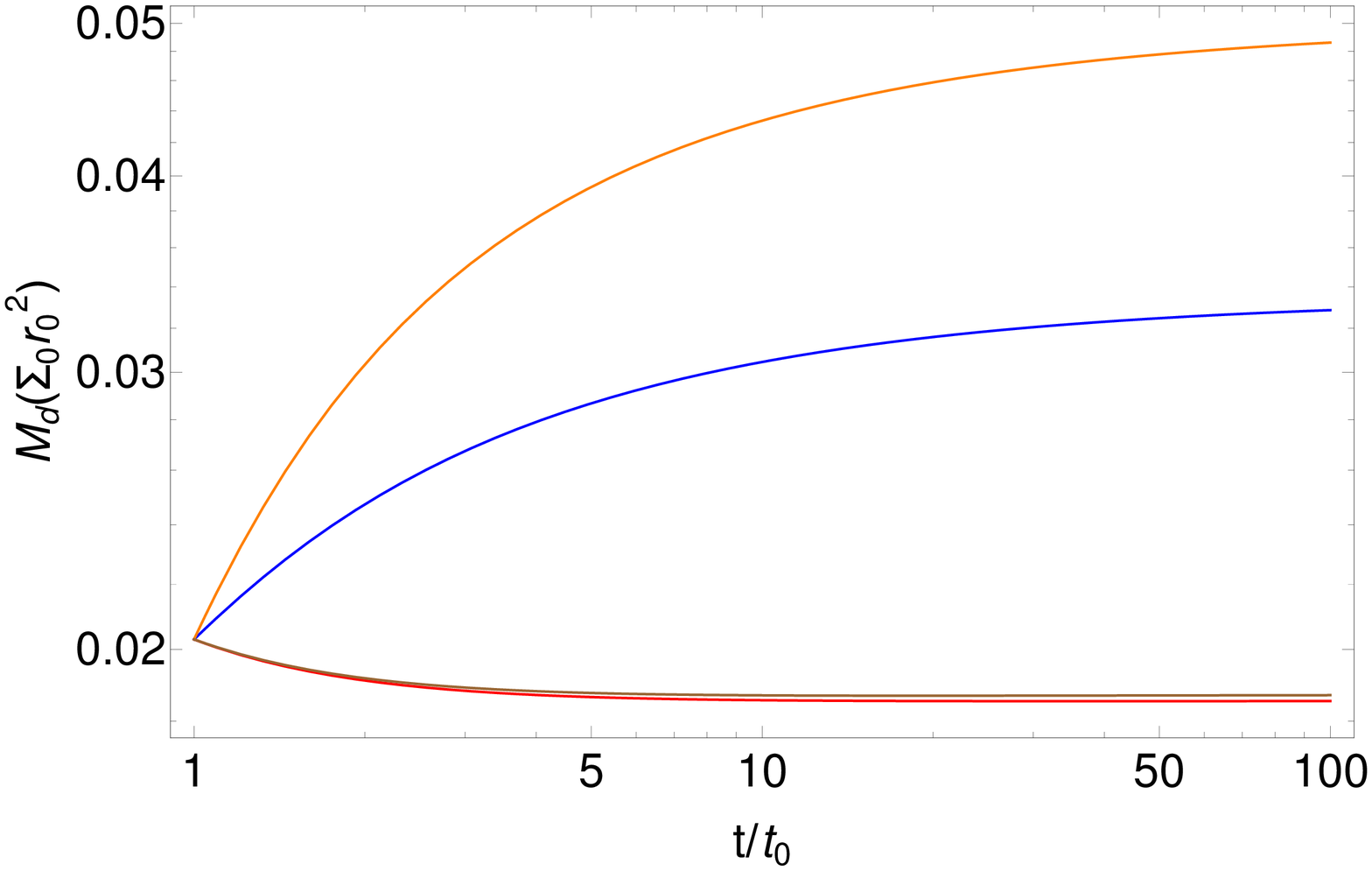}}
\subfigure[]{\includegraphics[scale=0.4]{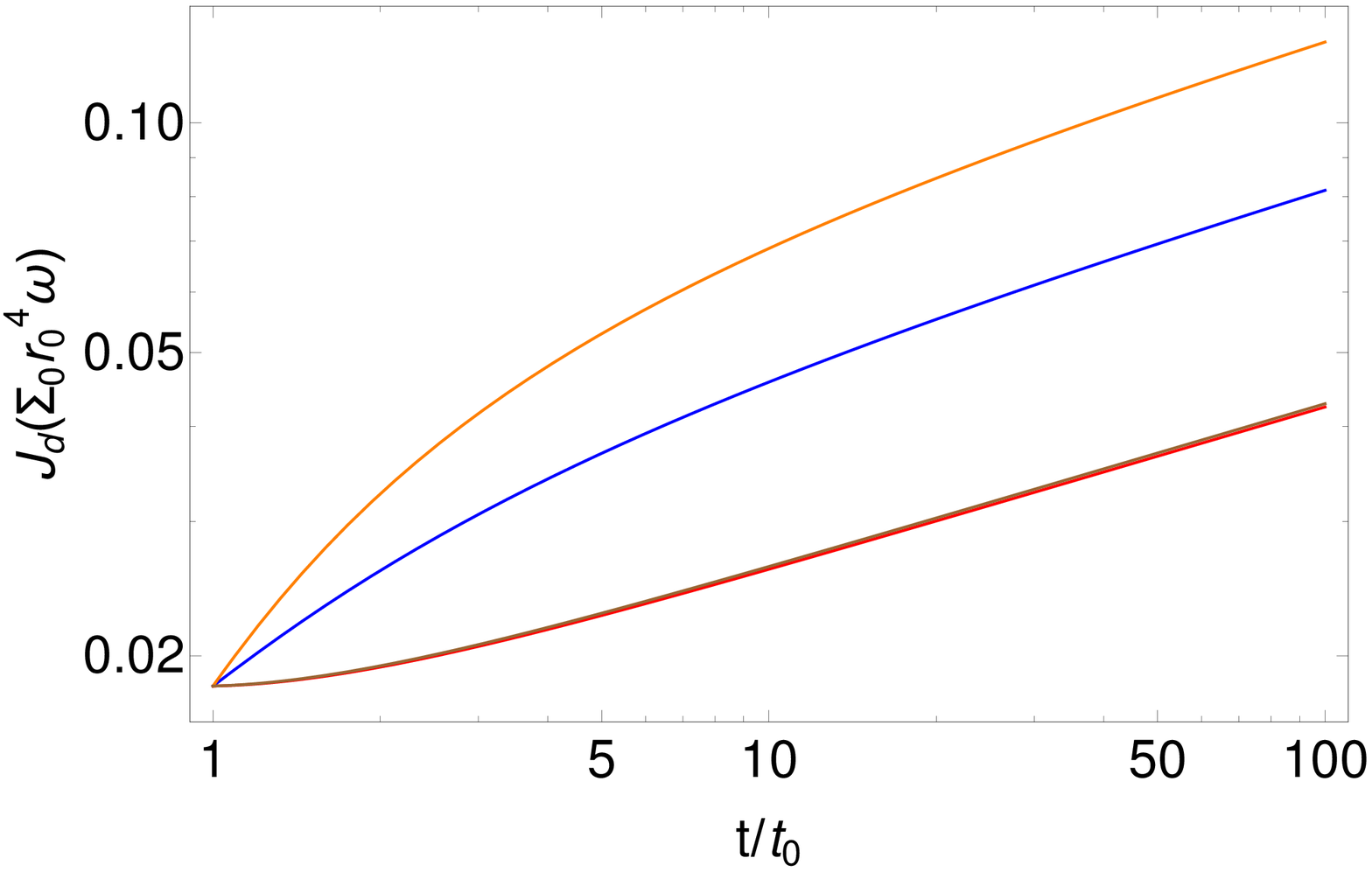}}
\end{center}
\caption{(a) The evolution of the mass of disk and (b) the angular momentum is shown for model A2 and the parameter set I1 (blue), I3 (red), I5 (orange) and I7 (brown) with the $\Sigma_0$ and $r_0$ values given in Table \ref{svalsub}. See \S\ref{modelA2}.}
\label{mjdsg}
\end{figure}

\begin{figure*}
\begin{center}
\subfigure[Bolometric]{\includegraphics[scale=0.41]{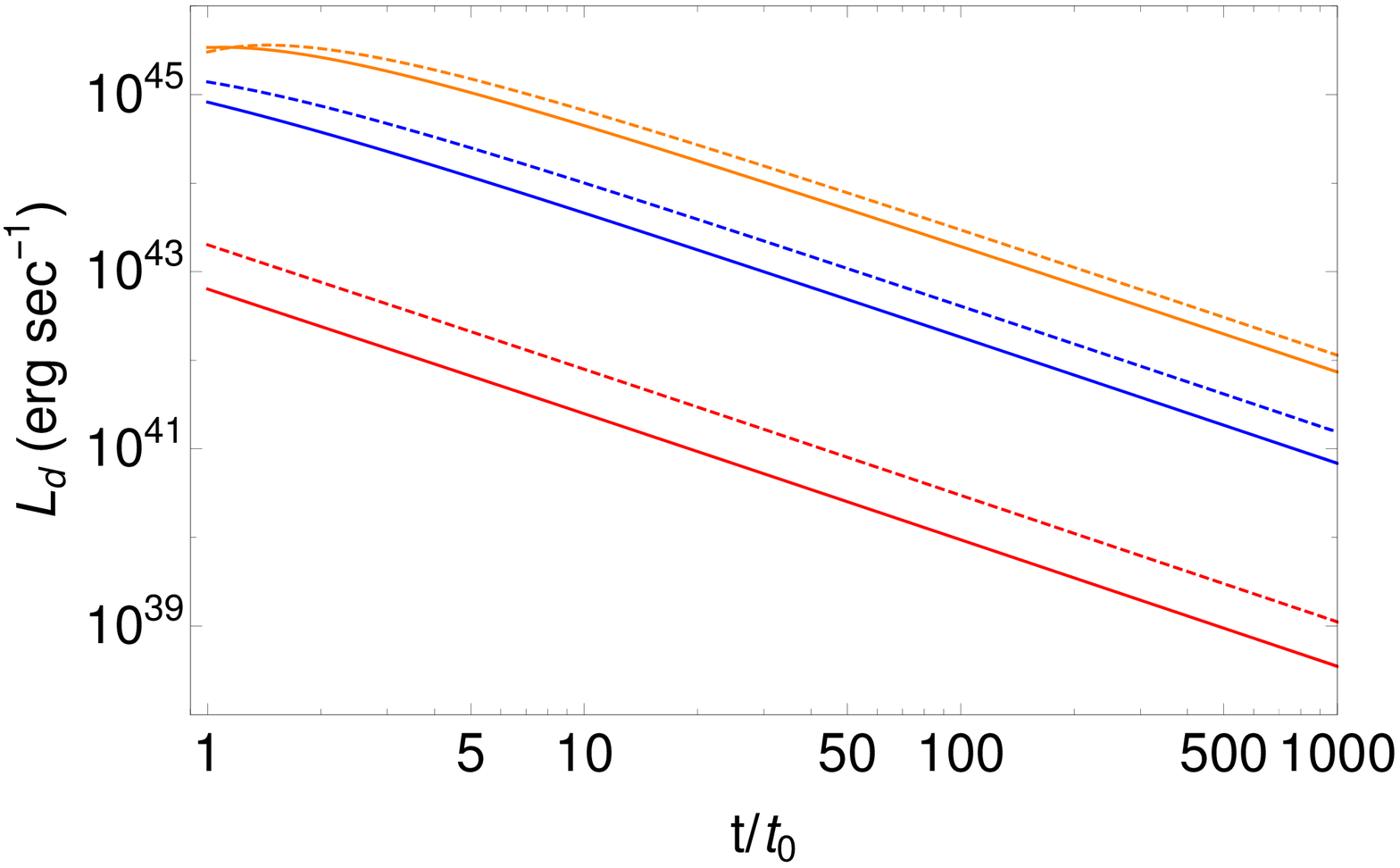}}
\subfigure[$n= \diff \ln L_d/\diff \ln t$]{\includegraphics[scale=0.4]{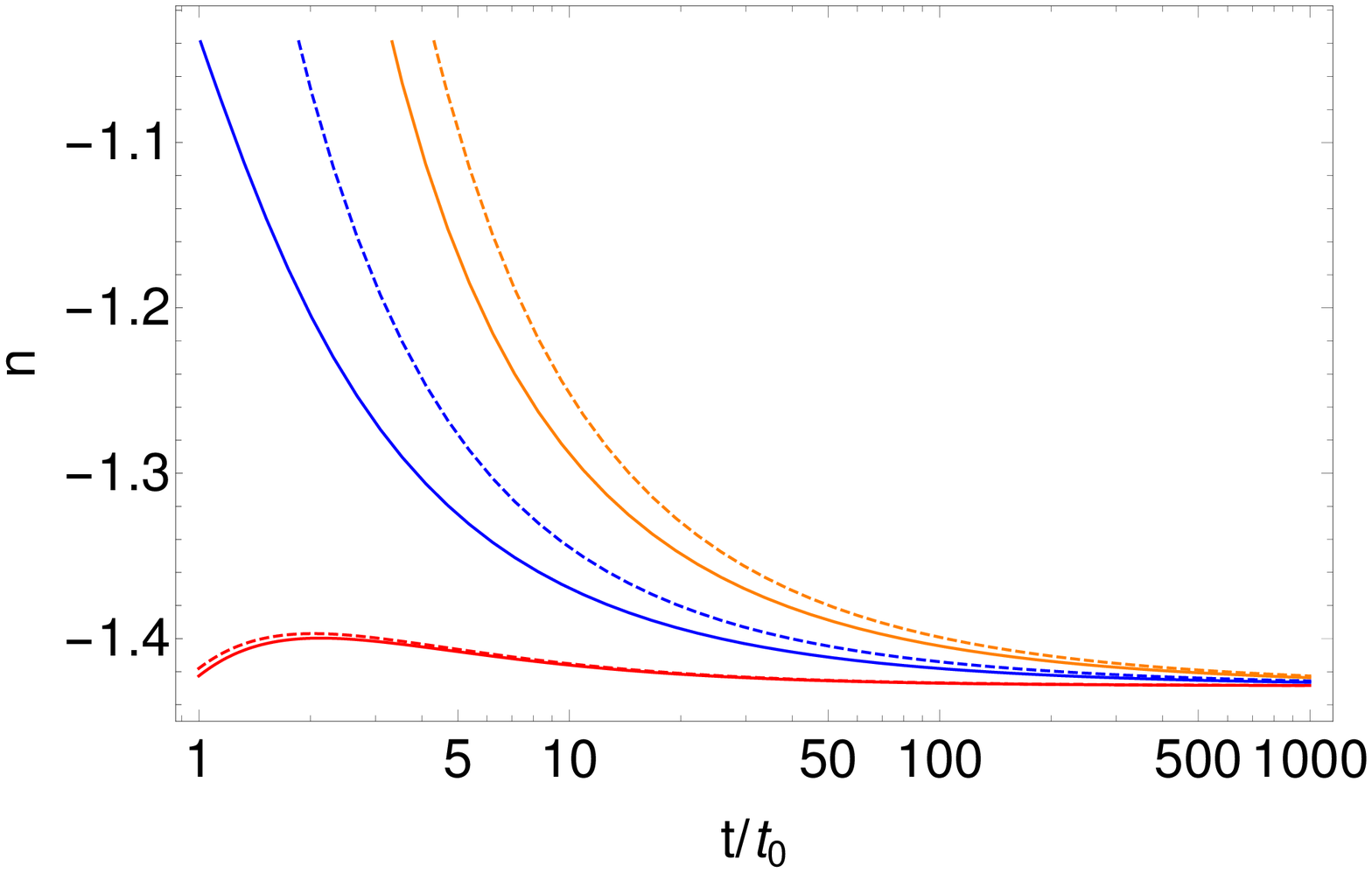}}
\subfigure[X-ray band]{\includegraphics[scale=0.42]{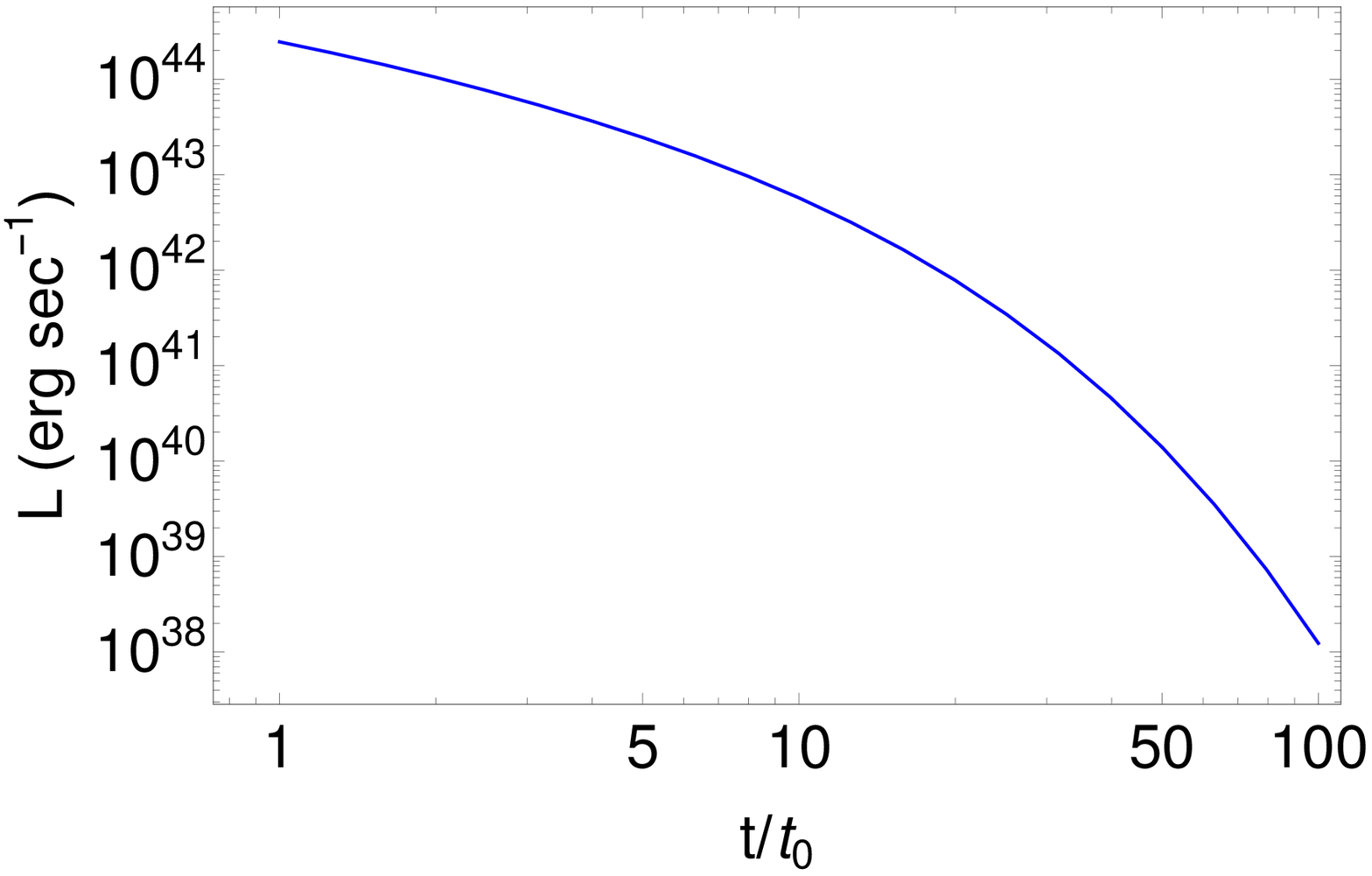}}
\subfigure[UV bands]{\includegraphics[scale=0.42]{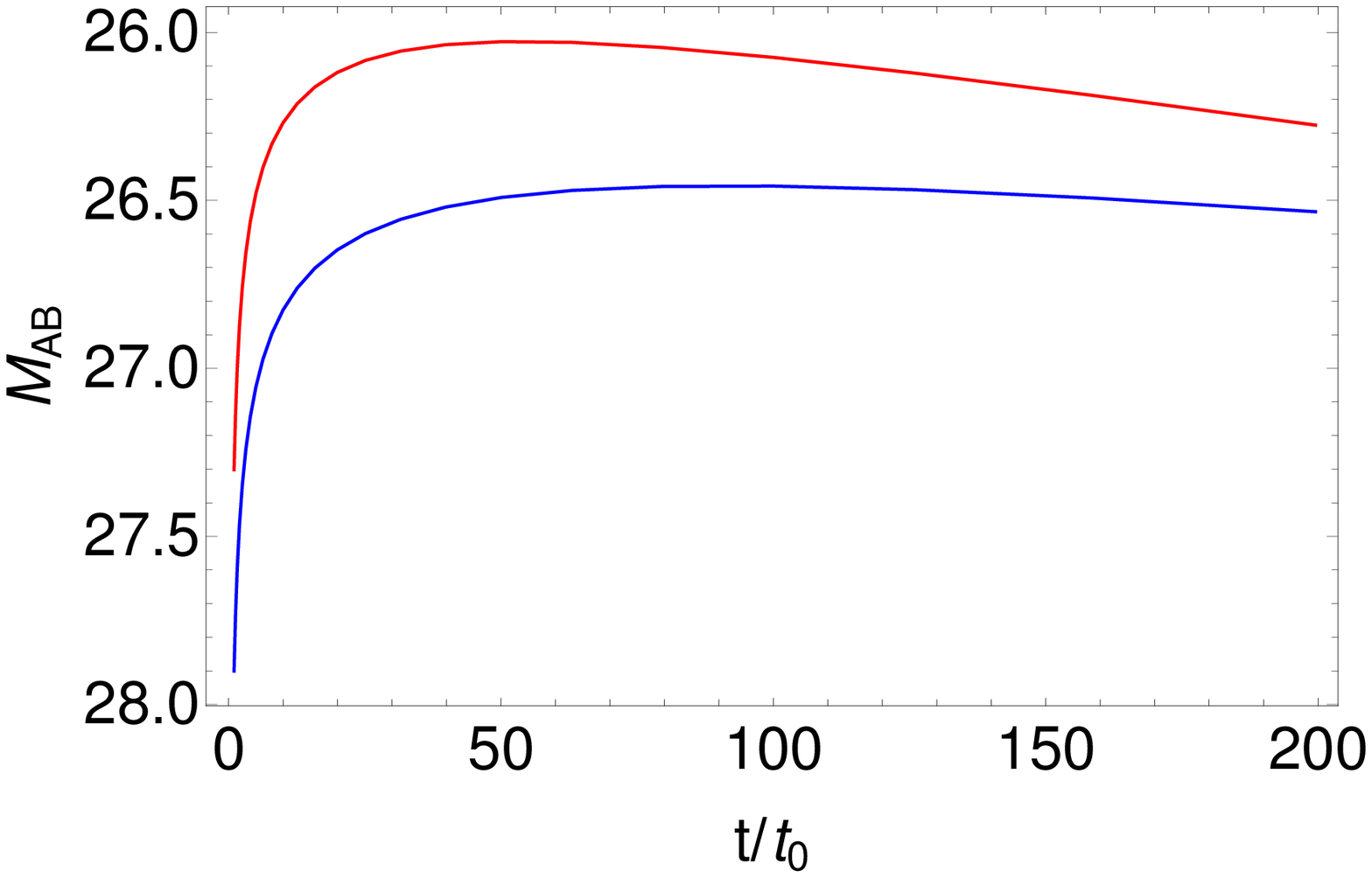}}
\subfigure[Optical bands]{\includegraphics[scale=0.42]{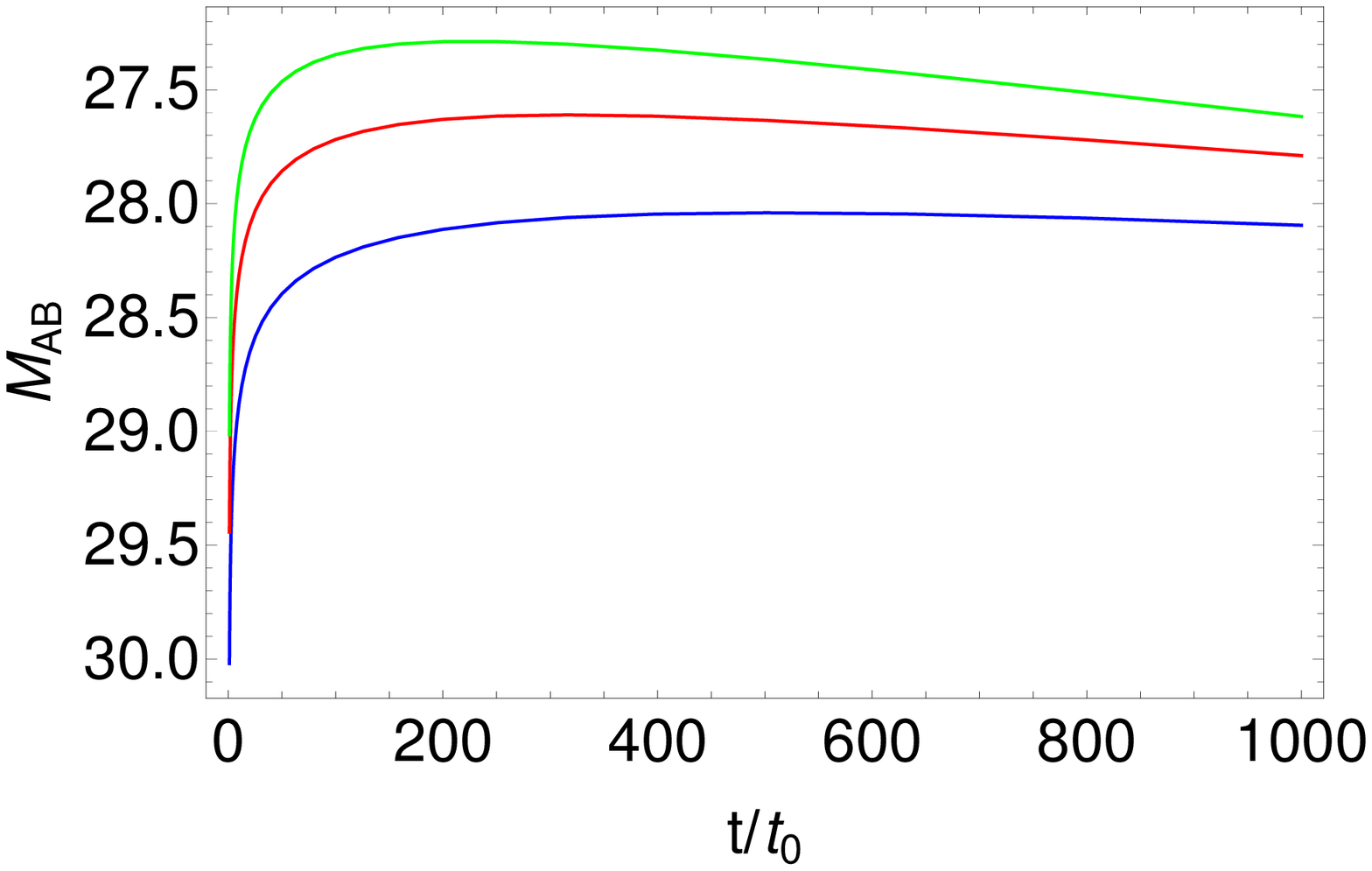}}
\end{center}
\caption{(a) The bolometric luminosity (eqn \ref{dislum}) as a function of $t$ obtained for model A2 with the parameter set I1 (blue), I2 (orange), I3 (red), I5 (blue dashed), I6 (orange dashed) and I7 (red dashed). (b) The time evolution of $n= \diff \ln L_d/\diff \ln t$  is shown for the luminosity curves shown in (a). The spectral luminosity simulated in various bands for the run I1 and redshift $z=0.05$ for soft X-ray in (c), UV where Swift UVM2 (1800-3000 $A^{\circ}$) (blue) and UVW2 (1500-2500 $A^{\circ}$) (red) in (d) and optical in (e) where the curves for V band (blue), B Band (red) and U Band (green) are indicated. See \S\ref{modelA2}. }
\label{lsubg}
\end{figure*}

From eqn (\ref{mamfg}) and Fig \ref{mamfsubg}, we see that the rate of mass loss by the disk due to accretion is lower than the mass gain by the addition of fallback debris at late times which results in the growth of the disk as shown in Fig \ref{mjdsg}. Since the mass fallback rate is also a declining function of time, the mass of the disk increases slowly with time at late times. Using eqns (\ref{hcwt}) and (\ref{visstres}), the disk height is given by 

\begin{equation}
\frac{H}{r}=6.5\times 10^{-3}~M_6^{-\frac{1}{6}}q^{\frac{1}{4}}\left(\frac{\alpha_s}{0.1}\right)^{\frac{1}{6}}\left(\frac{r_{in}}{R_s}\right)^{\frac{1}{4}}\left(\frac{\Sigma_0}{10^{10}~{\rm g~cm^{-2}}}\right)^{\frac{1}{3}} \left(\frac{t}{t_0}\right)^{-\frac{9}{28}}\xi^{\frac{3}{4}},
\end{equation}

\noindent where $M_6=M_{\bullet}/(10^6 M_{\odot})$. The bolometric luminosity is given by eqn (\ref{dislum}) and the spectral luminosity is given by eqn (\ref{splum}). The Fig \ref{lsubg}, shows the bolometric luminosity as a function of $t$ and the luminosity at late time decreases as $L_d \propto t^{-1.42}$. The increase in the mass fallback rate causes an enhancement in surface density, viscous heating, effective disk temperature and luminosity. The spectral luminosity simulated in various spectral bands for the set I1 are also shown.
 
\section{Model B: super-Eddington disk with radiative viscosity}
\label{modelB}

Following assumption similar to model A, we take $r_0=q~r_{in}$, at $t=t_0$, $\xi_{in}(t_0)=1/q$ and $\xi_{out}(t_0)=1$, so that by using eqns (\ref{diss1}, \ref{diss2}, \ref{masdis}), we obtain 

\begin{equation}
t_0^{-\frac{9}{8}} M_d(t_0)= \frac{2\pi}{2+p}A \left(\frac{\Psi_1}{\Psi_2}\right)^{\frac{1}{8}} r_0^{1/4} \left[\xi_{out}^{2+p}(t_0)-\xi_{in}^{2+p}(t_0)\right],
\label{t0supt}
\end{equation}

\noindent where 

\begin{eqnarray}
\Psi_1 &=& \frac{96e}{a}\delta_0\frac{m_p}{\sigma_T c} \left(\frac{c^2}{GM_{\bullet}}\right)^{-1}\left(\frac{k_B}{\mu m_p}\right)^4 (G M_{\bullet})^{-3}\\
\Psi_2  &=& \left[\frac{3}{\sqrt{8}}\left(\frac{k_B}{\mu m_p}\right)^{\frac{1}{2}} (G M_{\bullet})^{-\frac{7}{8}} a^{-\frac{1}{8}} \frac{\kappa^{\frac{7}{8}}}{\mathcal{W}}\right]^8.
\end{eqnarray}

Using eqn (\ref{t0supt}), we calculate $t_0$ which is then used in the eqn (\ref{diss1}) to calculate $\Sigma_0$.  Using eqn (\ref{diss1}) and $K$ from eqn (\ref{radviscosity}), we obtain

\begin{equation}
\frac{\beta_g^4}{1-\beta_g}=\frac{96e}{a}\delta_0\frac{m_p}{\sigma_T c} \left(\frac{c^2}{GM_{\bullet}}\right)^{-1}\left(\frac{k_B}{\mu m_p}\right)^4 (G M_{\bullet})^{-3} t_0.
\label{beta1}
\end{equation}

\begin{figure}
\begin{center}
\includegraphics[scale=0.35]{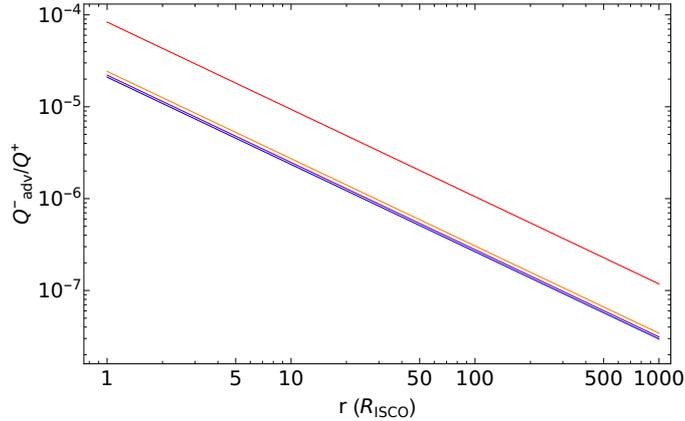}
\end{center}
\caption{The ratio of advective rate to heating rate for the parameter sets I1 (blue), I2 (red), I4 (orange), and I5 (purple) shown in Table \ref{supar}. The most of energy generated due to viscosity is radiated in the form of radiation as advection is too small to be neglected. See \S\ref{modelB}.}
\label{adhrt}
\end{figure}

Using eqn (\ref{advisr}) and Fig \ref{adhrt}, we can see that the energy loss due to advection is very small compared to heating rate and this implies that the most of energy generated due to viscosity is radiated in the form of radiation. Thus, using eqn (\ref{dislum}) and neglecting the advection, it is seen that the disk luminosity is given by $L_b^d \propto (t/t_0)^{-5/3-\alpha/4}\left[\left(r_{out}/r_0\right)^{1/4-\delta_0}-\left(r_{in}/r_0\right)^{1/4-\delta_0}\right]$ such that at late times, the disk luminosity $L_b^d \propto t^{-5/3}$ for $\alpha=0$. Thus, we take  $\alpha=0$ in model B's subsequent calculations. 

\begin{figure}
	\begin{center}
		\includegraphics[scale=0.45]{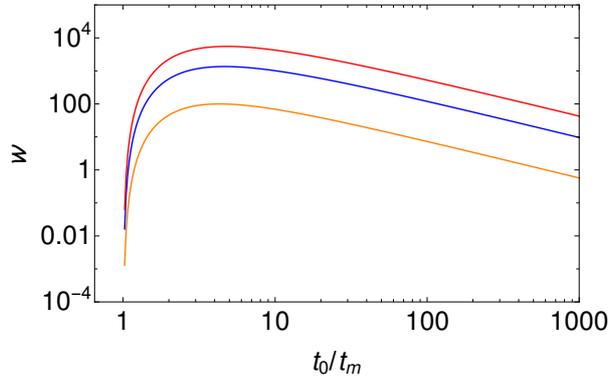}
	\end{center}
	\caption{The $\mathcal{W}$ as a function of parameter $t_0$ in terms of $t_m$ for the set I1 (blue), I2 (red) and I3 (orange) given in Table \ref{supar}. See \S\ref{modelB}.}
	\label{w1c}
\end{figure}

We have numerically solved the eqns given in Table \ref{modtab} to obtain the maximum value of $\mathcal{W}$ denoted by $\mathcal{W}_{max}$ as shown in Fig \ref{w1c} and normalized the $\mathcal{W}$  with respect to its maximum given by $\mathcal{W}_n=\mathcal{W}/\mathcal{W}_{max}$. The $\mathcal{W}_n$ is taken to be a free parameter and is used to obtain $t_0$ and then $\Sigma_0$ is given by eqn (\ref{t0supt}). For a given value of $\mathcal{W}_n$, there are two values of $t_0$ and we have considered the smaller value because the super-Eddington phase dominates early. The free parameters are $\bar{e},~\ell,~M_6,~m,~\alpha$ and $\mathcal{W}_n$. The value for $\delta_0$ is taken to be 0.05 which is typically the mid-value in the range, $0.02<\delta_0<0.1$, as shown in appendix \ref{drv}.

\begin{figure}
	\begin{center}
		\subfigure[]{\includegraphics[scale=0.38]{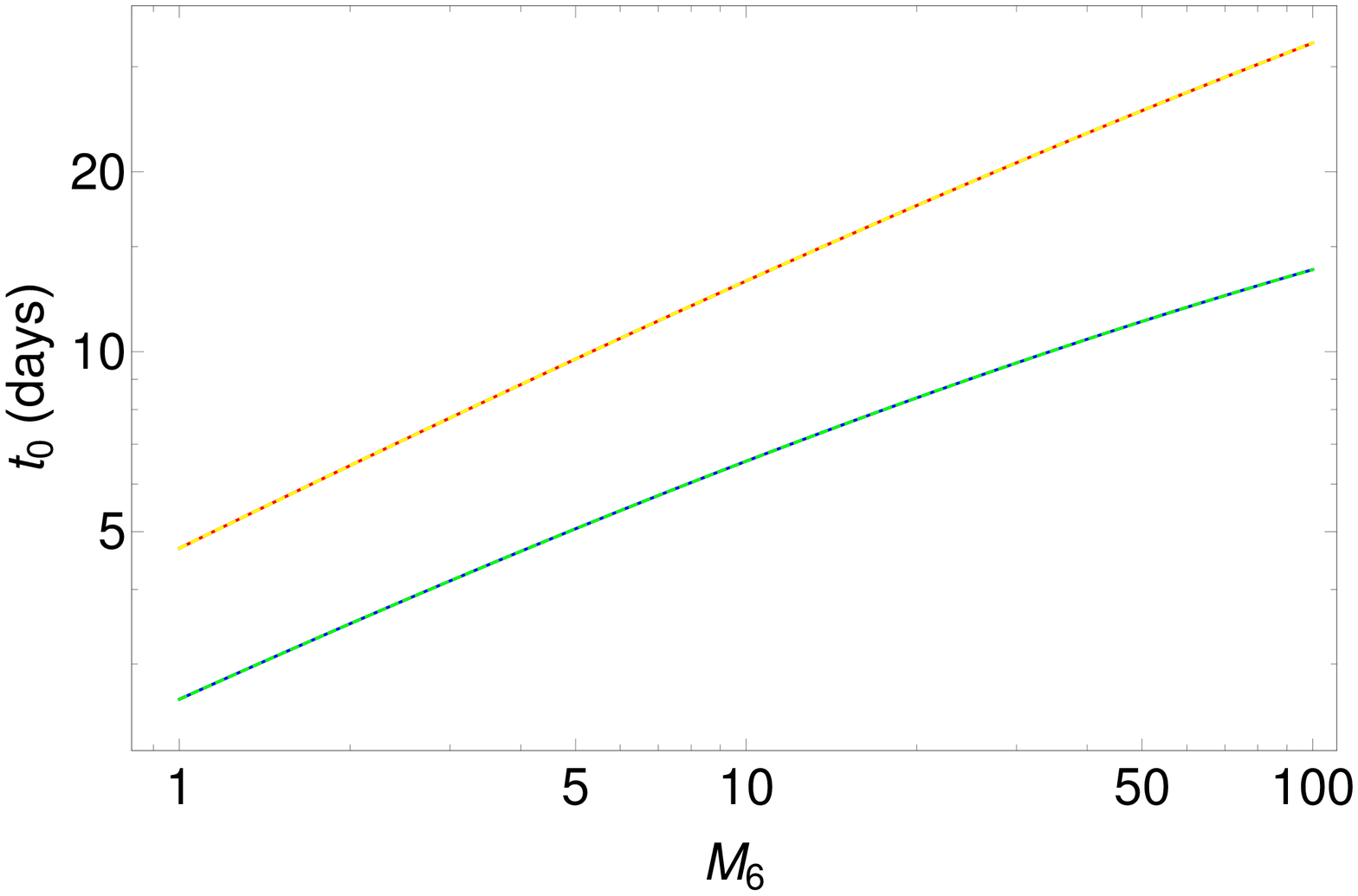}}
		\subfigure[]{\includegraphics[scale=0.40]{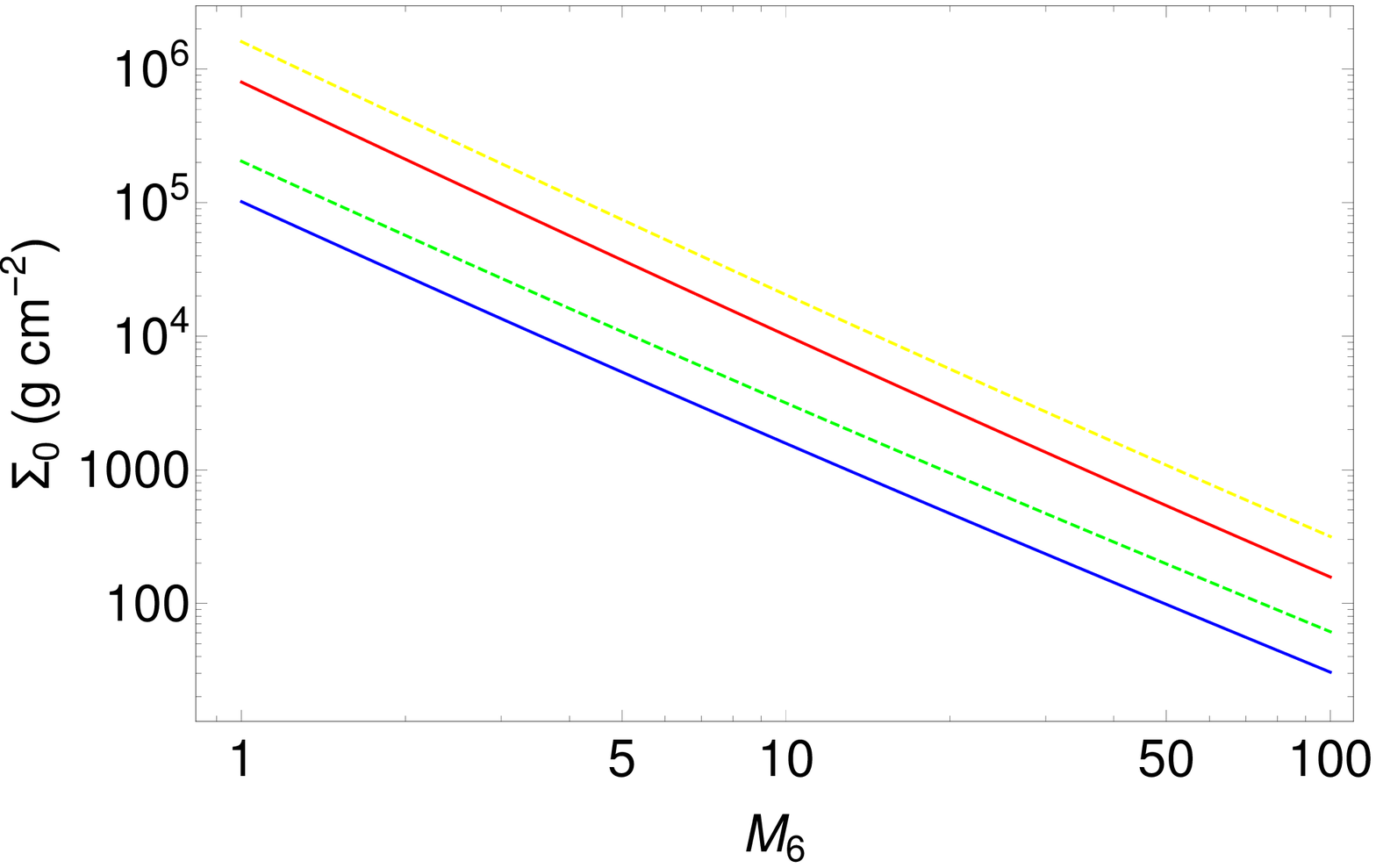}}
	\end{center}
	\caption{(a) The parameters $t_0$ in units of $t_m$ and (b) $\Sigma_0$ as a function of $M_6$ with other parameters given in set I1 (blue), I2 (red), I5 (green dashed) and I6 (orange dashed). See \S\ref{modelB}.}
	\label{t0sup}
\end{figure}

\begin{figure}
	\begin{center}
		\includegraphics[scale=0.4]{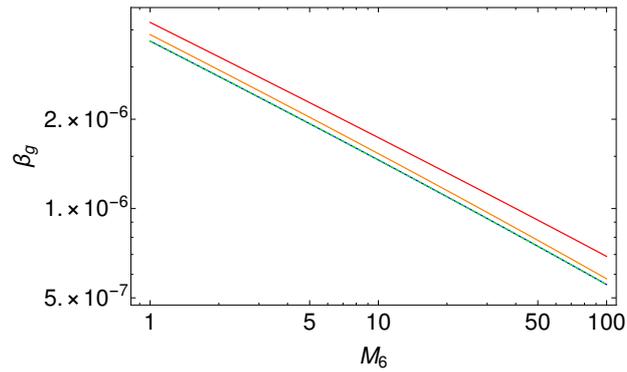}
	\end{center}
	\caption{The value of $\beta_g$ given by eqn (\ref{beta1}) is shown as a function of $M_6$ for the parameter sets I1 (blue), I2 (red), I4 (orange) and I5 (green dashed) given in Table \ref{supar}. See \S\ref{modelB}.}
	\label{betac}
\end{figure}

\begin{figure}
\begin{center}
\subfigure[]{\includegraphics[scale=0.4]{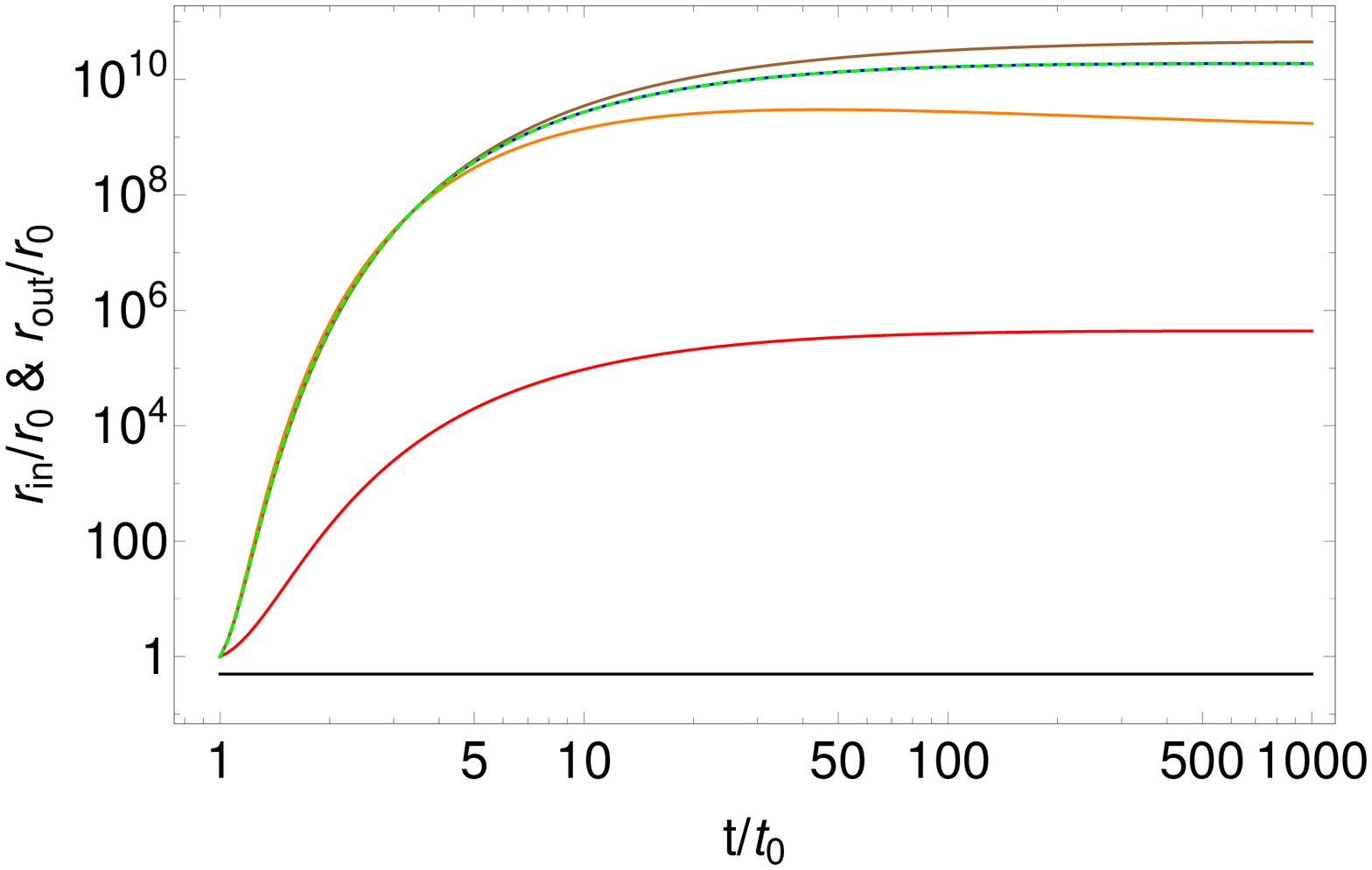}}~~~~
\subfigure[]{\includegraphics[scale=0.39]{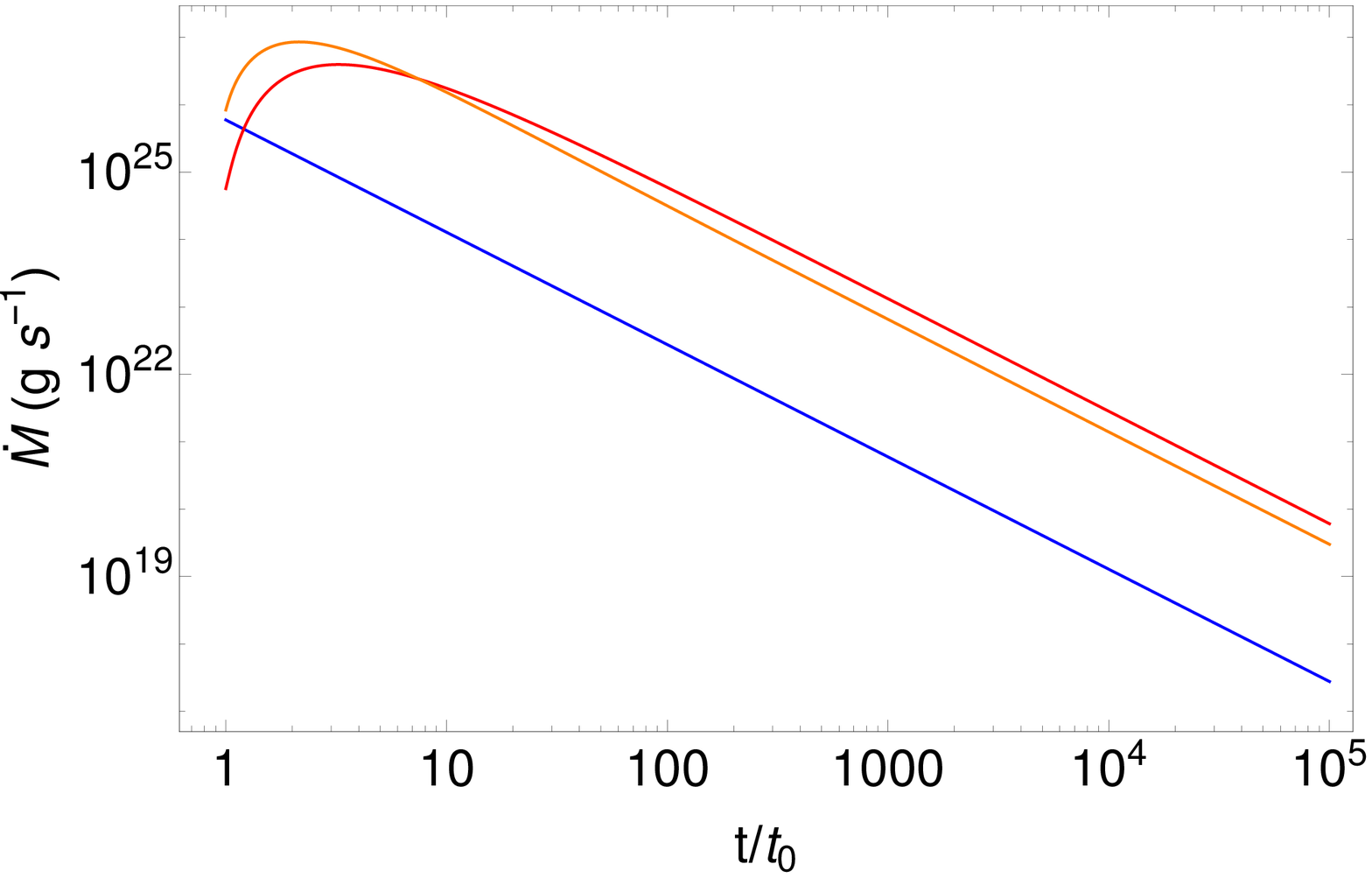}}
\end{center}
\caption{(a) The evolution of outer radius with time for the set I1 (blue), I4 (red), I3 (orange), I2 (brown) and I5 (green dashed). The black line shown the inner radius $r_{in}/r_0$. (b) The accretion rate $\dot{M}_a$ (blue), mass fallback rate $\dot{M}_{fb}$ (orange) and wind outflow rate $\dot{M}_w$ (red) for super-Eddington disk with time for the run I1. At the late stage, the wind loss rate dominates over fallback rate and accretion rate which implies that the disk mass will decrease at late stages. See \S\ref{modelB}.}
\label{r0supt}
\end{figure}

\begin{figure}
	\begin{center}
		\subfigure[]{\includegraphics[scale=0.42]{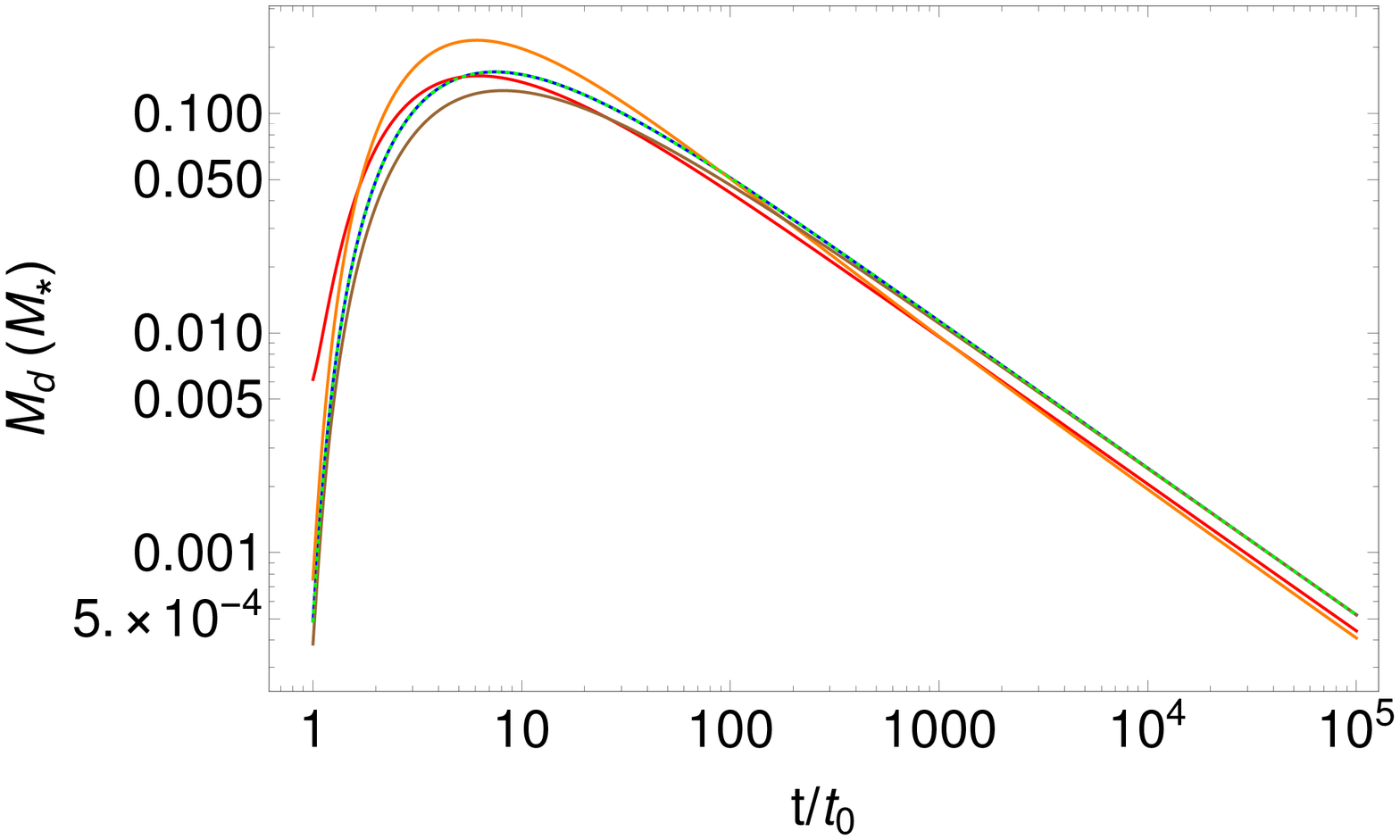}}~~~~
		\subfigure[]{\includegraphics[scale=0.39]{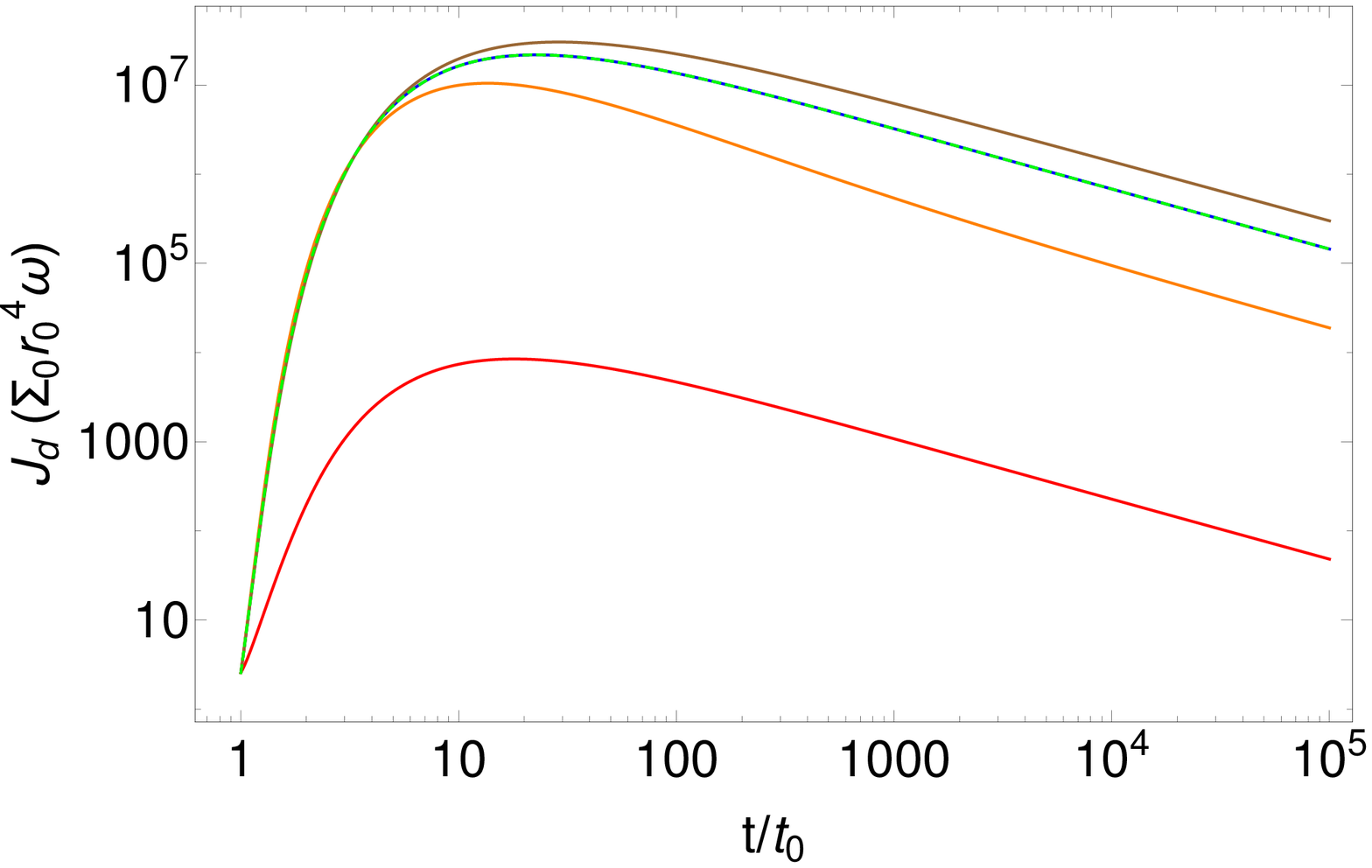}}
	\end{center}
	\caption{The evolution  of mass in (a) and angular momentum in (b) for super-Eddington disk with time for the set I1 (blue), I4 (red), I3 (orange), I2 (brown) and I5 (green dashed). The mass decreases at late stages because the mass loss rate due to out-flowing wind dominates over the mass fallback rate. See \S\ref{modelB}.}
	\label{mdjdsupp}
\end{figure}

The $t_0$ and $\Sigma_0$ values are shown in Fig \ref{t0sup} for model B. For higher $m$, the fallback rate of the debris is higher which results in the growth of the mass of the disk, enhanced surface density and viscous stress shortening  $t_0$. The pressure in the disk is dominated by radiation because of the small value of $\beta_g$ as shown in Fig \ref{betac}. The Fig \ref{r0supt} shows the evolution of outer and inner radius obtained using eqn (\ref{mceqnt}) and the radius decreases with $j$ and $\alpha$.  The Fig \ref{r0supt}b shows the evolution of $\dot{M}_a,~\dot{M}_w$ and $\dot{M}_{fb}$. In the initial stages, the mass fallback rate dominates and as the time progresses, the mass loss due to wind dominates over accretion and fallback which results in the reduction in the mass of the disk as shown in Fig \ref{mdjdsupp}. The outflowing wind also carries the angular momentum from the disk and the dominance of mass loss rate due to the wind later stages results in the reduction of disk angular momentum as shown in Fig \ref{mdjdsupp}. 

\begin{figure*}
	\begin{center}
		\subfigure[Bolometric]{\includegraphics[scale=0.4]{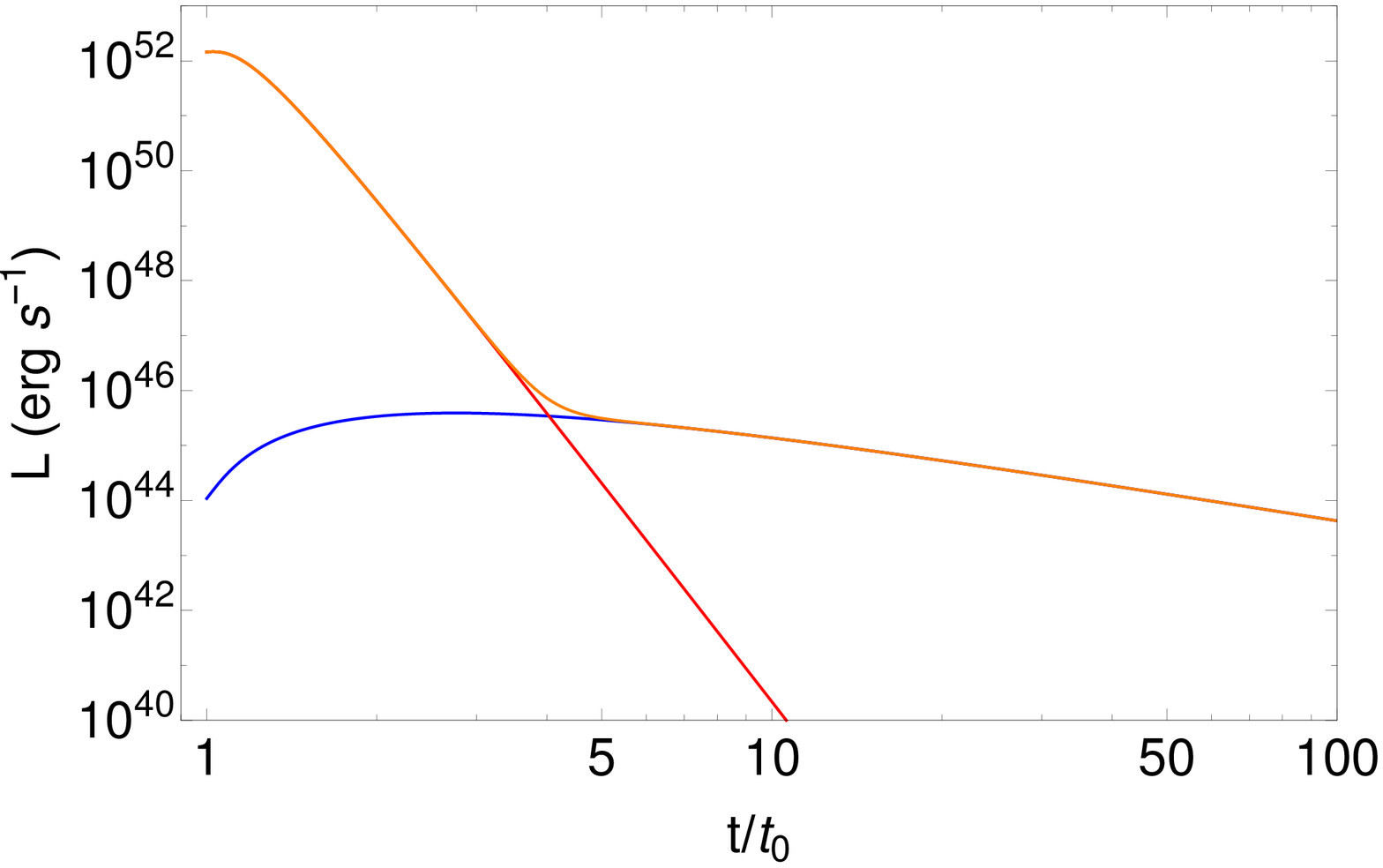}}
		\subfigure[X-ray band]{\includegraphics[scale=0.4]{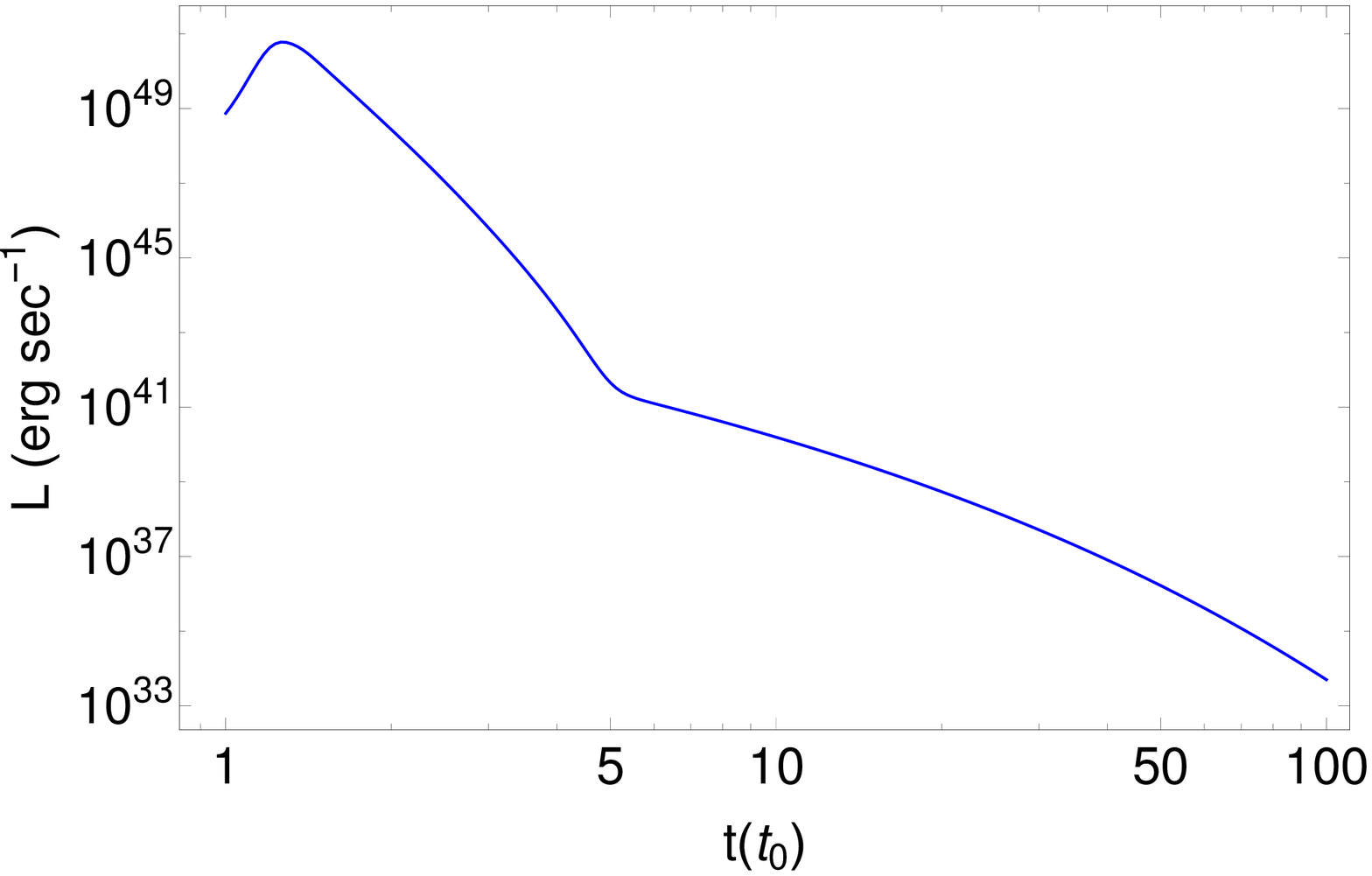}}
		\subfigure[UV bands]{\includegraphics[scale=0.4]{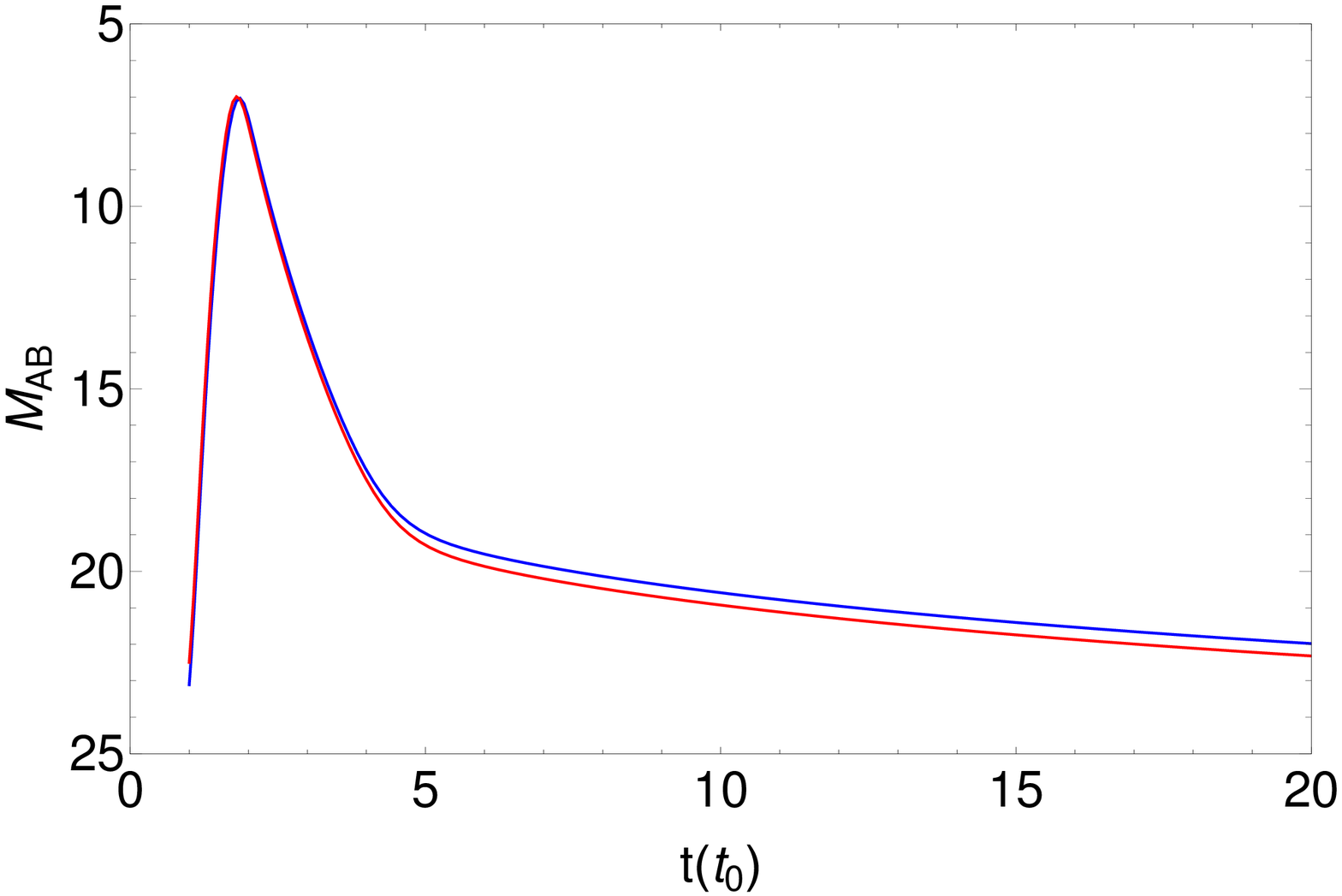}}
		\subfigure[Optical bands]{\includegraphics[scale=0.4]{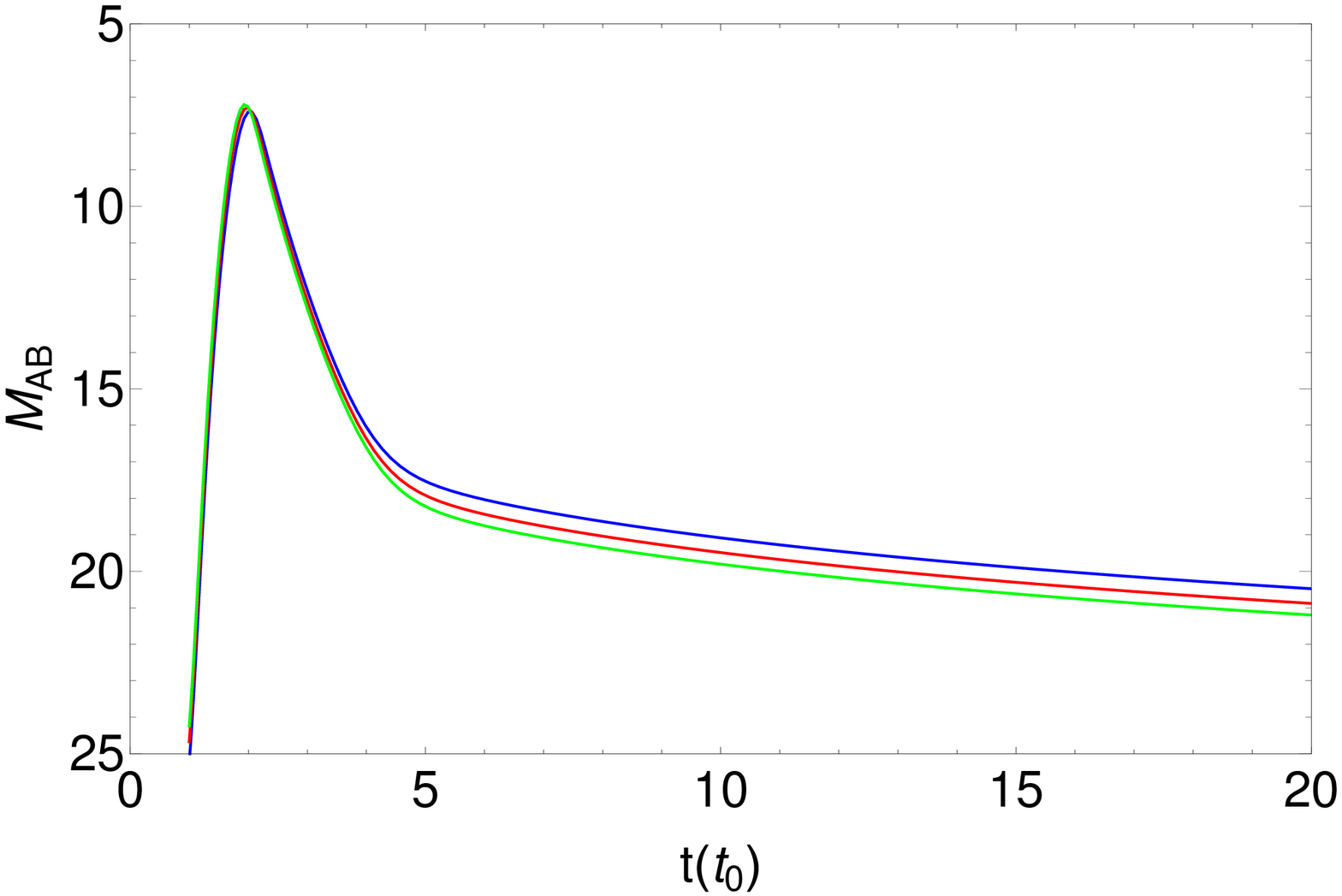}}
	\end{center}
	\caption{(a) The bolometric disk luminosity (blue), wind luminosity (red) and total luminosity (orange) of super-Eddington disk are shown for the run I1 and $c_2=1$. The spectral luminosity in various bands for redshift $z=0.1$ for soft X-ray in (b), UV where Swift UVM2 (1800-3000 $A^{\circ}$) (blue) and UVW2 (1500-2500 $A^{\circ}$) (red) in (c) and optical in (d) where the curves for V band (blue), B Band (red) and U Band (green) are indicated. See \S\ref{modelB}.}
	\label{lfsup}
\end{figure*}

The Fig \ref{lfsup} shows the bolometric luminosity obtained using eqn (\ref{dislum}) and spectral luminosity in various spectral bands using eqn (\ref{splum}). It shows that the wind luminosity declines faster compared to the disk luminosity which follows $t^{-5/3}$ law at the late stage. 

\section{Time behaviour of the models}
\label{tbeh}

We list the key findings of our time-dependent accretion models below.
	
\begin{enumerate}[1.]
	
\item By comparing runs I1 and I2, we find that the self-similar time parameter $t_0$ increases with black hole $M_6$ and star mass $m$ for both sub-Eddington model A1 and super-Eddington model B (see Figs \ref{t0plot}a, \ref{t0sup}a) but decreases with $m$ for sub-Eddington model A2 (see Fig \ref{t0plotg}a). The $\Sigma_0$ decreases with black hole mass $M_6$ and increases with star mass $m$ (see Figs \ref{t0plot}b, \ref{t0plotg}b, \ref{t0sup}b). The $t_0$ is high for model A2 compared to model A1 which implies that accretion begins late in case of sub-Eddington disk with gas pressure. The $t_0$ decreases with $j$ (runs I1 and I5) in case of model A1 and A2, but has insignificant effect in case of model B.
        
\item The accretion rate dominates over the fallback rate at late times in model A1 as shown in Fig \ref{xino}b, whereas the mass fallback dominates at late times in model A2 as shown in Fig \ref{mamfsubg}. The outflow rate dominates over accretion and fallback in case of model B as shown in Fig \ref{r0supt}b. The accretion rate increase with black hole mass (runs I1 and I3) and decreases with BH spin $j$ (runs I1 and I5).
        
\item In case of models A1 and A2, the disk mass, and angular momentum decreases with $M_6$ (runs I1 and I3) due to increase in accretion rate as shown in Figs \ref{mjds} and \ref{mjdsg}. In case of model A2, the disk mass increases with time for low mass black holes and decreases for high mass black holes. In case of model B, the disk mass and angular momentum decreases with $M_6$ (runs I1 and I3) and has an insignificant variation with $j$ (runs I1 and I5) as shown in Fig \ref{mdjdsupp}.
        
\item In case of model A1, the bolometric luminosity increases with $M_6$ (runs I1 and I3) and $m$ (runs I1 and I2), however, the increase is higher with $m$ as can be seen from Fig \ref{lsub}a. For model A2, the bolometric luminosity decreases with $M_6$ (runs I1 and I3) and increases with $m$ (runs I1 and I2) as can be seen from Fig \ref{lsubg}a. The wind luminosity drops faster in model B than the disk luminosity as shown in Fig \ref{lfsup}a, suggesting that the disk luminosity dominates at late stages. 
        
\item The $\beta_g$ for model B decrease with $M_6$ and increases with $m$ (runs I1 and I2) and $\mathcal{W}_n$ (runs I1 and I4) but shows no variation with $j$ (runs I1 and I5) as can be seen from Fig \ref{betac}. 
		
\end{enumerate}

\section{Fit to observations}
\label{obsfits}

The dynamics of TDEs depend on various physical parameters such as $M_6$, $m$, $\bar{e}$, $\ell$ and the multiwavelength observations of TDEs are useful in constraining the parameters. The TDE candidates used for the model fits are shown in Table \ref{ob} along with their redshift and inferred black hole mass. We have used the PS1-10jh observations to fit the models in optical and UV bands that also includes the initial rising phase of TDE emission. We have also taken the Swift J1644+57, XMMSL1 J061927.1-655311 and SDSS J120136.02+300305.5 observations in X-rays.

\begin{table*}
\caption{Observations considered for our model fits with redshift $z$ and inferred black hole mass. The black hole mass inferred for Swift J1644+57 is further modified by \citet{2014MNRAS.437.2744T} through their TDE magnetic accretion disk (MAD) model which constrained the TDE to a disruption of main sequence star by $M_{\bullet} \sim 10^5-10^6~M_{\odot}$. \citet{2016ApJ...819...51L} through the study of the host galaxy photometry obtained the black hole mass to be $3\times 10^6~M_{\odot}$.
}
\label{ob}
\begin{center}
\begin{tabular}{|c|c|c|c|c|}
\hline
&&&&\\
Observations & Band & Redshift & Black hole mass & References  \\
&& $z$ & $M_{\bullet} (10^6 M_{\odot})$ &\\
\hline
&&&&\\
XMMSL1 J061927.1-655311 &  X-ray & 0.146 & 60 $\pm$ 30  & {\citet{2014A&A...572A...1S}} \\
&&&&\\
SDSS J120136.02+300305.5 & X-ray & 0.0729 & 20 & {\citet{2012A&A...541A.106S}} \\
&&&&\\
PS1-10jh & Optical and UV & 0.1696 & 2.8 & {\citet{2012Natur.485..217G}} \\
&&&&\\
Swift J1644+57 & X-ray & 0.354 & 6-20 & {\citet{2011Natur.476..421B}} \\
&&&&\\
\hline
\end{tabular}
\end{center}
\end{table*}

\begin{figure*}
\begin{center}
\scalebox{0.9}
{
\tikzstyle{decision} = [diamond, draw,  
    text width=6em, text badly centered, node distance=5cm, inner sep=0pt]
\tikzstyle{block} = [rectangle, draw, 
    text width=12em, text centered, rounded corners, minimum height=2em]
\tikzstyle{line} = [draw, -latex', line width=0.102em]
\tikzstyle{cloud} = [draw, ellipse,node distance=3cm,
    minimum height=2em]

\begin{tikzpicture}[node distance = 1.4cm, auto]
\node [block] (para) {Physical parameters \\ $M_{\bullet}$, $m$, $\bar{e}$, $\ell,~q$ and $j$};
\node [block, below of=para] (obs) {Observational data };
\node [block, left of=obs,node distance=6cm] (sub) {Sub Eddington model\\ (Model A) };
\node [block, below of=sub] (fit) {Fit to data};
\node [block, below of=fit] (obt) {Obtain the parameters};
\node [block, below of=obt] (bol) {Bolometric luminosity ($L_b$)\\ Eddington luminosity ($L_E^s$)};
\node [decision, right of=bol, node distance=6cm] (chk) {Is \\ $L_b \leq L_E^s$ ?};
\node [block, right of=chk,node distance=6cm] (sup) {Super Eddington Model\\ (Model B)};
\node [block, below of=sup] (fitsup) {Fit to data};
\node [block, below of=chk,fill=yellow!20,node distance=3cm] (derpar) {Physical parameters Derived};
\draw [->,thick] (para) -| (sub);
\draw [->,thick] (sub) -- (fit);
\draw [->,thick] (obs) -- (sub);
\draw [->,thick] (fit) -- (obt);
\draw [->,thick] (obt) -- (bol);
\draw [->,thick] (bol) |- (chk);
\draw [->,thick] (chk) -- node {Yes} (derpar);
\draw [->,thick] (chk) -- node {No} (sup);
\draw [->,thick] (sup) -- (fitsup);
\draw [->,thick] (obs) -| (sup);
\draw [->,thick] (para) -| (sup);
\draw [->,thick] (fitsup) |- (derpar);
\end{tikzpicture}
}
\end{center}
\caption{ The flow chart of the procedure we have adopted in fitting the model to the observations.}
\label{fc}
\end{figure*}
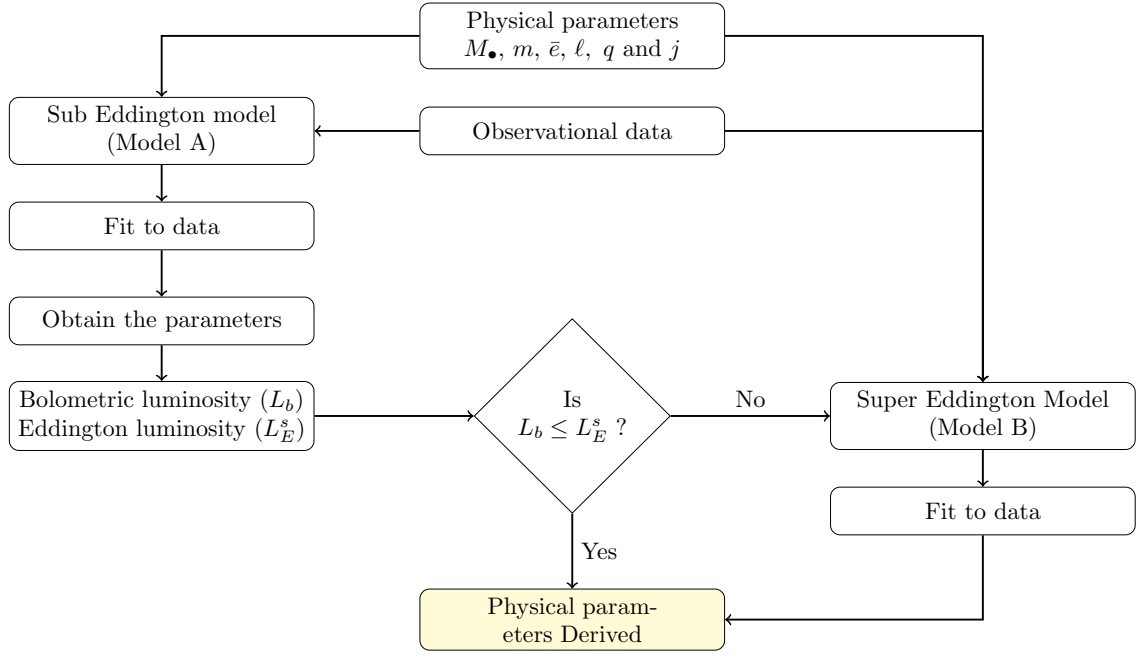

\begin{figure*}
	\begin{center}
		\subfigure[]{\includegraphics[scale=0.38]{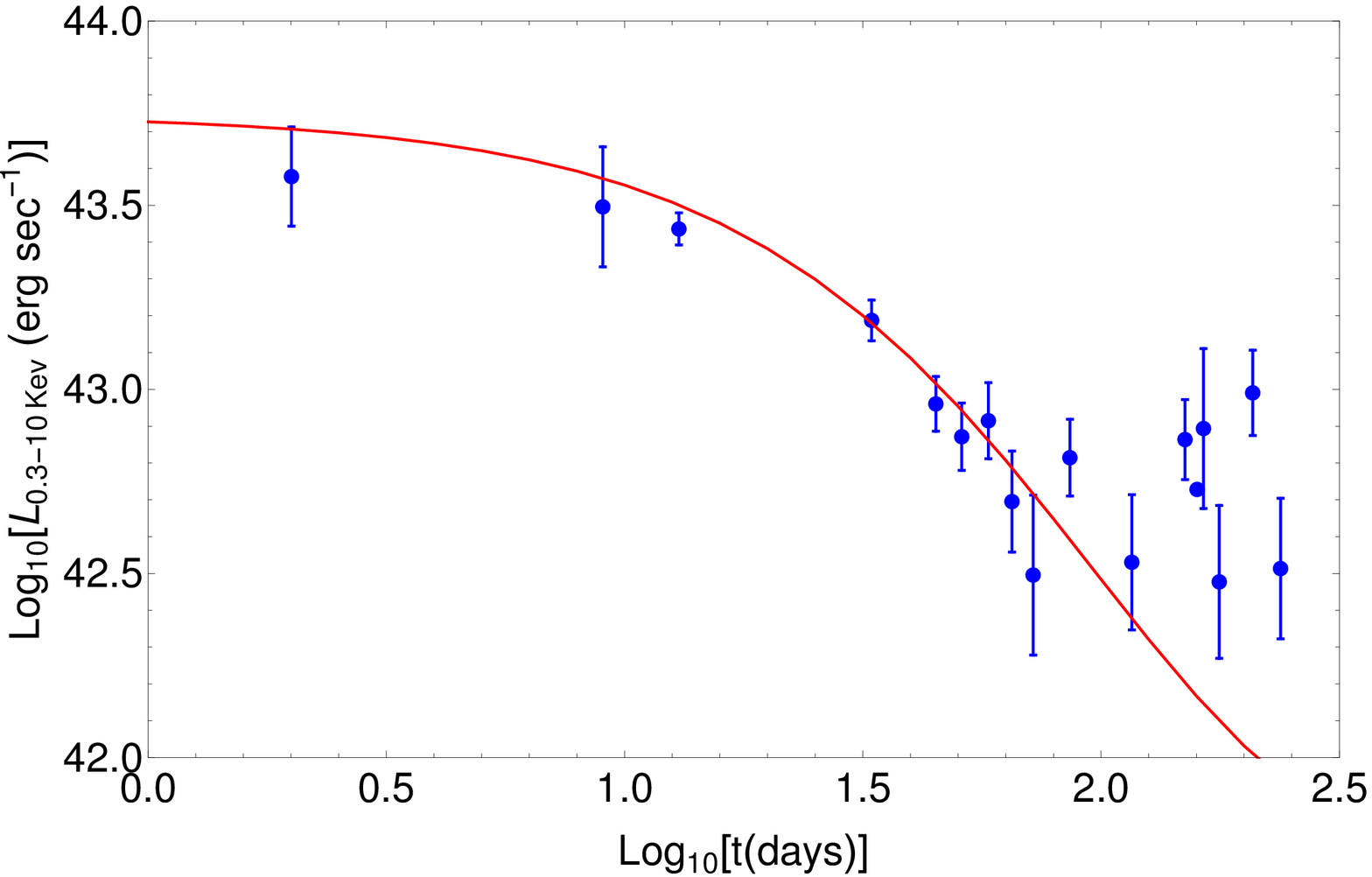}}
		\subfigure[]{\includegraphics[scale=0.40]{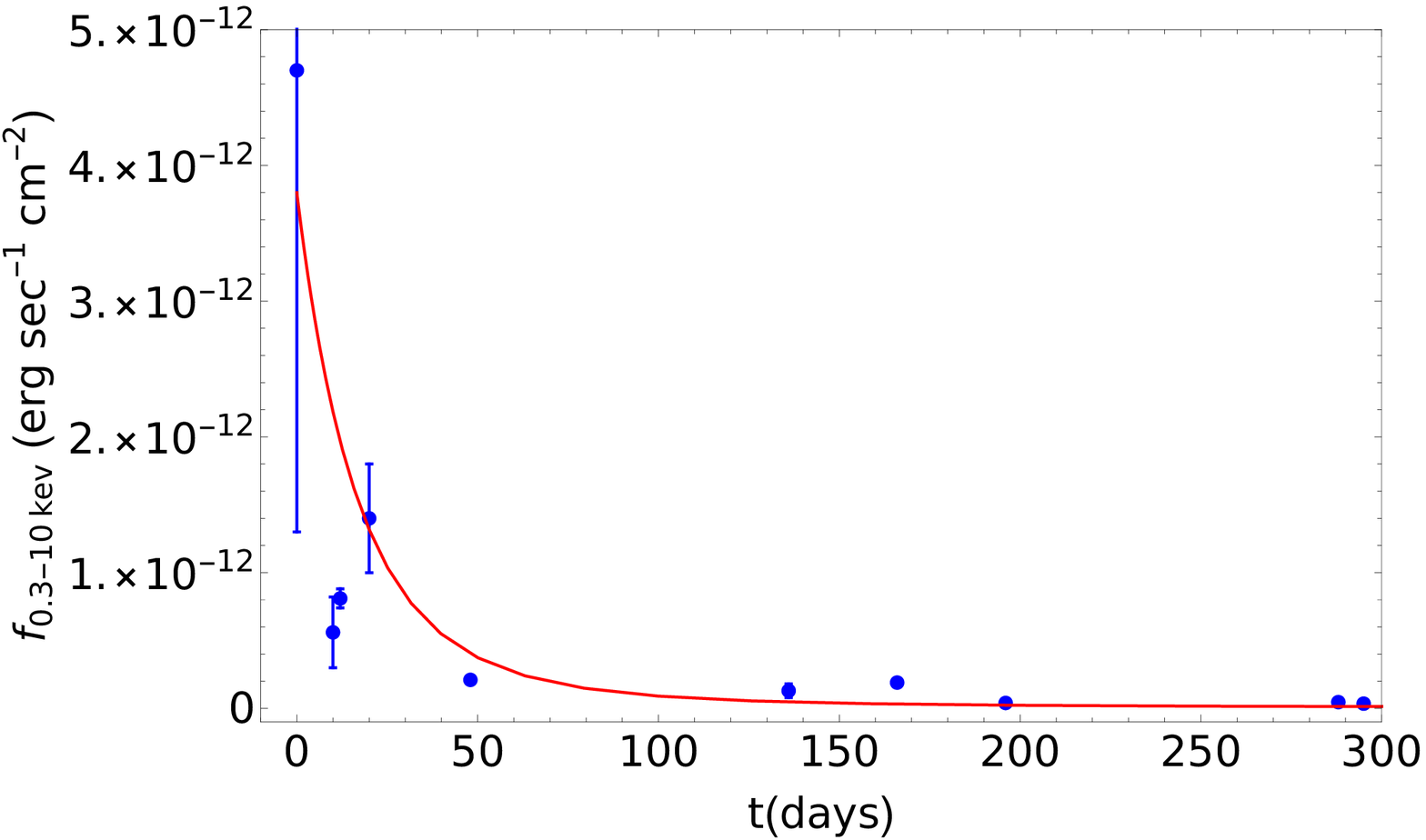}}
	\end{center}
	\caption{The model B (red) fits to the X-ray observations of XMMSL1 J061927.1-655311 (Left) \citet{2014A&A...572A...1S} and SDSS J120136.02+300305.5 (Right) \citet{2012A&A...541A.106S}. The derived parameters are given in Table \ref{ps1tab}.}
	\label{subobs}
\end{figure*}

\begin{figure*}
\begin{center}
\subfigure[PS1-10jh]{\includegraphics[scale=0.35]{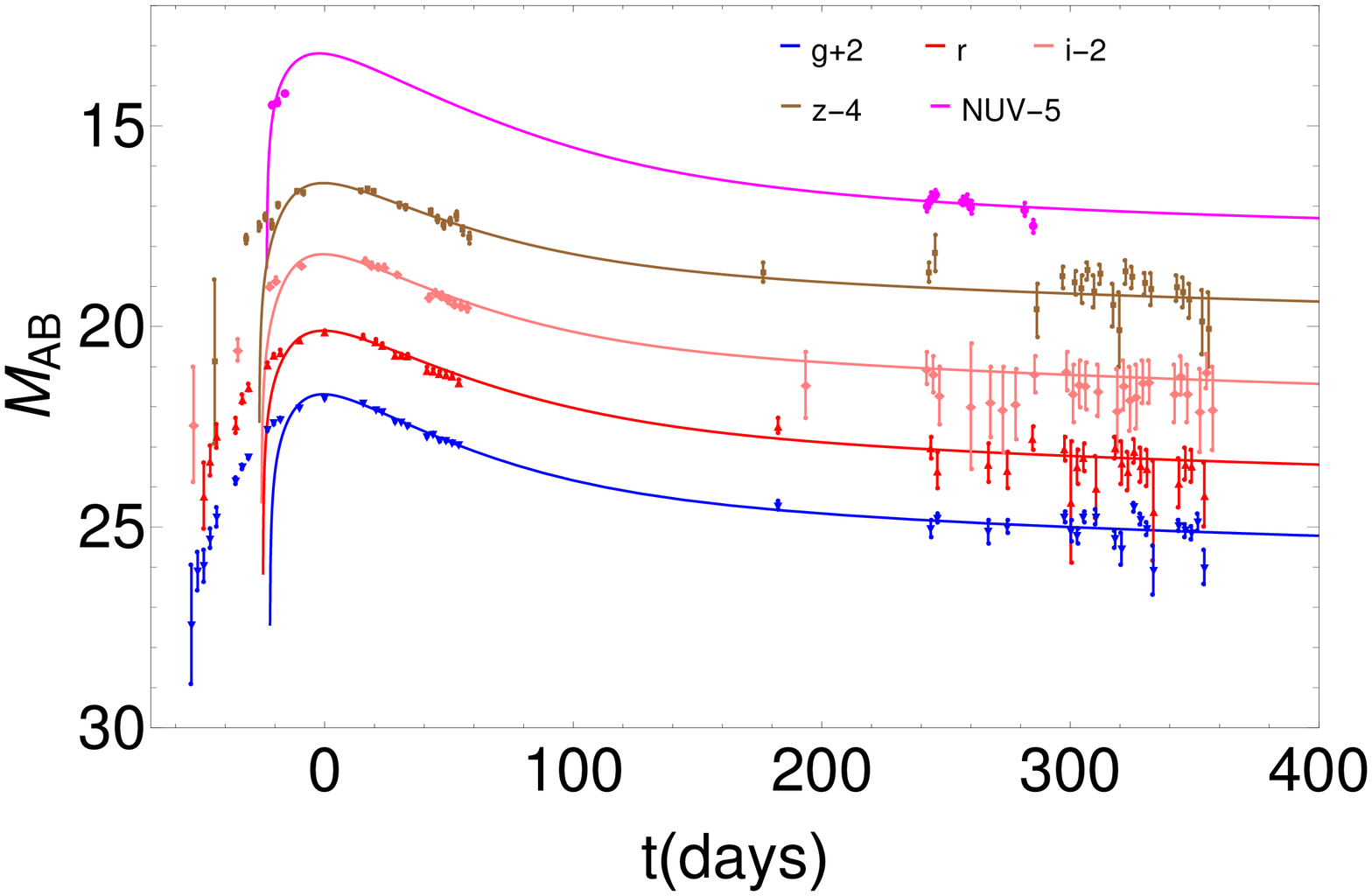}}
\subfigure[Swift J1644+57]{\includegraphics[scale=0.39]{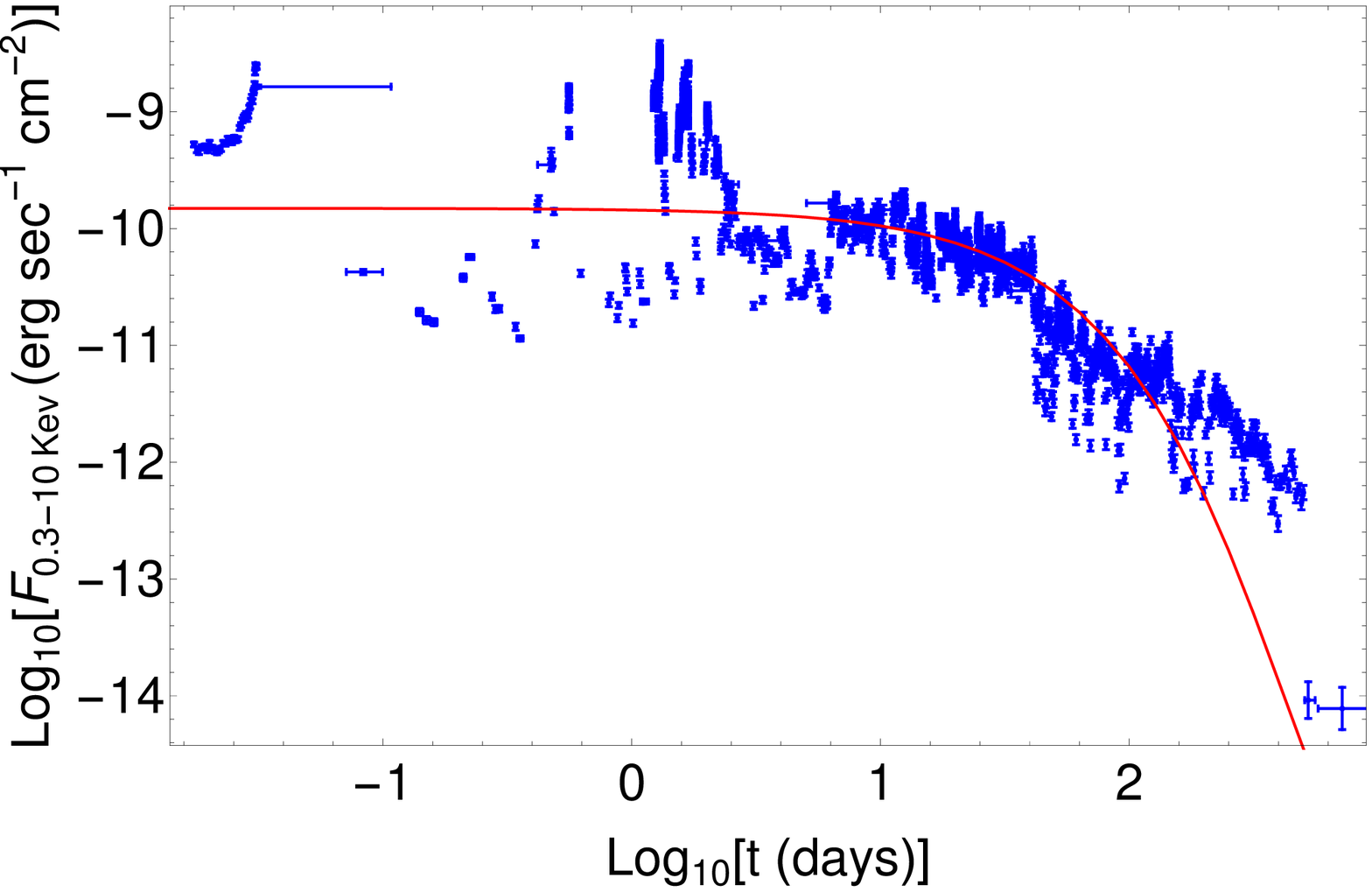}}
\end{center}
\caption{The model B (brown) fit to the PS1-10jh \citep{2012Natur.485..217G} and Swift J1644+57 \citep{2011Natur.476..421B} observations with the deduced parameters are given in Table \ref{ps1tab}. The fit to PS1-10jh can be improved by including model A in the rise part and then matching to model B.}
\label{ps1fit}
\end{figure*}

We have fit our time-dependent models A and B to the observed TDEs and derived the physical parameters such as dimensionless orbital energy $\bar{e}$ and angular momentum $\ell$, star mass $m$, black hole mass $M_6$, $q$ and $\delta t$, where $\delta t$ is the shift in time $t$ to fit the observations considering the starting time of accretion to be $t=t_0$. We first fit model A to the observations to obtain the physical parameters and then compared the bolometric luminosity obtained using the derived physical parameters with the Eddington luminosity. If the bolometric luminosity is less than the Eddington luminosity, the disk is sub-Eddington and we obtain the required parameters. If not, we fit with model B to obtain the physical parameters. The Fig \ref{fc} shows the flow chart of the procedure adopted in fitting the model to the observational data. Figs \ref{subobs} and \ref{ps1fit} shows the model B fit to the X-ray observations XMMSL1 J061927.1-655311, SDSS J120136.02+300305.5, Swift J1644+57 and to the optical and UV observations of PS1-10jh with the obtained parameters are shown in Table \ref{ps1tab}.

\begin{table*}
\caption{The physical parameters obtained from reduced chi-square $\chi^2$ fit with models B to the observations with a time shift of $\delta t$, mass of star $m$, star's initial dimensionless energy $\bar{e}$, angular momentum $\ell$ and black hole mass $M_6$ and spin $j$ along with the $\chi$ values are shown above. The model fit to observations XMMSL1 J061927.1-655311, SDSS J120136.02+300305.5 and PS1-10jh are shown in Figs \ref{subobs}a, \ref{subobs}b and \ref{ps1fit} respectively.}
\label{ps1tab}
\scalebox{0.8}{
\begin{tabular}{|c|c|c|c|c|c|c|c|c|c|c|c|c|c|}
\hline
&&&&&&&&&&&&&\\
Model & Observation & Band & $\alpha$ & $\bar{e}$ & $\ell$ & $M_6$ & $m$ & $q$ & $j$ & $w_n$ & $c_2$ & $\delta t$ (days) & $\chi$   \\
\hline
&&&&&&&&&&&&&\\
B &  PS1-10jh & g band & 0 & 0.01 & 1 & 6.8 & 1.0 & 1.119 & 0.4 & 0.101 & 1 & 225 & 4.1 \\
\hline
&&&&&&&&&&&&&\\
B &  PS1-10jh & r band & 0 & 0.01 & 1 & 6.8 & 1.25 & 1.119 & 0.4 & 0.1 & 1 & 28 & 3.66 \\
&&&&&&&&&&&&&\\
\hline
&&&&&&&&&&&&&\\
B & PS1-10jh  & i band & 0 & 0.01 & 1 & 6.8 & 1.1 & 1.119 & 0.4 & 0.091 & 1 & 28 & 3.78\\
&&&&&&&&&&&&&\\
\hline
&&&&&&&&&&&&&\\
B & PS1-10jh  & z band & 0 & 0.01 & 1 & 6.8 & 1.19 & 1.119 & 0.4 & 0.091 & 1 & 29 & 2.1 \\
&&&&&&&&&&&&&\\
\hline
&&&&&&&&&&&&&\\
B &  PS1-10jh & nuv band & 0 & 0.01 & 1 & 6.8 & 1.15 & 1.119 & 0.4 & 0.11 & 0.61 & 27 & 3.8\\
&&&&&&&&&&&&&\\
\hline
&&&&&&&&&&&&&\\
B & Swift J1644+57  & X-ray & 0 & 0.01 &  1 & 1 & 21 & 1.2 & 0.4 & 0.09 & 0.1 & 400 & 3.8 \\
&&&&&&&&&&&&&\\
\hline
&&&&&&&&&&&&&\\
B &  XMMSL1 J061927.1-655311 & X-ray & 0 & 0.001 & 1 & 8 & 8.3 & 5 & 0.5 & 0.44 & 0.1 & 200 & 1.6 \\
&&&&&&&&&&&&&\\
\hline
&&&&&&&&&&&&&\\
B &  SDSS J120136.02+300305.5 & X-ray & 0 & 0.001 & 1 &  7 & 8.3 & 5 & 0.6 & 0.425 & 0.1 &  180 & 1.94\\
&&&&&&&&&&&&&\\
\hline
\end{tabular}
}
\end{table*}

\section{Discussion}
\label{discussion}

We have constructed a self-similar model of the time-dependent and non-relativistic accretion disk with fallback from disrupted debris and viscosity prescription $\Pi_{r\phi}=K \Sigma_d^b r^d$ and derived $b$ and $d$ for an assumed pressure and density structure of the disk. We have considered two models which include the $\alpha$ viscosity in the sub-Eddington regime (model A) and the radiative viscosity in super-Eddington (model B). The radiative viscosity dominates when the radiation pressure is strong for which $\beta_g \ll 1$. The radiative viscosity model for a super-Eddington disk is given in \S \ref{drv}. Using $\delta_p(r)$ given by eqn (\ref{lldp}) in the limit $\delta_p(r) \ll 1$, the physical solution of eqn (\ref{eec1}) for $\omega=v_{\phi}/r$ is given by  

\begin{equation}
\omega(r)=\frac{v_0}{r} \exp\left\{\frac{\delta_0}{2 s} \left(\frac{c^2 r_s}{G M_{\bullet}}\right)^{-s} \left(1-\left(\frac{r}{r_s}\right)^{-s}\right) \right\},
\end{equation}

\noindent where $v_{\phi}(r_s)=v_0$ and $r_s=r_{ISCO}$. For a typical value of $\delta_0=0.1$ and $s=1$ as shown in eqn (\ref{dps}), the range of $\omega$ is given by

\begin{equation}
\omega(r) = \frac{v_0}{r} \left[1 ~~{\rm to}~~\left\{
\begin{array}{ll}
1.0084, & j=0 \\
& \\
1.05, & j=1
\end{array}
\right.  \right]
\label{}
\end{equation}

\noindent which shows that the variation in $\omega(r)$ is extremely small and with increase in $s$, the variation becomes even smaller resulting in $\omega(r) \propto r^{-1}$. We have shown in appendix \ref{drv} that for $\delta_p$ constant, the angular velocity is given by $\omega \propto r^{-e}$, where $e=1+\delta_0/2$ and approaches to unity for $\delta_0 \ll 1$ implying that the rotational velocity of the disk is nearly constant which is consistent with the above result and validates our assumption of constant $\delta_p$.   

\citet{2014ApJ...784...87S} considered the surface density of out flowing wind $\dot{\Sigma}_w \propto \Sigma_d~ \nu /r^2 \propto \Sigma_d~ r^{-3/2}$, which is derived on the basis that for non-radiative advective disks the steady accretion rate is $\dot{M}_a \propto r^s$, where $s$ is constant. We have modeled the super-Eddington disk (model B) with the radiative structure and following vertical momentum equation obtained the surface density of the wind $\dot{\Sigma}_w \propto r^{-7/4}(t/t_0)^{\delta}$ where $\delta=-5/3$.
\citet{2009MNRAS.400.2070S} have developed the wind structure using spherical geometry and fraction of out flowing mass constant. \citet{2011MNRAS.413.1623D} using their porous model with steady accretion disk obtained the fraction of outflow $f_{out}=\dot{M}_w/\dot{M}_a$, numerically which is approximated by a formula in MM15 of the form $f_{out}=(2/\pi) \arctan \left[(1/4.5)(\dot{M}_{a}/\dot{M}_E-1)\right]$ where $\dot{M}_{E}$ is the Eddington mass accretion rate given by $L_E^S/(\eta c^2)$ and $\eta$ is radiative efficiency. In model B, the $f_{out}$ is the function of edge radius $\xi_{out}$ and $\xi_{in}$ given by $\displaystyle{f_{out}=4 (\xi_{out}^{1/4}-\xi_{in}^{1/4})\left[A(p+4-e)/(2-e)\xi_{in}^{p+4-e}+\xi_{in}^{1/4}/(2-e)\right]^{-1}}$ grows as $\xi_{out}/\xi_{in}=r_{out}/r_{in}$ evolves. Thus the mass loss rate by outflows is higher than that of accretion and is even higher than the fallback rate as shown in Fig \ref{r0supt}b which results in reduction in disk mass at late times as shown in Fig \ref{mdjdsupp}. 

\citet{2009MNRAS.400.2070S} have taken the edge radius of the disk to be steady with an inner radius as $r_{ISCO}$ and outer radius to be circularization radius $\sim 2~r_p$. The self-similar model by \citet{2001A&A...379.1138M} and \citet{1990ApJ...351...38C} for sub-Eddington disks have shown that the outer radius increases with time as $r_{out} \propto t^{3/8}$. We have used the mass conservation to derive the evolution of outer radius $r_{out}$ which increases with time for both super and sub-Eddington disks. The evolution of the outer radius is faster for model B due to a smaller timescale of accretion $t_0$. 

The mass and angular momentum of the disk in Model A enhances at late stages whereas it diminishes for model B because of the wind outflow rate which dominates at late times. The previous studies have taken the spherical Eddington luminosity $L_E^S=1.48 \times 10^{44}~M_6~{\rm erg~sec^{-1}}$ and the disk is considered to be sub-Eddington if the luminosity is less than the $L_E^S$. The wind is launched if photosphere temperature is higher than the Eddington temperature which is derived using the vertical momentum equation given by eqn (\ref{vzr}) and the Eddington luminosity can exceed the spherical Eddington value at late times. Therefore Eddington temperature is useful in defining the wind structure locally and we have assumed the ratio $T_{ph}/T_E$ to be the function of time alone so that the entire disk is super-Eddington with the wind as seen in eqn (\ref{tphc}).  

We have considered that the fallback debris forms a seed disk that evolves due to mass loss due to accretion by the black hole and the outflowing wind and mass gain through fallback debris assuming that the self-similar structure of accretion disk remains same. If the matter enters at the outer radius, this induces a high density at the outer radius which results in a jump in the self-similar structure at the outer radius. We have assumed that the added matter is quickly distributed in the disk since the disk evolutionary time scale is longer than the accretion time scale. 

The bolometric luminosity of the model A1 is given by $L_b \propto t^{-1} (\xi_{out}^{3/4}-\xi_{in}^{3/4})$ (see eqn \ref{dislum}) and the net bolometric luminosity at late time is found to be $L_b \propto t^{-0.7}$. The bolometric luminosity obtained for model A2 at the late time is given by $L \propto t^{-1.42}$, and thus, the luminosity in a sub-Eddington disk with gas pressure declines faster than a sub-Eddington disk with total pressure. \citet{1990ApJ...351...38C} obtained the bolometric luminosity of the sub-Eddington accretion disk dominated by gas pressure and without fallback to be $L_b \propto t^{-1.2}$. Thus, the inclusion of mass fallback rate results in a faster decline in the luminosity. \citet{2009MNRAS.400.2070S} and \citet{2011MNRAS.410..359L} have assumed the accretion rate to be $\dot{M} \propto t^{-5/3}$ and for sub-Eddington disk without fallback and constant outer edge radius, they found $L_b \propto \dot{M} \propto t^{-5/3}$ and for a super-Eddington disk with slim disk structure, they obtained the bolometric luminosity from the outflowing wind as $L_b \propto t^{-5/9}$ and from the disk as $L_b \propto t^{-5/3}$. \citet{2011ApJ...736..126M} performed the numerical simulation of an accretion disk without wind with edge radius constant and showed that the bolometric luminosity $L_b \propto t^{-5/3}$ at late stages. Our accretion model without wind given by models A1 and A2 have an evolving outer radius and results in a slow decline in luminosity compared to $L \propto t^{-5/3}$.

The bolometric disk luminosity for model B is $L_b^d \propto t^{-5/3+(3-2e)\alpha} (\xi_{out}^{9/4-2e}-\xi_{in}^{9/4-2e})$ and the outflow luminosity $L_b^w \propto (\mathcal{W}^2(t/t_0)^{(-10/3-\alpha/2)}+1-c_2)^4 {\rm \ln}(r_{out}/r_{in})$ where $\alpha$, $e$ and $c_2$ are considered to be a free parameters. The bolometric luminosity derived in previous TDE models are for tidally captured stars on parabolic orbits. The inclusion of angular momentum and energy of the star in the calculation by MM15  have given a deeper understanding of disruption dynamics by modifying the mass fallback rate of debris and two more free parameters to fit for the observations. We have fit our time dependent accretion models A and B to the observations given in Table \ref{ob} as shown in Fig \ref{subobs} and Fig \ref{ps1fit} with the obtained parameters are given in Table \ref{ps1tab}. 

The ratio of radiation pressure ($P_r= (1/3) a T^4$) to gas pressure ($P_g = k_B/(\mu m_p) \rho T$) in model A2 is given by 

\begin{equation}
\frac{P_r}{P_g}= \frac{3 \kappa}{16 c} \left(\frac{\mu m_p}{k_B}\right)^{\frac{1}{2}} \left(\frac{32}{9} \frac{a c}{\kappa}\right)^{\frac{1}{8}} (G M_{\bullet})^{-\frac{1}{16}} K^{\frac{3}{40}} \Pi_{r\phi}^{\frac{4}{5}} r^{\frac{3}{20}}.
\label{prpg}
\end{equation}

\noindent For steady accretion, the viscous stress is given by $\Pi_{r\phi} = \dot{M}_a \omega(r)/(2 \pi)$ \citep{2001A&A...379.1138M}. By incorporating the viscous stress in eqn (\ref{prpg}), the ratio of pressures is given by 

\begin{equation}
\frac{P_r}{P_g}= 422.2 \left(\frac{\alpha_s}{0.1}\right)^{\frac{1}{10}} M_6^{\frac{1}{10}} \left(\frac{r}{R_s}\right)^{-\frac{21}{20}} \left(\frac{\dot{M}_a}{\dot{M}_E}\right)^{\frac{4}{5}},
\end{equation}

\noindent where $\dot{M}_E = 4 \pi G M_{\bullet} /(\eta c \kappa)$ is the Eddington rate calculated for radiative efficiency $\eta = 0.1$. The radiation pressure dominates for accretion rate close to the Eddington rate and thus the model A2 is self-inconsistent near the Eddington rate. However, the model A2 is applicable for $\dot{M}_a \ll \dot{M}_E$. Our model A1, where the sub-Eddington disk is formulated with total pressure is valid near the Eddington rate and model B is applicable for a super-Eddington disk. \citet{1974ApJ...187L...1L} showed that a sub-Eddington disk with radiation pressure results in viscous stress $\Pi_{r\phi} \propto \Sigma^{-1}$ for a steady disk and such a disk will result in thermal instability (provided the timescale of the instability is lower than the accretion timescale) as disk height $H \propto \Sigma^{-1}$. Our sub-Eddington disk with total pressure given by model A1 has a similar viscous form. The stability analysis for all three models is shown in \ref{stable}. The model A1 is found to be thermally unstable whereas the models A2 and B are thermally stable. But a study of the time evolution of thermal instability of a time-dependent disk will require a detailed time-dependent perturbation analysis that compares the radial infall rate and growth rate of the instability and is beyond the scope of the paper. The time evolution of the mass accretion rate in model A1 at late times decreases as $\dot{M}_a \propto t^{-1/2}$ and the disk will eventually transit to a sub-Eddington disk with gas pressure when $\dot{M}_a \ll \dot{M}_E$. Thus, a super-Eddington disk given by model B evolves with time and transits to the model A1 when the accretion rate is lower than the Eddington rate. With time, the accretion rate in model A1 decreases, and for $\dot{M}_a \ll \dot{M}_E$, the disk transits from model A1 to model A2. We have studied the time-evolution of individual models here and a detailed analysis of disk transition will be taken up in the future. 

\citet{2014ApJ...781...82C} modeled a super-Eddington disk in spherical coordinates using the self-similar form of \citet{2004MNRAS.349...68B}. For a steady accretion disk, they calculated the extent of the radii at an initial time assuming all the debris has fallen to form the accretion disk. However, they have not done detailed modeling of the super-Eddington disk with varying outer radius. Our time-dependent models also include similar mass conservation where along with the mass fallback rate and accretion, we have also included the mass outflowing rate due to the wind and showed that the outer radius increases with time. We have included the radiative viscosity in our momentum equation whereas they do not include viscosity in their momentum equations and focus on calculating the mass accretion rate to obtain the jet luminosity using $L_j \propto \dot{M} c^2$. In our models, we have constructed a more detailed self-similar model including the viscous mechanism to calculate the radiation from the disk due to viscous heating.

The super-Eddington model B is two-dimensional as the density and the temperature of the disk are functions of $r$ and $z$. We have also used the vertical momentum equation to obtain the velocity of the outflow that occurs at the photosphere $z_{ph}$, where the luminosity is calculated.  Our simplistic semi-analytic model of the super-Eddington disk which includes the fallback $\dot{M}_{fb}$, accretion $\dot{M}_a$, and outflow $\dot{M}_w$, is able to produce reasonable fits to the observations as compared to earlier steady accretion models. A more general multi-dimensional model with a detailed radiative transfer which entails intensive high-performance computing will be taken up in the future. However, we expect a multi-dimensional model of the disk to have a structure similar to our models.

\section{Summary}
\label{summary}

We have constructed the self-similar models of the time-dependent and non-relativistic accretion disk in both sub and super-Eddington disk with fallback from disrupted debris. We have obtained the following results.

\begin{enumerate}
	
\item  The formulary for the models A1, A2, and B is presented in Table \ref{modtab}. 
    
\item The parameter $t_0$ is smaller in model B (Fig \ref{t0sup}) compared to models A1 (Fig \ref{t0plot}) and A2 (Fig \ref{t0plotg}). This implies that the accretion process starts earlier in a super-Eddington disk.

\item Using the mass conservation equation, we have found that the outer radius increases for a super-Eddington model (model B) as both the accretion rate and the wind outflow decrease with time whereas the outer radius increases initially and then decreases for the sub-Eddington disk with total pressure (model A1) as the mass accretion rate dominates over fallback at late times. The outer radius increases for a sub-Eddington disk dominated by gas pressure (model A2).
    
\item The disk mass in models A1 and B increases initially and decreases as accretion rate dominates over the mass fallback rate at late times in model A, whereas the mass outflow rate dominates over the fallback rate in model B. The disk mass in model A2 increases and decreases for low and high mass black holes respectively. At late times, the time evolution of disk mass is weak.
    
\item The angular momentum as shown in Fig \ref{mjds} for model A1 and in Fig \ref{mdjdsupp} for model B increases initially and decreases at late stages due to the dominance of mass accretion and outflow rates respectively. The angular momentum of the disk shown in Fig \ref{mjdsg} increases for model A2 as the mass fallback dominates over the mass accretion rate.
    
\item The bolometric luminosity of the model A1 is given by $L_b \propto t^{-1} (\xi_{out}^{3/4}-\xi_{in}^{3/4})$ (see eqn \ref{dislum}) and the net bolometric luminosity at late time is found to be $L_b \propto t^{-0.7}$ (see Fig \ref{lsub}). The bolometric luminosity of model A2 decreases at late time as $L_b \propto t^{-1.42}$ (see Fig \ref{lsubg}). The bolometric disk luminosity for model B is $L_b^d \propto t^{-5/3+(3-2e)\alpha} (\xi_{out}^{9/4-2e}-\xi_{in}^{9/4-2e})$ and the outflow luminosity $L_b^w \propto (\mathcal{W}^2(t/t_0)^{(-10/3-\alpha/2)}+1-c_2)^4 \ln(r_{out}/r_{in})$ where $\alpha$, $e$ and $c_2$ are considered to be a free parameters which are obtained by fit to the observations.
    
\item We have fit our time-dependent accretion model to the X-ray observations of XMMSL1 J061927.1-655311, SDSS J120136.02 + 300305.5 as shown in Fig \ref{subobs}, Swift J1644+57 in Fig \ref{ps1fit}b and in optical and UV observations of PS1-10jh as shown in Fig \ref{ps1fit}a. The best fit physical parameters to the observations are shown in \ref{ps1tab}.
    
\end{enumerate} 

\begin{table*}
\caption{Formulary for the various quantities calculated for models A1, A2, and B. The Eddington luminosity $L_E$ is given by eqn (\ref{edlum}). The viscous stress is given by $\Pi_{r\phi}=K\Sigma_d^b r^d (t/t_0)^{\delta_1}$ where the self similar forms are $\Sigma_d=\Sigma_0 (t/t_0)^{\beta} g(\xi)$ and $\xi=(r/r_0)(t/t_0)^{-\alpha}$ with $g(\xi)=A \xi^p$. The parameters $t_0$ and $\Sigma_0$ are obtained by solving eqns (\ref{diss}, \ref{diss1}).}
	\label{modtab}
\scalebox{0.85}{
	\begin{tabular}{|l|}
		\hline
		{\bf Model A1: sub-Eddington disk with $\alpha$ viscosity and total pressure} \\
		\hline
		\\
		
		$b=-1,~d=0,~\beta=2/3,~\alpha=-2/3,~\delta_1=0,~p=-1/4$, $A=\sqrt{63/4}$ and $\omega=\sqrt{GM_{\bullet}/r^3}$. \\
		\\
		$\Sigma_0=\frac{2+p}{2\pi}\frac{1}{A} \frac{M_d(t_0)}{r_0^2} \frac{1}{\xi_{out}^{2+p}(t_0)-\xi_{in}^{2+p}(t_0)}$, $r_{out}(t_0)=r_0=q r_{in}$ and $r_{in}=r_{ISCO}$.\\
		\\
		$M_d(t_0) \equiv M_d(\bar{e},~\ell,~M_6,~m,~t_0)= M_{\star} \int^{\frac{t_0}{t_m}}_1 \frac{\diff \mu_m}{\diff \tau_m^{'}}\, \diff \tau_m^{'}$,~~~~$\frac{M_d(t_0)}{\sqrt{t_0}}=\frac{64\pi}{(2+p)\sqrt{18}}A \frac{r_0^{7/4}(1-q^{-2-p})}{(G M_{\bullet})^{1/4}}\frac{1-\beta_g}{\sqrt{\alpha_s}} \frac{c}{\kappa}$.\\
		\\
		$M_d(t)=\frac{2\pi}{2+p} A \Sigma_0 r_0^2 \left(\frac{t}{t_0}\right)^{\beta+2\alpha}(\xi_{out}^{2+p}(t)-\xi_{in}^{2+p}(t))$,~~~~ $J_d(t)=\frac{2\pi A}{p+5/2}\sqrt{G M_{\bullet}}\Sigma_0 r_0^{\frac{5}{2}} \left(\frac{t}{t_0}\right)^{\beta+\frac{5}{2}\alpha} \left(\xi_{out}^{p+\frac{5}{2}}-\xi_{in}^{p+\frac{5}{2}}\right)$.\\
		\\
		$\dot{M}_a = 4\pi \frac{\Sigma_0r_0^2}{t_0}\left(\frac{t}{t_0}\right)^{\beta+2\alpha-1}A^b (p~b+2+d)\xi_{in}^{p~b+d+3/2}$, ~~~~$L_b^d=\frac{3\pi G M_{\bullet}}{4}\frac{\Sigma_0 r_0}{t_0}\frac{A^b}{p~b+d+1/2} \left(\frac{t}{t_0}\right)^{\beta+\alpha-1} \left[\xi_{out}^{p~b+d+1/2}-\xi_{in}^{p~b+d+1/2}\right].$\\
		\\
		\hline
{\bf Model A2: sub-Eddington disk with $\alpha$ viscosity and gas pressure} \\
\hline
\\

$b=5/3,~d=-1/2,~\beta=-8/7,~\alpha=5/21,~\delta_1=0,~p=3/2$ and $A=0.0124$. \\
\\
$\Sigma_0=\frac{2+p}{2\pi}\frac{1}{A} \frac{M_d(t_0)}{r_0^2} \frac{1}{\xi_{out}^{2+p}(t_0)-\xi_{in}^{2+p}(t_0)}$, $r_{out}(t_0)=r_0=q r_{in}$ and $r_{in}=r_{ISCO}$.\\
\\
$M_d(t_0) \equiv M_d(\bar{e},~\ell,~M_6,~m,~t_0)= M_{\star} \int^{\frac{t_0}{t_m}}_1 \frac{\diff \mu_m}{\diff \tau_m^{'}}\, \diff \tau_m^{'} $,~~~~$t_0 M_d(t_0)^{b-1}=\frac{\sqrt{GM_{\bullet}}}{K}\left[\frac{2+p}{2\pi A}\right]^{1-b} \frac{r_0^{1/2-d-2(1-b)}}{(\xi_{out}^{2+p}-\xi_{in}^{2+p})^{1-b}}$.\\
\\
$M_d(t)=\frac{2\pi}{2+p} A \Sigma_0 r_0^2 \left(\frac{t}{t_0}\right)^{\beta+2\alpha}(\xi_{out}^{2+p}(t)-\xi_{in}^{2+p}(t))$,~~~~ $J_d(t)=\frac{2\pi A}{p+5/2}\sqrt{G M_{\bullet}}\Sigma_0 r_0^{\frac{5}{2}} \left(\frac{t}{t_0}\right)^{\beta+\frac{5}{2}\alpha} \left(\xi_{out}^{p+\frac{5}{2}}-\xi_{in}^{p+\frac{5}{2}}\right)$.\\
\\
$\dot{M}_a = 4\pi A^b (p~b+2+d) \frac{\Sigma_0r_0^2}{t_0}\left(\frac{t}{t_0}\right)^{\beta+2\alpha-1}\xi_{in}^{p~b+d+3/2}$,~~~~
$L_b^d=\frac{G M_{\bullet}\Sigma_0 r_0}{t_0} \left(\frac{t}{t_0}\right)^{\beta+\alpha-1}\frac{3\pi}{4(p~b+d+1/2)} A^b \left[\xi_{out}^{p~b+d+1/2}-\xi_{in}^{p~b+d+1/2}\right].$\\
\\
\hline
		{\bf Model B: super-Eddington disk with radiative viscosity}\\
		\hline
		\\
		
		$b=1,~d=3-e-s,~\beta=-2/3-\alpha,~\delta_1=-1,~s=1,~p=-7/4,~A=\left(\frac{e-7/4}{2-e}\right)\left[\beta+\frac{7}{4}\alpha-\frac{1}{16}\frac{9-4e}{2-e}\right]^{-1}$ and $\omega=\omega_s (r/r_s)^{-e}$. \\
		\\
		$\mathcal{W}=\frac{3}{\sqrt{8}}\left(\frac{k_B}{\mu m_p}\right)^{\frac{1}{2}} (G M_{\bullet})^{-\frac{7}{8}} a^{-\frac{1}{8}}\kappa^{\frac{7}{8}} \frac{(1-\beta_g)^{1/8}}{\beta_g^{1/2}} \frac{r_0^{7/4}\Sigma_0}{t_0}$,~~~~$\frac{\beta_g^4}{1-\beta_g}=\frac{96e}{a}\delta_0\frac{m_p}{\sigma_T c} \left(\frac{c^2}{GM_{\bullet}}\right)^{-1}\left(\frac{k_B}{\mu m_p}\right)^4 (G M_{\bullet})^{-3} t_0$.\\
		\\
		$\Sigma_0=\frac{2+p}{2\pi}\frac{1}{A} \frac{M_d(t_0)}{r_0^2} \frac{1}{\xi_{out}^{2+p}(t_0)-\xi_{in}^{2+p}(t_0)}$, $r_{out}(t_0)=r_0=q r_{in}$ and $r_{in}=r_{ISCO}$. \\
		\\
		$M_d(t)=\frac{2\pi}{2+p} A \Sigma_0 r_0^2 \left(\frac{t}{t_0}\right)^{\beta+2\alpha}(\xi_{out}^{2+p}(t)-\xi_{in}^{2+p}(t))$,~~~~$J_d(t)=\frac{2\pi A}{p+5/2}\sqrt{G M_{\bullet}}\Sigma_0 r_0^{\frac{5}{2}} \left(\frac{t}{t_0}\right)^{\beta+\frac{5}{2}\alpha} \left(\xi_{out}^{p+\frac{5}{2}}-\xi_{in}^{p+\frac{5}{2}}\right)$.\\
		\\
		$\dot{M}_a=2\pi \frac{\Sigma_0r_0^2}{t_0}\left(\frac{t}{t_0}\right)^{\beta+2\alpha-1} \left[\frac{A^b}{2-e} (p~b+2+d)\xi_{in}^{p~b+d+e}+\frac{\xi_{in}^{1/4}}{2-e}\right]$,~~~~$\dot{M}_w = 8 \pi \frac{\Sigma_0r_0^2}{t_0}\left(\frac{t}{t_0}\right)^{\beta+2\alpha-1} \left(\xi_{out}^{\frac{1}{4}}-\xi_{in}^{\frac{1}{4}}\right)$.\\
		\\
		$L_b^d= \frac{3\pi \omega_s^2 r_s^{2e}}{4}\frac{\Sigma_0 r_0^{4-2e}}{t_0}\frac{A^b}{p~b+d-e+2} \left(\frac{t}{t_0}\right)^{\beta-2\alpha(e-2)-1}\left[\xi_{out}^{p~b+d-e+2}-\xi_{in}^{p~b+d-e+2}\right]$.\\
		\\
		$L_b^w=\left(\mathcal{W}^2 \left(\frac{t}{t_0}\right)^{2\delta}+1-c_2\right)^4 L_E$,~~~~$L_E=\frac{\pi}{2} \frac{G M_{\bullet} c}{\kappa} (1-\beta_g)~ \ln \left(\frac{r_{out}(t)}{r_{in}}\right)$.\\
		\\
		\hline
	\end{tabular}
}
	\end{table*}

We discuss the distinct and useful features of the model below.

\section{Conclusions}
\label{conclusion}

We have been able to construct TDE light curves in both super-Eddington and sub-Eddington regimes by demonstrating that our models A and B are viable and useful as a companion test bed for simulations. The conclusions are listed below.

\begin{enumerate}[1.] 

\item \textit{\underline{Validity of models A1, A2, and B:}} The time-dependent model A1 is valid when the accretion rate is close to the Eddington accretion rate whereas the model A2 is valid when the accretion rate is much smaller than the Eddington rate. The model B is valid when the radiation pressure dominates in the disk ($\beta_g \ll 1$). The model B contains self-consistently the evolution of accretion disk and the wind. 

\item \textit{\underline{Choice of viscosity:}} It is shown that luminosity is higher for a super-Eddington disk with a radiative viscosity over $\alpha$ viscosity when the disk is dominated by radiation pressure ($\beta_g \ll 1$). The sub-Eddington disk with $\alpha$ viscosity, total pressure, and Thomson opacity are valid close to the Eddington rate whereas the $\alpha$ viscosity with gas pressure is valid only when the accretion rate is much smaller than the Eddington rate.

\item \textit{\underline{Validity of the energy flow:}} We have shown in \S \ref{modelB} through Fig \ref{adhrt} that for a self-similar super-Eddington disk, the advection rate is much smaller than the heating rate which implies that energy generated due viscous heating is radiated in form of radiation.

\item
There is likely to be an $\alpha$ viscous sub-Eddington disk which would be operative during the rising phase. The typical time scale of
this phase will be given by either the typical rise time in the $\dot{M}$ profile or the time taken to reach the Eddington luminosity.  
\item
If there is a super-Eddington phase to follow, then the typical time scale would be given by eqn (\ref{taradsup}). The super-Eddington disk will be effected by radiative viscosity with a wind launched from the photosphere.

\item
The TDE physics we have employed include all the essentials of accretion, fall back, and the wind; we
have presented hydrodynamic criteria for the operation of the wind with $v_z^2 \propto (T -T_E)$ given by eqn (\ref{vzr}). We have demonstrated in appendix \ref{vitress} that the timescales of evolution and magnitude of the bolometric luminosity are in good agreement with typical observed values. The detailed fits produced by models A and B in \S \ref{obsfits} produce good $\chi^2$ values; they are better than the $\chi^2$ of the quasi-steady models of MM15. The fit to PS1-10jh can be improved by including model A in the rise part and then matching to model B. The details of the transition conditions is a subject for future work.

\end{enumerate}

While our time-dependent models are reasonably successful in producing fits to the four diverse TDE sources chosen here, we plan on producing fits to a larger sample of light curves available in the literature with a higher resolution search in parameter space. In the fits produced thus far, we have been able to extract the mass of the star and its orbital elements, the black hole mass, and the initial accretion disk radius. The parameter search was limited by numerical resources; in the future, we plan on doing more extensive simulations. The basic paradigm is sufficiently elaborate in terms of essential physics; it also transparently and adequately demonstrates the existence of two different phases with a transition model that produces reasonably good fits.  
In the future, we plan to add an atmosphere to predict the details of the spectrum. One can with large statistics of detection that will soon become available, infer the basic parameters using our models and study the demographics of the black hole mass and stellar properties such as mass and the evolutionary state as a function of redshift (\citealp{2016MNRAS.461..371K},\citealp{2016MNRAS.455..859S}). It is also desirable and possible in the future to add black spin and mass evolution to predict the jet phase. 

We thank the anonymous referee for an insightful and helpful report. We acknowledge support from the DST-SERB grant CRG/2018/003415. We also acknowledge the use of and support from the HPC facility at IIA. We thank the staff of the Vainu Bappu Observatory, Kavalur (Indian Institute of Astrophysics) for the hospitality provided during our visits where some of the work was done.   

\bibliographystyle{elsarticle-num-names} 
\bibliography{reference} 

\begin{appendix} 

\section{Disk with radiative viscosity }
\label{drv}

The radial and vertical momentum equations for $v_{r},~v_{z} \ll v_{\phi}$, are given by

\begin{subequations}
\begin{align}
v_{\phi}^2 &=\frac{1}{\rho}\frac{\partial P}{\partial r}+\frac{G M_{\bullet}r}{(r^2+z^2)^{3/2}}\\
\frac{1}{\rho}\frac{\partial P}{\partial z} &= -\frac{G M_{\bullet}r}{(r^2+z^2)^{3/2}}.
\end{align}
\label{rveq}
\end{subequations} 

\citet{1992ApJ...384..115L} used a radiative viscosity $\eta=(8/27)\epsilon_{\gamma} m_p/(\rho c \sigma_t)$ and obtained the corresponding conservation equation that is given by

\begin{equation}
\eta \left[\left(\frac{\partial v_{\phi}}{\partial r}-\frac{v_{\phi}}{r}\right)^2+\left(\frac{\partial v_{\phi}}{\partial z}\right)^2\right]=\vec{\nabla}\cdot \vec{F},
\end{equation}

\noindent where radiative heat flux $\vec{F}$ for Thomson scattering is given by

\begin{equation}
F=-\frac{c}{n_e \sigma_T}\vec{\nabla}P= -\frac{m_p c}{\sigma_T} \frac{1}{\rho}\nabla P.
\label{rafl}
\end{equation}

Using the Euler equation given by

\begin{equation}
\frac{\partial v}{\partial t}+ v \cdot\nabla v -\frac{v_{\phi}^2}{r} \hat{r}=-\frac{1}{\rho} \nabla P + g,
\end{equation}

\noindent where $g$ is the gravity due to the black hole and the divergence of eqn (\ref{rafl}) is given by

\begin{equation}
\nabla \cdot F= \frac{m_p c}{\sigma_T} \frac{1}{r} \frac{\partial v_{\phi}^2}{\partial r},
\end{equation}

\noindent for a space dependent velocity field. Thus, assuming $\vec{v}\cdot\vec{\nabla} \vec{v}=(v_r \partial_r +v_{\phi} \partial_{\phi}+v_z \partial_{z})\vec{v}$, where $v_r$ and $v_z$ are small and $\partial_{\phi}=0$, the energy conservation equation is given by

\begin{equation}
\eta \left[\left(\frac{\partial v_{\phi}}{\partial r}-\frac{v_{\phi}}{r}\right)^2+\left(\frac{\partial v_{\phi}}{\partial z}\right)^2\right]=\frac{m_p c}{\sigma_T} \frac{1}{r} \frac{\partial v_{\phi}^2}{\partial r}.
\label{eec}
\end{equation}

Using the density structure given in eqn (\ref{densupt}) that is obtained using the pressure $P=\mathcal{K} \rho^{\gamma}$, the $\delta_p=\epsilon_{\gamma}/(\rho c^2)$ for $\gamma=4/3$ is given by

\begin{equation}
\delta_p(r)=\frac{3}{8}\frac{1-\beta_g}{Z(j)}\left(\frac{r}{r_{ISCO}}\right)^{-1},
\label{dps}
\end{equation}

\noindent which is in the range $0.0625-0.375$ for $j=0-1$ and decreases with radius. The disk extent is few times $r_{ISCO}$ initially and the variation in $\delta_p$ is small. 

Using $\eta=(8/27)(m_p c/\sigma_T)\delta_p(r)$, eqn (\ref{eec}) is given by

\begin{equation}
\frac{8}{27} \delta_p(r) \left[\left(\frac{\partial v_{\phi}}{\partial r}-\frac{v_{\phi}}{r}\right)^2+\left(\frac{\partial v_{\phi}}{\partial z}\right)^2\right]=\frac{1}{r} \frac{\partial v_{\phi}^2}{\partial r}.
\label{eec1}
\end{equation}

We consider the solution of eqn (\ref{eec1}) for $v_{\phi}=v_0 (r/r_s)^{-f}$ with $f>0$, assuming the variation in $\delta_p$ to be small which leads to 

\begin{equation}
f(r)=-1+\frac{27}{8\delta_p(r)} \pm \frac{27}{8\delta_p(r)}\sqrt{1-\frac{16\delta_p(r)}{27}}.
\end{equation}

Assuming that the $\delta_p(r)\ll 1$ and neglecting $\mathcal{O}(\delta_p^3)$, we obtain

\begin{equation}
f(r)=-1 \mp 1+\frac{27}{8 \delta_p(r)}\left(1\pm 1\right) \mp \frac{4\delta_p(r)}{27}.
\label{frtt}
\end{equation}

The first solution of eqn (\ref{frtt}) is given by $f(r)=-2-4/(27\delta_p(r))+(27/4)\delta_p(r)$, is very high and negative for small $\delta_p(r)$; so we take the second solution which is a useful one and is given by $f(r)=4 \delta_p(r)/27$. This solution produces $f$ in the range of 0.01 to 0.05 so that $v_0$ is nearly constant. Following \citet{1992ApJ...384..115L}, who assumed 

\begin{equation}
\delta_p = (27/8)\delta_0 (c^2 r/(G M_{\bullet}))^{-s},
\label{lldp}
\end{equation}

\noindent where $\delta_0$ is a constant and applying the radial momentum equation given in eqn (\ref{rveq}) at the mid-plane plane, we integrate it to obtain the density structure 

\begin{equation}
\ln\left(\frac{\rho}{\rho_0}\right) = \left\{
\begin{array}{ll}
s\neq 1: \\
\\
\frac{8}{9 \delta_0 c^2} \left(\frac{c^2 r_s}{GM_{\bullet}}\right)^s\left[\frac{v_s^2}{s-2f}\left\{\left(\frac{r}{r_s}\right)^{s-2f}-1\right\}-\right. \\
\left. \frac{1}{s-1}\frac{GM_{\bullet}}{r_s}\left\{\left(\frac{r}{r_s}\right)^{s-1}-1\right\}\right]+s \ln\left(\frac{r}{r_s}\right)
& \\
\\
s=1:\\
\\
\frac{8}{9 \delta_0 c^2} \left(\frac{c^2 r_s}{GM_{\bullet}}\right)\left[\frac{v_0^2}{1-2f}\left\{\left(\frac{r}{r_s}\right)^{1-2f}-1\right\}-\right. \\
\left. \frac{GM_{\bullet}}{r_s}\ln\left(\frac{r}{r_s}\right)\right]+\ln\left(\frac{r}{r_s}\right) 
\end{array}
\right. 
\label{rcc}
\end{equation}

Since $v_0$ is independent of radius, we take $r_s=r_{ISCO}=(GM_{\bullet}/c^2)Z(j)$ and we assume that $\rho(r_q)=\rho_0$ where $r_{q}=q r_s$, so that $v_0$ is given by 

\begin{equation}
v_0^2 = \left\{
\begin{array}{ll}
s\neq 1: \\
\\
\frac{s-2f}{q^{s-2f}-1}\left[\frac{1}{s-1}\frac{GM_{\bullet}}{r_s}(q^{s-1}-1)-\frac{9\delta_0 c^2}{8}s \left(\frac{G M_{\bullet}}{c^2 r_s}\right)^s\right],
& \\
\\
s=1:\\
\\
\left(1-\frac{9\delta_0}{8}\right)\frac{(1-2f)\ln q}{q^{1-2f}-1} \frac{G M_{\bullet}}{r_s}, 
\end{array}
\right. 
\label{rcc1}
\end{equation}

The range of $\delta_0=(8/27)\delta_p $ is $0.018-0.11$ and the range of $f(r)=(4/27)\delta_p$ is $0.01-0.05$. Thus, we neglect the variation of $v_{\phi}\propto r^{-f(r)}$ in radius and consider it to be a constant. We take $v_{\phi}=\omega(r) r=v_0 (r/r_s)^{-f}$, where $f=\delta_0/2$, $\delta_0=0.05$ and $\omega=\omega_s (r/r_s)^{-e}$, where 

\begin{equation}
\omega_s=\frac{v_0}{r_s}~~~{\rm and}~~~e=1+\frac{\delta_0}{2}.
\end{equation}

\section{Viscous stress }
\label{vitress}

The accretion timescale for a steady flow is given by \citep{2001A&A...379.1138M}

\begin{equation}
t_a \simeq \frac{M_d \omega}{2\pi \Pi_{r\phi}}
\label{acctime}
\end{equation}

\noindent where $M_d$ is the disk mass and the luminosity to be

\begin{equation}
L=\int_{r_{in}}^{r_{out}} Q_{rad}^{-}~ 2\pi r \, \diff r,
\label{lint}
\end{equation}

\subsection{Sub-Eddington disk dominated with total pressure and Thomson opacity (model A1)} 

The viscous stress is given by eqn (\ref{gascon}) and the accretion time, heating flux and the luminosity using eqns (\ref{acctime}, \ref{radlum}), \ref{lint}), are estimated to be 

\begin{equation}
t_{\alpha}^{sub}=8.2 \times 10^6~{\rm yr} \left(\frac{\alpha_s}{0.1}\right)^{-1} M_6^{-1} \left(\frac{M_d}{M_{\odot}}\right) \left(\frac{\Sigma_d}{10^6 ~{\rm g~cm^{-2}}}\right) \cdot \left(\frac{r}{R_s}\right)^{-\frac{3}{2}} (1-\beta_g)^{-2}
\label{tagasub}
\end{equation}

\begin{equation}
Q^{+}=1.2 \times 10^{17} \left(\frac{\alpha_s}{0.1}\right)^{-1} \left(\frac{\Sigma_d}{10^6~{\rm g~cm^{-2}}}\right)^{-1}\left(\frac{r}{R_s}\right)^{-\frac{3}{2}}M_6^{-1} (1-\beta_g)^2~~ {\rm erg~sec^{-1}~cm^{-2}}
\end{equation}

\begin{equation}
L_{\alpha}^{sub}=1.3 \times 10^{40} \left(\frac{\alpha_s}{0.1}\right)^{-1} \left(\frac{\Sigma_d}{10^6~{\rm g~cm^{-2}}}\right)^{-1} M_6 (1-\beta_g)^2 \left(\left(\frac{r_{out}}{R_s}\right)^{\frac{1}{2}}-\left(\frac{r_{in}}{R_s}\right)^{\frac{1}{2}}\right) ~~{\rm erg~ sec^{-1}}
\label{lalpha}
\end{equation}

\subsection{Sub-Eddington disk dominated with gas pressure and Thomson opacity (model A2)}
\label{subgt}

The viscous stress is given by eqns (\ref{gascon1}), (\ref{acctime}), (\ref{radlum}) and (\ref{lint}), the accretion time, heating flux and the luminosity are estimated to be 

\begin{equation}
t_{\alpha}^{sub}=1.9 \times 10^4 ~{\rm yr} \left(\frac{\alpha_s}{0.1}\right)^{-\frac{4}{3}} M_6^{-\frac{2}{3}} \left(\frac{M_d}{M_{\odot}}\right) \left(\frac{\Sigma_d}{10^6 ~{\rm g~cm^{-2}}}\right)^{-\frac{5}{3}} \left(\frac{r}{R_s}\right)^{-1}
\label{tagasub1}
\end{equation}

\begin{equation}
Q^{+}=9.8 \times 10^{17} \left(\frac{\alpha_s}{0.1}\right)^{\frac{4}{3}} \left(\frac{\Sigma_d}{10^6~{\rm g~cm^{-2}}}\right)^{\frac{5}{3}}\left(\frac{r}{R_s}\right)^{-2}M_6^{-\frac{4}{3}}~~{\rm erg~sec^{-1}~cm^{-2}}
\end{equation}

\begin{equation}
L_{\alpha}^{sub}=5.4 \times 10^{41} \left(\frac{\alpha_s}{0.1}\right)^{\frac{4}{3}} \left(\frac{\Sigma_d}{10^6~{\rm g~cm^{-2}}}\right)^{\frac{5}{3}} M_6^{\frac{2}{3}}~ {\rm Log}\left(\frac{r_{out}}{r_{in}}\right)~~{\rm erg~ sec^{-1}}.
\label{lalpha1}
\end{equation}

\subsection{Super-Eddington disk with radiative viscosity (model B)} 

The viscous stress is given by eqn (\ref{radviscosity}) and the accretion time, heating flux and the luminosity using eqns (\ref{acctime}, \ref{radlum}, \ref{lint}), are estimated to be 

\begin{equation}
t_{R}^{sup}= 0.143~{\rm yr} \left(\frac{M_d}{M_{\odot}}\right) \left(\frac{r}{R_s}\right)^{-2} \left(\frac{\Sigma_d}{10^6~{\rm g~cm^{-2}}}\right)^{-1} \left(\frac{\beta_g}{10^{-6}}\right)^4\left(1-\beta_g\right)^{-2},
\label{taradsup}
\end{equation}

\begin{equation}
Q^{+}=1.16 \times 10^{24}\left(\frac{\Sigma_d}{10^6~{\rm g~cm^{-2}}}\right) M_6^{-2} (1-\beta_g)^2 \left(\frac{\beta_g}{10^{-6}}\right)^{-4} Z(j)^{-0.99} \left(\frac{r}{R_s}\right)^{-0.01}~~{\rm erg~sec^{-1}~cm^{-2}},
\end{equation}

\begin{equation}
L_R^{sup}=3.24 \times 10^{47}\left(\frac{\Sigma_d}{10^6~{\rm g~cm^{-2}}}\right)(1-\beta_g)^2 \left(\frac{\beta_g}{10^{-6}}\right)^{-4} Z(j)^{-0.99} \left[\left(\frac{r_{out}}{R_s}\right)^{0.99}-\left(\frac{r_{in}}{R_s}\right)^{0.99}\right]~~ {\rm erg~ sec^{-1}}.
\label{lumr}
\end{equation}

The smaller the value of $\beta_g$, the higher is the dominance of radiation pressure and the luminosity which implies that the radiation pressure dominated disk is more luminous compared to gas pressure dominated disk. 

\section{Other viscosity models}

\subsection{\textit{Sub-Eddington disk dominated with gas pressure with Kramer opacity}}
\label{subgk}

The Kramer's opacity given by

\begin{equation}
\kappa_R=\chi_k \rho T_c^{-\frac{7}{2}},
\label{kropc}
\end{equation}

\noindent where $\chi_k=5 \times 10^{20}~ {\rm m^5~kg^{-2}~K^{7/2}}$ in SI units calculated for $\rho = 1 ~{\rm kg~m^{-3}}$ and $T_c= 1 {\rm K}$ \citep{2002apa..book.....F}.

Using eqns (\ref{visstres}) and (\ref{kropc}), for a disk with pressure $P_g=\rho k_B T_c/(\mu m_p)$ and Thomson opacity, the viscous stress is given by

\begin{equation}
\Pi_{r\phi}= \left(\frac{\alpha_s}{2}\right)^{\frac{8}{7}} \left(\frac{9 \chi_k}{64 a c}\right)^{\frac{1}{7}} \left(\frac{k_B}{\mu m_p}\right)^{\frac{15}{14}} (G M_{\bullet})^{\frac{1}{7}} \Sigma^{\frac{10}{7}} r^{-\frac{3}{7}}.
\label{gaskr}
\end{equation}

Using eqns (\ref{acctime}), (\ref{radlum}) and (\ref{lint}), the accretion time, heating flux and the luminosity are estimated to be 

\begin{equation}
t_{\alpha}^{sub}=3.1 \times 10^4 ~{\rm yr} \left(\frac{\alpha_s}{0.1}\right)^{-\frac{8}{7}} M_6^{-\frac{5}{7}} \left(\frac{M_d}{M_{\odot}}\right) \left(\frac{\Sigma_d}{10^6 ~{\rm g~cm^{-2}}}\right)^{-\frac{10}{7}} \left(\frac{r}{R_s}\right)^{-\frac{15}{14}}
\label{tagakrsub}
\end{equation}

\begin{equation}
Q^{+}=6.2 \times 10^{17} \left(\frac{\alpha_s}{0.1}\right)^{\frac{8}{7}} \left(\frac{\Sigma_d}{10^6~{\rm g~cm^{-2}}}\right)^{\frac{10}{7}}\left(\frac{r}{R_s}\right)^{-\frac{27}{14}}M_6^{-\frac{9}{7}}~~{\rm erg~sec^{-1}~cm^{-2}}
\end{equation}

\begin{equation}
L_{\alpha}^{sub}=4.8 \times 10^{42} \left(\frac{\alpha_s}{0.1}\right)^{\frac{8}{7}} \left(\frac{\Sigma_d}{10^6~{\rm g~cm^{-2}}}\right)^{\frac{10}{7}} M_6^{\frac{5}{7}} \left[\left(\frac{r_{out}}{R_s}\right)^{\frac{1}{14}}-\left(\frac{r_{in}}{R_s}\right)^{\frac{1}{14}}\right]~~{\rm erg~ sec^{-1}}.
\label{lalphakr}
\end{equation}

\subsection{Sub-Eddington disk with total pressure and Kramer opacity}
\label{subdisk}

For a disk with total pressure $P=P_g/\beta_g$ and Kramer's opacity, the viscous stress is given by

\begin{equation}
\Pi_{r\phi}= \alpha_s P H = \left[\frac{3}{128}\frac{\chi_k}{c}\right]^{\frac{16}{7}}(2\alpha_s)^{\frac{23}{7}} \left(\frac{a}{3}\right)^{2} (1-\beta_g)^{-\frac{30}{7}}(G M_{\bullet})^{\frac{1}{7}} \cdot  \nonumber\Sigma^{\frac{25}{7}} r^{-\frac{3}{7}}.
\end{equation}

The height of the disk is given by

\begin{equation}
\frac{H}{r}=\sqrt{\frac{2}{\alpha_s}}\left[\frac{3}{128}\frac{\chi_k}{c}(2\alpha_s)^{\frac{23}{16}}\left(\frac{3}{a}\right)^{-\frac{7}{8}}\right]^{\frac{8}{7}} (1-\beta_g)^{-\frac{15}{7}} (G M_{\bullet})^{-\frac{3}{7}} \Sigma^{\frac{9}{7}} r^{\frac{2}{7}}= 0.014 \left(\frac{\alpha_s}{0.1}\right)^{\frac{8}{7}} M_6^{-\frac{1}{7}} \left(\frac{\Sigma}{10^6~{\rm g~cm^{-2}}}\right)^{\frac{9}{7}} \left(\frac{r}{R_s}\right)^{\frac{2}{7}}.
\end{equation}

Using eqns (\ref{acctime}), (\ref{radlum}) and (\ref{lint}), the accretion time, heating flux and the luminosity are estimated to be 

\begin{equation}
t_{\alpha}^{sub}= 1.85 \times 10^6 ~{\rm yr} \frac{M_d}{M_{\odot}} \left(\frac{\alpha_s}{0.1}\right)^{-\frac{23}{7}} \left(\frac{\Sigma}{10^6~{\rm g~cm^{-2}}}\right)^{-\frac{25}{7}} (1-\beta_g)^{\frac{30}{7}} M_6^{-\frac{5}{7}} \left(\frac{r}{R_s}\right)^{-\frac{15}{14}}
\label{tagasub1}
\end{equation}

\begin{equation}
Q^{+}=6.6 \times 10^{17}\left(\frac{\alpha_s}{0.1}\right)^{\frac{23}{7}}\left(\frac{\Sigma}{10^6~{\rm g~cm^{-2}}}\right)^{\frac{25}{7}}(1-\beta_g)^{-\frac{30}{7}} M_6^{-\frac{9}{7}} \left(\frac{r}{R_s}\right)^{-\frac{27}{14}} ~~{\rm erg~sec^{-1}~cm^{-2}}
\end{equation}

\begin{equation}
L_{\alpha}^{sub}=5.1 \times 10^{41}\left(\frac{\alpha_s}{0.1}\right)^{\frac{23}{7}}\left(\frac{\Sigma}{10^6~{\rm g~cm^{-2}}}\right)^{\frac{25}{7}}(1-\beta_g)^{-\frac{30}{7}} M_6^{\frac{5}{7}} \left[\left(\frac{r_{out}}{R_s}\right)^{\frac{1}{14}}-\left(\frac{r_{in}}{R_s}\right)^{\frac{1}{14}}\right]~~{\rm erg~ sec^{-1}}.
\label{lalpha1}
\end{equation}

\subsection{Super-Eddington disk: $\alpha$ viscous stress}

The $\alpha$ viscous stress for a gas pressure dominated disk is given by

\begin{equation}
\Pi_{r\phi}=\alpha_s P H.
\end{equation}

The $H$ is obtained using eqn (\ref{yphcal}) and $P=\mathcal{K}\rho_0^{4/3}$ with $\mathcal{K}$ given is by eqn (\ref{kcc}) and density $\rho_0=B/r^3$ where $B$ is given below eqn (\ref{densupt}). Thus the viscous stress is given by

\begin{equation}
\Pi_{r\phi}=\frac{\alpha_s}{8} G M_{\bullet} \Sigma r^{-1}.
\label{gassup}
\end{equation}

Taking $\delta_0=0.01$, eqns (\ref{acctime}), (\ref{radlum}) and (\ref{lint}), the accretion time, heating flux and luminosity are estimated to be 

\begin{equation}
t_{\alpha}^{sup}=0.018~{\rm yr}~\left(\frac{M_d}{M_{\odot}}\right)M_6^{-1}\left(\frac{\Sigma_d}{10^6~{\rm g~cm^{-2}}}\right)^{-1} \left(\frac{\alpha_s}{0.1}\right)^{-1} \left(\frac{r}{R_s}\right)^{-0.005} Z(j)^{-\frac{1}{2}}
\label{tagasup}
\end{equation}

\begin{equation}
Q^{+}=2.1 \times 10^{23}\left(\frac{\Sigma_d}{10^6~{\rm g~cm^{-2}}}\right)\left(\frac{\alpha_s}{0.1}\right)M_6^{-1}\left(\frac{r}{R_s}\right)^{-2.005} Z(j)^{-\frac{1}{2}}~~{\rm erg~sec^{-1}~cm^{-2}}
\end{equation}

\begin{equation}
L_{\alpha}^{sup}=2.35 \times 10^{48}\left(\frac{\alpha_s}{0.1}\right) \left(\frac{\Sigma_d}{10^6~{\rm g~cm^{-2}}}\right)  M_6 Z(j)^{-\frac{1}{2}} \left[\left(\frac{r_{in}}{R_s}\right)^{-0.005}-\left(\frac{r_{out}}{R_s}\right)^{-0.005}\right]~~{\rm erg~ sec^{-1}}.
\label{lgsup}
\end{equation}

\section{Self-similar disk solution}
\label{sssolt}

The matter added to the disk at outer radius $r_{out}$ results in a jump in the density at outer radius so that the eqn (\ref{diseqn}) is given by

\begin{equation}
\frac{\partial}{\partial t}(\Sigma_d(r_{out}^{+},~t)-\Sigma_d(r_{out}^{-},~t))=\frac{1}{r}\frac{\partial}{\partial r}\left[\frac{\partial_r(r^2\Pi_{r\phi})}{\partial_r(r^2\omega(r))}\right]_{r_{out}^{-}} +\omega(r)r\frac{\partial}{\partial r}\left[\frac{\dot{\Sigma}_w r}{\partial_r(r^2\omega(r))}\right]_{r_{out}^{-}}+S(r_{out}^{+},~t).
\end{equation}

In the $r_{out}^{-}$ region, $S(r_{out}^{-},~t)=0$ and thus the equation reduces to

\begin{equation}
\frac{\partial}{\partial t} \Sigma_d(r_{out}^{+},~t)=S(r_{out}^{+},~t).
\end{equation} 

The infall mass rate at $r_{out}$ can be written as $\dot{M}(r_{out}^{+},~t)=2\pi \Sigma_d(r_{out}^{+},~t)r_{out}^{+} v_f(r_{out}^{+})=\dot{M}_{fb}$, where $v_f(r_{out}^{+})$ is the infall velocity of the debris. The source function reduces to

\begin{equation}
S(r_{out}^{+},~t)=\frac{s_c}{2\pi A}\frac{\dot{M}_{fb}}{t~r_{out}^{+}~v_f(r_{out}^{+}) }\delta \left(\frac{r}{r_{out}}-1\right);
\end{equation} 

\noindent further, if the infall velocity $v_f(r_{out}^{+})$ is constant and $\dot{M}_{fb}\propto t^{-5/3}$, the source function $S(r_{out}^{+},~t) \propto t^{-8/3}$.

We neglect the jump in density at outer radius by assuming that the matter added is instantaneously (quickly compared to the disk evolution time) distributed on the disk so that the self-similar solution holds at all radii. For $r < r_{out}$, the eqn (\ref{acalt}) reduces to

\begin{equation}
(\beta - \alpha p)A-\frac{A^{b}}{2-e}(p~b+d+2)(p~b+d+e)\xi^{p~(b-1)+d+e-2}- \frac{e-7/4}{2-e}\xi^{-7/4-p} =0,
\label{acal}
\end{equation}

\noindent and for the equation to be $\xi$ independent, $p=(2-e-d)/(b-1)$ and $-7/4$. We consider a seed disk whose initial mass is $M_d(t_0)$, and the mass conservation equation is given by

\begin{equation}
\dot{M}_d=\dot{M}_{fb}-\dot{M}_a-\dot{M}_w,
\label{mcons}
\end{equation} 

\noindent where $M_d$ is the disk mass, $\dot{M}_a$ is the accretion rate onto the black hole and $\dot{M}_w$ is the mass outflow rate leaving the disk. The mass and angular momentum of the disk are 

\begin{equation}
M_d(t) \int_{r_{in}}^{r_{out}}2\pi \Sigma_d r \, \diff r =\frac{2\pi}{2+p} A \Sigma_0 r_0^2 \left(\frac{t}{t_0}\right)^{\beta+2\alpha}(\xi_{out}^{2+p}(t)-\xi_{in}^{2+p}(t)),
\label{masdis}
\end{equation} 

\begin{equation}
J_d(t) =\int_{r_{in}}^{r_{out}}2\pi \Sigma_d r \sqrt{G M_{\bullet}} \, \diff r=\frac{2\pi A}{p+5/2}\sqrt{G M_{\bullet}}\Sigma_0 r_0^{\frac{5}{2}} \left(\frac{t}{t_0}\right)^{\beta+\frac{5}{2}\alpha} \left(\xi_{out}^{p+\frac{5}{2}}-\xi_{in}^{p+\frac{5}{2}}\right),
\end{equation} 

The accretion rate to the black hole $\dot{M}_a=2\pi r \Sigma_d v_r|_{r_{in}}$, and the mass outflow rate are given by 

\begin{equation}
\dot{M}_a = 2\pi \frac{\Sigma_0r_0^2}{t_0}\left(\frac{t}{t_0}\right)^{\beta+2\alpha-1} \left\{
\begin{array}{ll}
{\rm sub-Eddington}:
& \\
\frac{A^b}{2-e} (p~b+2+d)\xi_{in}^{p~b+d+e},\\
& \\
{\rm super-Eddington}:
& \\
\frac{A^b}{2-e} (p~b+2+d)\xi_{in}^{p~b+d+e}+\frac{\xi_{in}^{1/4}}{2-e}, 
\end{array}
\right. 
\label{macceqn}
\end{equation}

\begin{equation}
\dot{M}_w = 2\pi \frac{\Sigma_0r_0^2}{t_0}\left(\frac{t}{t_0}\right)^{\beta+2\alpha-1}\left\{
\begin{array}{ll}
{\rm sub-Eddington}:
& \\
0, \\
& \\
{\rm super-Eddington}:
& \\
4\left(\xi_{out}^{\frac{1}{4}}-\xi_{in}^{\frac{1}{4}}\right),
\end{array}
\right. 
\label{mweqn}
\end{equation}

Using eqns (\ref{heat}, \ref{lint}), the viscous heating rate is given by

\begin{equation}
Q^{+}=\frac{e}{4}\frac{\omega_s^2 r_s^{2e} \Sigma_0 r_0^{2-2e} A^b}{t_0}\xi^{p~b+d-e}\left(\frac{t}{t_0}\right)^{\beta-2\alpha(e-1)-1}.
\label{heatr}
\end{equation}

For a super-Eddington disk, the accretion rate using eqn (\ref{macceqn}) is given by

\begin{equation}
\dot{M}_a=\frac{2 \pi }{2-e} \frac{\Sigma_0 r_0^2}{t_0} \left[\left(\frac{9}{4}-e\right)A+1\right] \left(\frac{t}{t_0}\right)^{\beta+2\alpha-1} \xi^{1/4}.
\end{equation}

Using total pressure $P=\mathcal{K} \rho^{\gamma}$ and using eqn (\ref{densupt}) for mid-plane, the advection rate is given by

\begin{equation}
Q_{adv}^{-}=\frac{3}{16 \pi} \beta_g \frac{G M \dot{M}_a}{r^3}= \frac{3}{8} \frac{\beta_g}{2-e} \frac{G M_{\bullet} \Sigma_0}{r_0 t_0}\left[\left(\frac{9}{4}-e\right)A+1\right] \left(\frac{t}{t_0}\right)^{\beta-\alpha-1} \xi^{-11/4}.
\label{adve2}
\end{equation}

Using eqns (\ref{heatr}, \ref{adve2}), the ratio of advective to heating rate is given by

\begin{equation}
\frac{Q_{adv}^{-}}{Q^{+}}= \frac{3}{2} \frac{\beta_g}{e(2-e)} \frac{G M_{\bullet}}{\omega_s^2 r_s^{2e}} \left[\frac{9}{4}-e+\frac{1}{A}\right] r^{2e-3}.
\label{advisr}
\end{equation} 

The self-similar models for sub-Eddington disk (model A) and super-Eddington disk (model B) are summarized in Table \ref{modtab}. 

\end{appendix}

\section{Stability of the disk}
\label{stable}

Accretion disks with the phases: (1) radiation pressure dominated advective disk, (2) radiation pressure dominated radiative disk and (3) gas pressure dominated radiative disk have been studied earlier \citep{1995ApJ...438L..37A,1999AdSpR..23.1065M}. The heating rate in the disk is higher than the radiative rate in phase 2 that leads to a temperature runaway and thus the disk is thermally unstable \citep{1999AdSpR..23.1065M,2008bhad.book.....K}. The transition between the phases depends on the accretion rate $\dot{M}_a$ and surface density $\Sigma_d$ and they decide the temperature of the disk; when $\dot{M}_a$ decreases with $\Sigma_d$, it becomes thermally unstable (phase 2). A super-Eddington disk is initially in phase 1 where $\dot{M}_a$ and $\Sigma_d$ decrease with time and the disk transits to phase 3 via the unstable phase 2. If there is no addition of mass to the disk, then $\Sigma_d$ decreases and the disk settles into phase 3, else the mass addition will increase $\Sigma_d$ if $\dot{M}_{fb}>\dot{M}_a$, which results in heating and a sufficient rise in $\Sigma_d$ will make the disk to transit to phase 1 \citep{2014ApJ...784...87S}. This is called as the limit cycle where disk oscillates between phase 1 and 3 until the $\dot{M}_{fb}$ reduce below the $\dot{M}_a$ (see Fig 10.4 of \citealp{2008bhad.book.....K}). 

We have three disk models, (1) sub-Eddington disk with $\alpha$ viscosity and total pressure (model A1), (2) sub-Eddington disk with $\alpha$ viscosity and gas pressure (model A2), and (3) super-Eddington radiative viscosity with wind (model B). Using the surface density given by eqn (\ref{ssol}) and the mass accretion rate given by eqn (\ref{macceqn}) is written as

\noindent and

\begin{equation}
\dot{M}_a =\frac{2\pi}{2-e} \frac{\Sigma_0 r_0^2}{t_0} \left[\frac{1}{\Sigma_0 A}\right]^{\frac{p\alpha-\beta+1}{p\alpha-\beta}} \left[\frac{r}{r_0}\right]^{2+\frac{p}{\beta-p\alpha}} \Sigma_d^{\Gamma_1} \left[A^b (p~b+2+d)+\left\{
\begin{array}{ll}
0, & {\rm no~ wind} \\
& \\
1, & {\rm with~ wind} 
\end{array}
\right\}
\right],
\label{dotma}
\end{equation}     

\noindent where $\Gamma_1$ is given by

\begin{equation}
\Gamma_1=\frac{p \alpha - \beta +1}{\alpha p-\beta}.
\end{equation}

\begin{table}
\caption{The self-similar parameters for various accretion models are shown along with $\Gamma_1$ so that the accretion rate $\dot{M}_a \propto \Sigma_d^{\Gamma_1}$.}
\label{thstab}
\scriptsize
\center
\scalebox{1.2}{
\begin{tabular}{|c|c|c|c|c|c|c|}
\hline
&&&&&&\\
Model & $b$ & $d$ & $\beta$ & $\alpha$ & $p$ & $\Gamma_1$  \\
\hline
&&&&&&\\
A1 &  -1 & 0 & 2/3 & -2/3 & -1/4 & -1 \\
&&&&&&\\
A2& 5/3 & -1/2 & -8/7 & 5/21 & 3/2 & 5/3 \\
&&&&&&\\
B & 1 & $2-e$ & -2/3 & 0 & -7/4 & 5/2 \\
&&&&&&\\
\hline
\end{tabular}
}
\end{table}

The disk is said to be thermally stable if accretion rate increases with a surface density which implies $\Gamma_1 > 0$ \citep{1974ApJ...187L...1L,2014ApJ...784...87S}. The slope $\Gamma_1$ shown in Table \ref{thstab} is positive for the models A2 and B, and is negative for model A1. This implies that the disk is thermally stable in the case of models A2 and B, but is thermally unstable in model A1. This is valid only for steady accretion disks. For a time-dependent disk, the thermal instability of the disk depends on the accretion rate and growth rate of instability. The time evolution of thermal instability will require a detailed analysis that will be taken in the future.

\end{document}